\documentclass[a4paper,11pt]{article}
\pdfoutput=1 

\usepackage{jcappub} 
\usepackage{natbib}
\usepackage{graphicx,color}
\usepackage[T1]{fontenc} 
\title{Interpretation of AMS-02 electrons and positrons data 
\footnote{LAPTH--008/14, SACLAY--T14/010}}

\author[a,b,c]{M. Di Mauro,}
\author[a,b]{F. Donato,}
\author[a,b]{N. Fornengo,}
\author[d]{R. Lineros,}
\author[a,b,e]{and A. Vittino}

\affiliation[a]{Dipartimento di Fisica, Universit\`a di Torino, via P. Giuria 1, 10125 Torino, Italy}
\affiliation[b]{Istituto Nazionale di Fisica Nucleare, Sezione di Torino, Via P. Giuria 1, 10125 Torino, Italy}
\affiliation[c]{Laboratoire d'Annecy-le-Vieux de Physique Th\'eorique (LAPTh), Univ. de
Savoie, CNRS, B.P.110, Annecy-le-Vieux F-74941, France}
\affiliation[d]{AHEP Group, Institut de F\'isica Corpuscular, C.S.I.C. Universitat de Val\`encia, Edificio Institutos de Paterna, Apt 22085, E-46071 Valencia, Spain}
\affiliation[e]{Institut de Physique Th\'eorique, CNRS, URA 2306 \& CEA/Saclay, F-91191 Gif-sur-Yvette, France}

\emailAdd{mattia.dimauro@to.infn.it}
\emailAdd{donato@to.infn.it}
\emailAdd{fornengo@to.infn.it}
\emailAdd{rlineros@ific.uv.es}
\emailAdd{vittino@to.infn.it}

\abstract{We perform a combined analysis of the recent AMS-02 data on electrons, positrons, electrons plus positrons and positron fraction, in a self-consistent framework where we realize a theoretical modeling of all the astrophysical components that can contribute to the observed fluxes in the whole energy range. The primary electron contribution is modeled through the sum of an average flux from distant sources and the fluxes from the local supernova remnants in the Green catalog. 
The secondary electron and positron fluxes originate from interactions on the interstellar medium of primary cosmic rays, for which we derive a novel determination by using AMS-02 proton
and helium data. Primary positrons and electrons from pulsar wind nebulae in the ATNF catalog 
are included and studied in terms of their most significant (while loosely known) properties and
under different assumptions  (average contribution from the whole catalog, single dominant pulsar, a few dominant pulsars).
We obtain a remarkable agreement between our various modeling and the AMS-02 data for all 
types of analysis, demonstrating that the whole AMS-02 leptonic data admit a self-consistent interpretation in terms of astrophysical contributions.}

\begin{document}
\maketitle
\flushbottom

\section{Introduction}
\label{sec:intro}
A huge experimental effort undertaken in the last decades has led to
increasingly accurate measurements
of cosmic rays (CRs) by means of space borne detectors.
In particular, excellent data have been provided
for the positron fraction ($e^+/(e^+ + e^-)$) and for the absolute positron
($e^+$),  electron ($e^-$) and total
($e^+ + e^-$) CR spectra by the Pamela
\citep{2009Natur.458..607A,2013arXiv1308.0133P,2011PhRvL.106t1101A}
and {\it Fermi}-LAT \citep{2012PhRvL.108a1103A,2010PhRvD..82i2004A}
Collaborations.
Very recently, also the AMS-02 Collaboration has provided its first data
on the positron fraction spectrum
measured by the Alpha Magnetic Spectrometer installed on the International
Space Station
\citep{Aguilar:2013qda}, and preliminary results for the other leptonic
observables \citep{electrons_AMS02,totalelectrons_AMS02}.
The data cover an energy range spanning from about one GeV up to hundreds of
GeV, depending on the species.

The most direct interpretation of these data refers to secondary
production from nuclear
collisions with primary CRs and the atoms of the interstellar medium (ISM), as well as to
galactic astrophysical sources injecting
fresh primary leptons into the ISM (see e.g. Ref. \cite{2012APh....39....2S} and references therein, and Refs.
\cite{Hooper:2008kg,Profumo:2008ms,2009AA...501..821D,2009PhRvL.103h1104M,2010AA...524A..51D,
2013arXiv1312.3483D,2013APh....49...23E,Gaggero:2013nfa,Gaggero:2013oga,Gaggero:2013rya,Grasso:2013nyx,Yuan:2013eba,Cholis:2013lwa}).
The observed raise of the positron fraction, firstly reported by the
Pamela Collaboration
and confirmed with higher precision by the AMS-02 data, has stimulated
an extensive speculation on a possible dark matter (DM) contribution at
high energies
\citep{2013arXiv1305.0084I,2013arXiv1307.6204B,2013arXiv1312.7841S,2013arXiv1309.2570I,2013PhRvD..88j3509D,2014PhLB..728...58H,Cirelli:2008pk,Zhao:2014nsa,Baek:2014awa,Kopp:2014tsa,Hryczuk:2014hpa,Choi:2013oaa,Ibarra:2013cra,Bergstrom:2013jra,Jin:2013nta,DeSimone:2013fia,Feng:2013zca,Masina:2013yea}.

In this paper, we explore at which level the new AMS-02 data on the whole
leptonic observables
may be accommodated in a purely astrophysical ($i.e.$ without invoking
contributions from exotic sources such as DM
annihilation) scenario, which counts the contribution from powerful
stellar sources as well as
from secondary reactions among primary CRs and atoms in the ISM.
Specifically, we study the possible contribution that supernova remnants
(SNRs) can give to high energy
electron fluxes, and pulsar wind nebulae (PWN) to both positron and
electron spectra.
Whenever available, independent data on these sources are taken into
account as constraints
on their emission properties.
We also employ the new preliminary AMS-02 data for proton and helium
cosmic fluxes,
in order to obtain a new evaluation of the secondary  $e^+$ and $e^-$ fluxes.
All these components are added together and propagated in the Milky Way.
A key point of our analysis is the requirement for any theoretical model
to fit
{\it simultaneously} all the four AMS-02 leptonic observables, namely
the $e^+/(e^+ + e^-)$, $e^+$, $e^-$ and $e^+ + e^-$ spectra.
As we will show in next Sections, we find several astrophysical models
compatible
with all the AMS-02 leptonic data. Our results imply that a consistent
and global picture of the Galaxy is possible, at least for the leptonic
sector.
Even more so, we show that high precision, low energy positron data, such
as the ones collected by AMS-02,
are on the verge of acting as a remarkable tool to constrain propagation
models,
and cooperate with the boron-to-carbon observations to this fundamental task.
Our analyses are all implemented by a thorough estimation of the underlying
possible uncertainties. This method allows us to derive  predictions
on the AMS-02 data at very high energy, expected after some increase in
the collected statistics.
As a final remark, we notice that our analysis indicates that there is no
particular need to invoke DM annihilation in the halo of the Milky Way in order to explain
high energy positron and electron data.

\section{Sources of galactic positron and electrons}\label{sec:sources}

As a first step of our analysis, we describe and model the various possible galactic sources 
for positrons and electrons. We can, as usual, identify two main categories: primary production, which refers to
electrons and positrons directly
injected in the galactic medium from astrophysical sources, like PWN and SNR; secondary production, which refers to electrons and positrons produced from a spallation reaction of a progenitor cosmic ray in the Galaxy.

\subsection{Primary electrons from SNR}
\label{sec:SNR}
SNR in our Galaxy are believed to be the major 
accelerators of charged particles up to
very high energies (at least $100$ TeV), via a first-type Fermi mechanism 
\cite{2007ApJ...661..879E,2009AA...499..191T,2013MNRAS.434.2748C,Blasi:2013rva}. 
Among accelerated species there are also electrons.
The mechanism of acceleration of cosmic rays  through non-relativistic expanding shock-waves,
activated by the star explosion, predicts power-law spectra with a cut-off at 
high energies:
\begin{equation}
     \label{Q}
         Q(E)=Q_0 \left(\frac{E}{E_0}\right)^{-\gamma}\exp{\left(-\frac{E}{E_c}\right)},
    \end{equation}
where $Q_0$ is the normalization of the spectrum, $\gamma$ is the power-law index, $E_c$ is 
the cut-off energy which in our analysis we fix at $E_c=2$ TeV, and $E_0=1$ GeV is just a reference
value. 
The values for the spectral index $\gamma$ for electrons are typically found around 2 \citep{2007ApJ...661..879E,2012SSRv..173..369H}, although they exhibit significant variations in analyses of radio data. Radio and gamma-ray observations also indicate that $E_c$ might 
be in the TeV range (see e.g. Ref.~\cite{2008PhRvL.101z1104A,2009ApJ...692.1500A,2010ApJ...714..163A,Aharonian:2001mz,Aharonian:2006ws}).
The value of $Q_0$ is by far non trivial to fix, but can in principle be estimated 
from radio data on  single sources, assuming that the radio flux $B_r^\nu$ 
at a specific frequency $\nu$ is entirely due to synchrotron emission of the ambient 
electrons in the SNR magnetic field $B$ \cite{2010AA...524A..51D,2013PASJ...65...69S,2014AA...561A.139S}:
\begin{equation}
     \label{Q0n}
         Q_0 = 1.2\cdot10^{47}\; {\rm GeV}^{-1}\; (0.79)^{\gamma} \left[\frac{d}{\rm{kpc}}\right]^2 \left[\frac{\nu}{\rm{GHz}}\right]^{(\gamma-1)/2}  \left[\frac{B}{100 \mu \rm{G}}\right]^{-(\gamma+1)/2}
	\left[\frac{B_r^{\nu}}{\rm{Jy}}\right]
    \end{equation}
where $d$ is the distance of the source from the observer. The well-known relation between the radio and electron flux index 
$\alpha_r = (\gamma -1)/2$ is here manifest. 

The most complete SNR catalog is provided by the Green catalog \cite{2009BASI...37...45G},  where 274 SNRs are listed. 
Among them, 88 objects have a distance measurement, and  209 have an observed radio spectral index. 
Following the procedure described in  Ref.~\cite{2010AA...524A..51D}, we can determine the average values for  the relevant parameters for those
88 SNRs with a clear distance information. We obtain: $\langle\alpha_r\rangle = 0.50 \pm 0.15$,  $\langle d^2 B_r^{1\rm{ Ghz}}\rangle = \exp{(7.1 \pm 1.7)}$ Jy kpc$^2$. From these results we then infer: 
$\langle\gamma\rangle = 2\cdot\langle\alpha_r\rangle +1 = 2.0 \pm 0.3$. Moreover, fixing a typical magnetic field of $B=30\,\mu$G (adopted in the 
following of our analysis) 
\cite{2000MNRAS.314...65L,2006ApJ...648L..33V,2009ApJ...692.1500A,2010ApJ...718..348A,2013MNRAS.431..415B,2013MNRAS.435.1174S,2009MNRAS.392..240M,2013IJMPS..23...82H,2014AA...561A.139S}, 
and employing Eq.~(\ref{Q0n}), we estimate  $\langle Q_0\rangle = 9.0\times10^{49}$ GeV$^{-1}$ for an index $\gamma = 2$, which implies a total emitted energy of
$\langle E_{\star}\cdot f \cdot{\Gamma_{\star}}\rangle = 8.9\times10^{50} \rm{\,GeV} = 1.4\times10^{48}$ erg (for a cut-off energy $E_c = 2$ TeV), 
where ${\Gamma_{\star}}\approx[2,4]$ is the SN explosion rate \cite{1998MNRAS.297L..17M,2006Natur.439...45D}, $E_{\star}$ is the kinetic energy released by the explosion, and $f\approx[10^{-5},10^{-4}]$ \cite{2009AA...499..191T} is the fraction of this energy
converted into electrons.

For the purposes of the analysis discussed in the next Sections, we divide the SNR population into a {\it near} component,  for sources lying at distances $d\leq3$ kpc from the Earth, and a {\it far} component, for sources located outside this region. For a discussion on the
choice of the separation distance between far and close sources, see Ref. \cite{2010AA...524A..51D}. In the catalog we find 41 near-SNRs, out of which only 35 have a measured distance, age, radio flux and spectral index. 
Therefore only these 35 sources have been taken into consideration in our analysis.
These sources are listed in Tab.~\ref{tab:snrn}, where we report  their characteristics: together with the  Green-catalog name and  (when available) the
association name, we list the radio flux at 1 GHz $B_r^{1\rm{ Ghz}}$, 
the radio index $\alpha_r$, the distance $d$ and the age $T$.
As done in Ref.~\cite{2010AA...524A..51D} (note that the critical distance, which separates
near from far SNR, is now assumed to be 
3 kpc instead of 2 kpc),  the near-SNRs are considered as single, independent sources, with their typical parameters 
fixed to the ones reported in Tab.~\ref{tab:snrn} or derived via Eq.~(\ref{Q0n}). 
The far-SNR population is instead treated as an average source population, with typical parameters
($Q_0$ and $\gamma$) fixed according to the analysis of Sect.~\ref{sec:fit}, and following the radial profile derived in Ref.~\cite{2004IAUS..218..105L}.

\begin{table}[t]
\center
\begin{tabular}{|c|c|c|c|c|c|c|}
\hline
Green	&   	Association  & $B_r^{1\rm{ GHz}} [\rm{Jy}]$&      $\alpha_r$	   		&   	
$d$ $[\rm{kpc}]$	&	$T$ $[\rm{kyr}]$  & 	 Refs.		\\
\hline
G006.4-00.1	 & 	W28	 	& 	287$\pm$27   	&   	-0.35	    	 		&	2.0$\pm$0.4	 	&	[33,150] 	
&  \cite{2009BASI...37...45G,1976MNRAS.174..267C,2002AJ....124.2145V}	  	\\
G018.9-01.1	 &		 	& 	37$\pm$2 		&    -0.39$\pm$0.03 		&	2  	 	 		&	[4.4,6.1]
&  \cite{2009BASI...37...45G,2011AA...536A..83S,1989AA...209..361F,1997AA...319..655F,2004ApJ...603..152H}	\\
G034.7-00.4    &	W44	 	&	213$\pm$11       	&	-0.33$\pm$0.05  		&	3.0  	 			&	[10,20]		
&  \cite{2009BASI...37...45G,2011AA...536A..83S,1989MNRAS.238..737G,1997AJ....113.1379G,1975AA....45..239C,
1991ApJ...372L..99W,1994ApJ...430..757R}	\\
G065.3+05.7	&			&	52$\pm$2  	    	&	-0.58$\pm$0.07  		&	0.9$\pm$0.1  		 &	26	
&  \cite{2009BASI...37...45G,2009AA...503..827X,1996ApJ...458..257G,2002AA...388..355}	\\
G065.7+01.2   &	DA495	&	4.88$\pm$0.25   	&	-0.57$\pm$0.01  		&	[1.0,1.8]	 	&	20	
&  \cite{2009BASI...37...45G,2011AA...536A..83S,2004ApJ...607..855K,2008ApJ...687..516K}	\\
G069.0+02.7   &	CTB80	&	60$\pm$10      	&	-0.36$\pm$0.02 		&	2  		 		&	20	
&  \cite{2005AA...440..171C,2006AA...457.1081K,2011AA...536A..83S,2000ASPC..202..509S,2003AJ....126.2114C}	\\
G074.0-08.5   &	Cygnus loop& 	175$\pm$30       	&	-0.40$\pm$0.06  		&	$0.58\pm0.06$  &	10	
&  \cite{2009BASI...37...45G,2006AA...447..937S,2005AJ....129.2268B,2009ApJ...692..335B}	\\
G078.2+02.1   &	DR4        	&	275$\pm$25       	&	-0.75$\pm$0.03 		&	1.5$\pm$0.1  		&	7	
&  \cite{2009BASI...37...45G,2008AA...490..197L,2004AA...427L..21B,2003AA...408..237M}	\\
G082.2+05.3   &	W63		&	105$\pm$10       	&	-0.36$\pm$0.08 		&	[1.6,3.3] 	 	&	[13,27]		
&  \cite{2009BASI...37...45G,2004AA...415.1051M,2006AA...457.1081K,1981RMxAA...5...93R}	\\
G089.0+04.7   &	HB21	&	200$\pm$15       	&	-0.27$\pm$0.07  		&	1.7$\pm$0.5  	 	&	5.6$\pm$0.3
&  \cite{2009BASI...37...45G,2006AA...457.1081K,2003AA...408..961R,2006ApJ...637..283B,2006ApJ...647..350L}\\
G093.3+06.9   &	DA530	&	7.0$\pm$0.5        	&	-0.45$\pm$0.04  		&	2.2$\pm$0.5  	 	&	[5.2,6.6]	
&  \cite{2009BASI...37...45G,2006AA...457.1081K,1999ApJ...527..866L,2003ApJ...598.1005F}\\
G093.7-00.2   &	DA551	&	42$\pm$7    		&	-0.52$\pm$0.12  		&	1.5$\pm$0.2  	 	&	[29,74]	
&  \cite{2009BASI...37...45G,2006AA...457.1081K,2002ApJ...565.1022U,1982AA...105..176M}\\
G109.1-01.0   &	CTB109	&	20.2$\pm$1.1      	&	-0.45$\pm$0.04  		&	3.0$\pm$0.5  	 	&	[13,17]
&  \cite{2009BASI...37...45G,2006AA...457.1081K,2011AA...536A..83S,2002ApJ...576..169K,1998AA...330..175P,1981ApJ...246L.127H}\\
G113.0+0.2	&			&	3.8$\pm$1.0    	&	-0.45$\pm$0.25		&	3.1	 		 	&	20	
&  \cite{2009BASI...37...45G,2011AA...536A..83S,2005AA...444..871K}\\
G114.3+00.3	&			&	6.4$\pm$1.4     	&	-0.49$\pm$0.25  		&	0.7		  	 	&	7.7	
&  \cite{2009BASI...37...45G,2006AA...457.1081K,2013IJMPS..23...82H,2004ApJ...616..247Y}\\
G116.5+01.1 	&			&	10.9$\pm$1.2      	&	-0.16$\pm$0.11  		&	1.6	 		 	&	[15,50]	
&  \cite{2009BASI...37...45G,2006AA...457.1081K,2013IJMPS..23...82H,2004ApJ...616..247Y}\\
G116.9+00.2   &	CTB 1	&	7.9$\pm$1.3        	&	-0.33$\pm$0.13  		&	1.6	  		 	&	[15,50]	
&  \cite{2009BASI...37...45G,2011AA...536A..83S,2006AA...457.1081K,2013IJMPS..23...82H,2004ApJ...616..247Y}\\
G119.5+10.2   &	CTA 1	&	42$\pm$3  		&	-0.57$\pm$0.06  		&	1.4$\pm$0.3  	 	&	[5,15]	
&  \cite{2009BASI...37...45G,1997AA...324.1152P,1993AJ....105.1060P}\\
G127.1+00.5   &     R5		&	12$\pm$1          	&	-0.43$\pm$0.10 		&	1.0$\pm$0.1  	 	&	[20,30]
&  \cite{2009BASI...37...45G,2006AA...457.1081K,2013IJMPS..23...82H,2006AA...451..251L,1989AA...219..303J}\\
G130.7+03.1   &	3C58	&	35$\pm$3          	&	-0.07$\pm$0.02 		&	3.0$\pm$0.2  	 	&	[2.7,5.4]	
&  \cite{2009BASI...37...45G,2011AA...536A..83S,2006AA...457.1081K,2008ApJS..174..379F,2008AA...486..273S}\\
G132.7+01.3   &	HB3		&	36$\pm$3	    	&	-0.59$\pm$0.14  		&	2.2$\pm$0.2  	 	&	30	
&  \cite{2009BASI...37...45G,2006AA...457.1081K,2008AA...487..601S,2006ApJ...647..350L,2006Sci...311...54X}\\
G156.2+05.7	&			&	5.0$\pm$0.8        	&	-0.53$\pm$0.17  		&	1.0$\pm$0.3  		 &	[15,26]
&  \cite{2009BASI...37...45G,2006AA...457.1081K,1992AA...256..214R,2009PASJ...61S.155K,2007AA...470..969X,1999PASJ...51...13Y}\\
G160.9+02.6   &	HB9		&	88$\pm$9        	&	-0.59$\pm$0.02  		&	0.8$\pm$0.4  		 &	[4,7]
&  \cite{2009BASI...37...45G,2006AA...457.1081K,2013IJMPS..23...82H,2003AA...408..961R,2007AA...461.1013L}\\
G180.0-01.7   &	S147		&	74$\pm$12        	&	-0.30$\pm$0.15  		&	1.47$^{+0.42}_{-0.27}$   &	[30,40]	
&  \cite{2009BASI...37...45G,2003AA...408..961R,2008AA...482..783X,1996ApJ...468L..55A,2007ApJ...654..487N,2003ApJ...593L..31K}\\
G184.6-05.8   &	Crab	&	1040  	    		&	-0.3  				&	2.0$\pm$0.5  	 &	[6,9]	
&  \cite{2009BASI...37...45G,2006AA...457..899A}\\
G189.1+03.0   &	IN 443	&	160$\pm$5       	&	-0.36$\pm$0.04	 	&	1.5$\pm$0.1 	 &	[20,30]	
&  \cite{2009BASI...37...45G,2003AA...408..961R,2013IJMPS..23...82H,2003AA...408..545W,2008AJ....135..796L}\\
G205.5+00.5   &	Monoceros&	156$\pm$20       	&	-0.47$\pm$0.06  		&	1.63$\pm$0.25  &	[29,150]	
&  \cite{2009BASI...37...45G,1982AA...109..145G,2009AN....330..741B,2009MNRAS.394.2127B,1982AA...109..145G,2001AA...372..516W}\\
G260.4-03.4   &	Puppis A	&	137$\pm$10       	&	-0.52$\pm$0.03 		&	2.2$\pm$0.3  	&	3.7	
&  \cite{2009BASI...37...45G,2006AA...459..535C,1988AAS...75..363D,1995AJ....110..318R,1988srim.conf...65W}\\
G263.9-03.3   &	Vela(XYZ)	&	2000$\pm$700	&	variable 				&	2.94$^{+0.76}_{-0.50}$	& 11.3	 
&  \cite{2009BASI...37...45G,2009BASI...37...45G,2001AA...372..636A,1999ApJ...515L..25C,2001ApJ...561..930C,
1993ApJS...88..529T,2008ApJ...676.1064M}\\
G266.2-01.2   &	Vela Jr	&	$\pm$4  	 	&	-0.3  				&	0.75  	 &	[1.7,4.3]	
&  \cite{2009BASI...37...45G,2008ApJ...678L..35K,2005MNRAS.356..969R,1998Natur.396..141A,1998Natur.396..142I}\\
G315.1+02.7	&			&	35$\pm$6	   	& 	-0.7  				&	1.7  		 &	[40,60]	
&  \cite{2009BASI...37...45G,1997MNRAS.287..722D,2007MNRAS.374.1441S}\\
G315.4-02.3   &	RCW 86	&	49  	    			&	-0.61  		&	2.3$\pm$0.2  		 &	10	
&  \cite{2009BASI...37...45G,2006ApJ...648L..33V,2001ApJ...546..447D,1996AA...315..243R,2003AA...407..249S}\\
G327.6+14.6   &	SN1006	&	16$\pm$2         	&	-0.6  				&	2.2$\pm$0.1  		 &	[0.9,1.3]	
&  \cite{2009BASI...37...45G,1988ApJ...332..940R,2006ApJ...640L..55K,2003ApJ...585..324W,2008PASJ...60S.153B}\\
G330.0+15.0   &	Lupus loop&	350		    		&	-0.5 				&	1.2$\pm$0.4  		 &	[20,50]  
&  \cite{2009BASI...37...45G,1991ApJ...374..218L,2006ApJ...644L.189S}\\
G347.3005      &			 &	4$\pm$1	    		&	-0.3 				&	1 		 &	[1.6,4.9]  
&  \cite{2009BASI...37...45G,2004ApJ...602..271L,2009MNRAS.392..240M}\\
\hline
\end{tabular}
\caption{Characteristic parameters for our sample of near ($\leq3$ kpc) SNRs: the columns report the Green-catalog name, the association name, the radio flux at 1 GHz $B_r^{1\rm{ GHz}}$, the radio index $\alpha_r$, the distance $d$ $[\rm{kpc}]$ and
 the SNR age $T$ $[\rm{kyr}]$.}
\label{tab:snrn}
\end{table}

\subsection{Primary electrons and positrons from PWN}
\label{sec:pulsar}
Pulsars, rapidly spinning neutron stars with a strong surface magnetic field,
are considered to be among the most powerful sources of electrons and positrons in the Galaxy
\citep{1970ApJ...162L.181S,1987ICRC....2...92H,1996SSRv...75..235A,2001AA...368.1063Z,Amato:2013fua,2013arXiv1312.3483D}. 
It is believed that the rotating magnetic field of the pulsar generates an intense 
electric field that can tear particles apart from the neutron star surface. 
These charged particles can then be accelerated and induce an electromagnetic cascade through 
  the emission of curvature radiation that, in turns, produces again particle/antiparticle pairs
  \cite{Ruderman:1975ju,1976ApJ...203..209C,Cheng:1986qt}.  
The star, and the wind of charged particles that surrounds it, are initially located inside the 
remnants of the supernova explosion that creates the pulsar. The impact of the relativistic wind 
produced by the pulsar on the much slower ejecta of the supernova usually creates a reverse shock 
(the so-called \emph{termination shock}) that propagates backwards, towards the pulsar 
\cite{1974MNRAS.167....1R}. 
In the region bound by the wind termination shock on one side and the 
ejecta on the other side, a bubble of relativistically hot magnetized plasma is created: this 
is the so-called pulsar wind nebula (PWN). The termination shock is also the place where the 
incoming pairs are accelerated to very high energies. After acceleration, these particles 
enter the PWN and then are trapped by the PWN magnetic field until it is disrupted. What is usually assumed is 
that the accelerated particles are completely released into the interstellar medium (ISM) after a time
 of about 50 kyr from the nebula formation. As stressed in Ref.~\cite{Grasso:2009ma},  since this injection is assumed to be quite fast 
 and the subsequent energy emission of the pulsar negligible, a mature pulsar can be treated as a burst-like source of $e^{\pm}$.
The emitted leptons can then reach the Earth with huge Lorentz factors (see, e.g., Ref.~\cite{1996MNRAS.278..525A} for the Crab Nebula). 
\\  
In order to determine the flux of  emitted electrons and positrons by a pulsar, we follow the model described in Ref.~\cite{2010AA...524A..51D} 
(and references therein) and remind here only the main ingredients relevant to our analysis. 
Nevertheless, we remark here that the actual process through which the electrons are injected from the PWN into the ISM is only 
very little known. As explained in Ref.~\cite{2009PhRvD..80f3005M}, the spectrum of electrons and positrons trapped
inside the PWN can be inferred by observing their broadband  emission which is due to synchrotron radiation (at low energies) 
and to inverse Compton (IC) scattering off background photons (at higher energies). This broadband spectrum shows a break between the radio 
and X-ray regimes which is believed to be the result of synchrotron cooling. However, even the time evolution of the electrons spectrum
inside the PWN is not known: this  means that the snapshot picture that one can derive from the observation of the broadband spectrum of the 
emitted radiation is not necessarily representative of the electron spectrum that 
eventually reaches the Earth. This is the reason for the large uncertainty that 
surrounds the parameters related to the $e^{\pm}$ flux produced by a PWN. 

For the computation of the flux of  $e^{\pm}$ emitted by pulsars, we consider a source spectrum of the same 
form as the one in Eq.~(\ref{Q}). As for the SNRs, the cutoff energy $E_c$ is expected to be in the TeV range (see Refs. \cite{Aharonian:2006xx,Atoyan11011996}); we fix $E_c=2$ TeV for most of our analysis of the PWN  (we will comment in Sect.~\ref{ecut} about the effect due to variation of the cut-off energy). For the spectral index, which we label $\gamma_{\rm PWN}$, we expect a value slightly smaller than 2 ({\it i.e.} in the range [1.3 - 2]) in agreement with the mean spectral index of the gamma-ray pulsars listed in the FERMI-LAT catalog \cite{TheFermi-LAT:2013ssa}.
The normalization of the spectrum, $Q_0$ can be fixed through the 
 total spin-down energy $W_0$ emitted by the pulsar \citep{ Blasi:2010de,2010AA...524A..51D}: 
\begin{equation}
\label{Wop}
\int_{E_{\rm min}}^{\infty}\,dE\,E\,Q(E)\,=\eta\,W_0
\end{equation}
The total spin-down energy $W_0$ can be expressed as:
\begin{equation}
W_0\approx\tau_0\dot{E}\left(1+\frac{t_*}{\tau_0}\right)^2
\end{equation}
and depends on the spin-down luminosity (i.e. the energy-loss rate) $\dot{E}$, 
the present age of the pulsar $t_*$ and the typical pulsar decay time $\tau_0$. 
The first two parameters are found in the pulsar ATNF catalog \cite{1993ApJS...88..529T}, 
while $\tau_0$ is fixed to 10 kyr for all the sources \cite{2010AA...524A..51D,Hooper:2008kg}. We fix $E_{\rm min}=0.1$ GeV, 
$\gamma_{\rm PWN}=1.9$ and $\eta=0.032$, if not differently stated. 
With these numbers, we can derive the spectral normalization $Q_0$ and therefore compute 
the $e^{\pm}$ spectrum produced and accelerated inside a PWN.

\subsection{Secondary positron and electrons}
\label{sec:secondary}

\begin{figure}[t]
\centering
\includegraphics[width=0.65\textwidth]{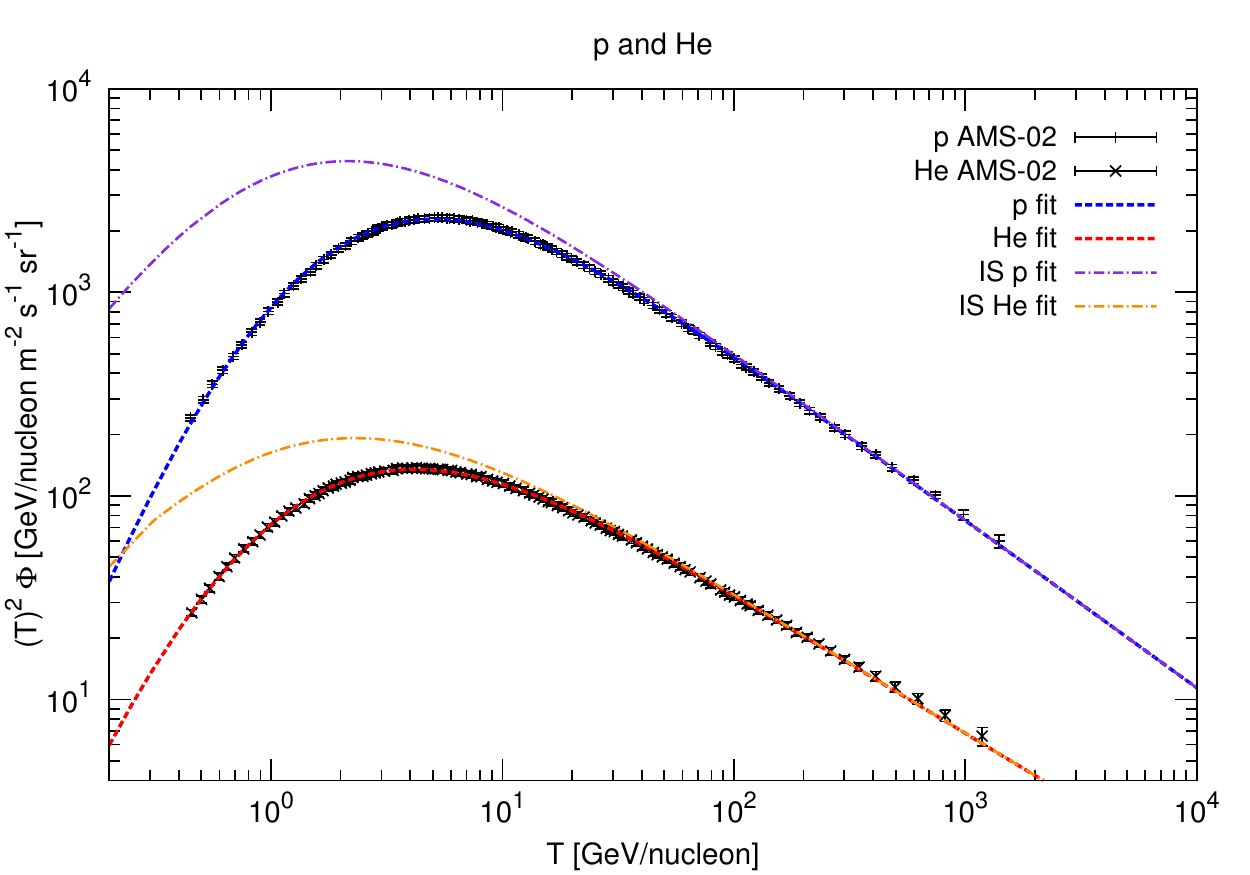}
\caption{Primary interstellar (dot-dashed) and solar modulated (dashed) fluxes of protons (upper flux) and helium (lower flux) in function of the kinematic energy per nucleon T. Data points refer to the recent AMS-02 measurements \citep{proton_AMS02,helium_AMS02}, dashed lines show out best fit to the data sets.}
\label{fig:AMSprimary} 
\end{figure}

Secondary electrons and positrons originate from the spallation reactions of hadronic CR species (mostly protons and $\alpha$ particles) 
with the interstellar material (mostly made of hydrogen and helium).
Since secondary positrons and electrons originate from positively 
charged ions, charge conservation implies a greater production of positrons with respect to 
electrons \cite{2006ApJ...647..692K}. 
We have thoroughly discussed the production of secondary electrons and positrons in Ref.~\citep{2009AA...501..821D,2010AA...524A..51D},  to which we refer for 
any detail. Here we only recall that the steady state source term for secondaries has the form:
\begin{eqnarray}
\label{eq:source}
  q_{e^{\pm}}(\mathbf{x},E_{e}) = 4 \pi \; n_{\rm ISM}(\mathbf{x}) \displaystyle 
\int dE_{\rm CR} \Phi_{\rm CR} \left( \mathbf{x} , E_{\rm CR} \right) \frac{d\sigma}{dE_{e}}(E_{\rm CR}, E_{e}) ,
\end{eqnarray}
where $n_{\rm ISM}$ is the interstellar gas density, the primary incoming CR fluxes are denoted by 
$\Phi_{\rm CR}$, and ${d\sigma}/{dE_{e}}$ refers to the leptonic part of the
 inclusive nucleon-nucleon cross section. 
 With respect to Ref.~\cite{2010AA...524A..51D}, we have computed Eq.~(\ref{eq:source}) by 
fixing here the proton and helium primary fluxes to the new measurements of AMS-02 \citep{proton_AMS02,helium_AMS02}. We fit the solar-modulated data by assuming interstellar proton
and He fluxes described by the function $\Phi=A \beta^{P_1}R^{-P_2}$, where $R=pc/Ze$ is the rigidity of the nucleus of charge number $Z$ and momentum $p$, and solar modulation
described by the force-field method.
We obtain: $A=22450 \pm  560$ m$^{-2}$s$^{-1}$sr$^{-1}$(GeV/n)$^{-1}$, $P_1=2.32 \pm 0.56$ and  $P_2=2.8232 \pm 0.0053$ for the
proton flux, and  $A= 5220\pm 110 $ m$^{-2}$s$^{-1}$sr$^{-1}$(GeV/n)$^{-1}$, $P_1= 1.34\pm 0.27$ and  $P_2= 2.6905\pm 0.0043$
for the helium flux (and for a Fisk solar modulation potential of $615 \pm 30$ MV).
The results of our fits on the primary proton and helium fluxes, compared to the AMS data, are 
shown in Fig.~\ref{fig:AMSprimary}.
The best-fit chi-squared value, for 236 data points and 7 degrees of freedom, is $\chi^2/{\rm d.o.f.}=0.17$.  
Let us mention that a determination of the interstellar proton spectrum free from solar modulation effects could be derived by using diffuse $\gamma$-ray data: this technique has been disucssed and undertaken, in a preliminary analysis, in Refs. \cite{2013arXiv1303.6482D,2013arXiv1307.0497D}, where a  break in the interstellar spectrum around a few GeV is found.

We consider the p-p cross section parameterization described in Ref.~\cite{2006ApJ...647..692K}, 
which includes additional processes (especially resonances other than the $\Delta$ at low interaction 
energies) and has been calibrated with recent data.  For reactions including helium, both as a 
target and as the incoming particle, we use the empirical 
prescription and the results described in Ref.~\cite{2009AA...501..821D}.

\begin{figure}[t]
\centering
\includegraphics[width=0.48\textwidth]{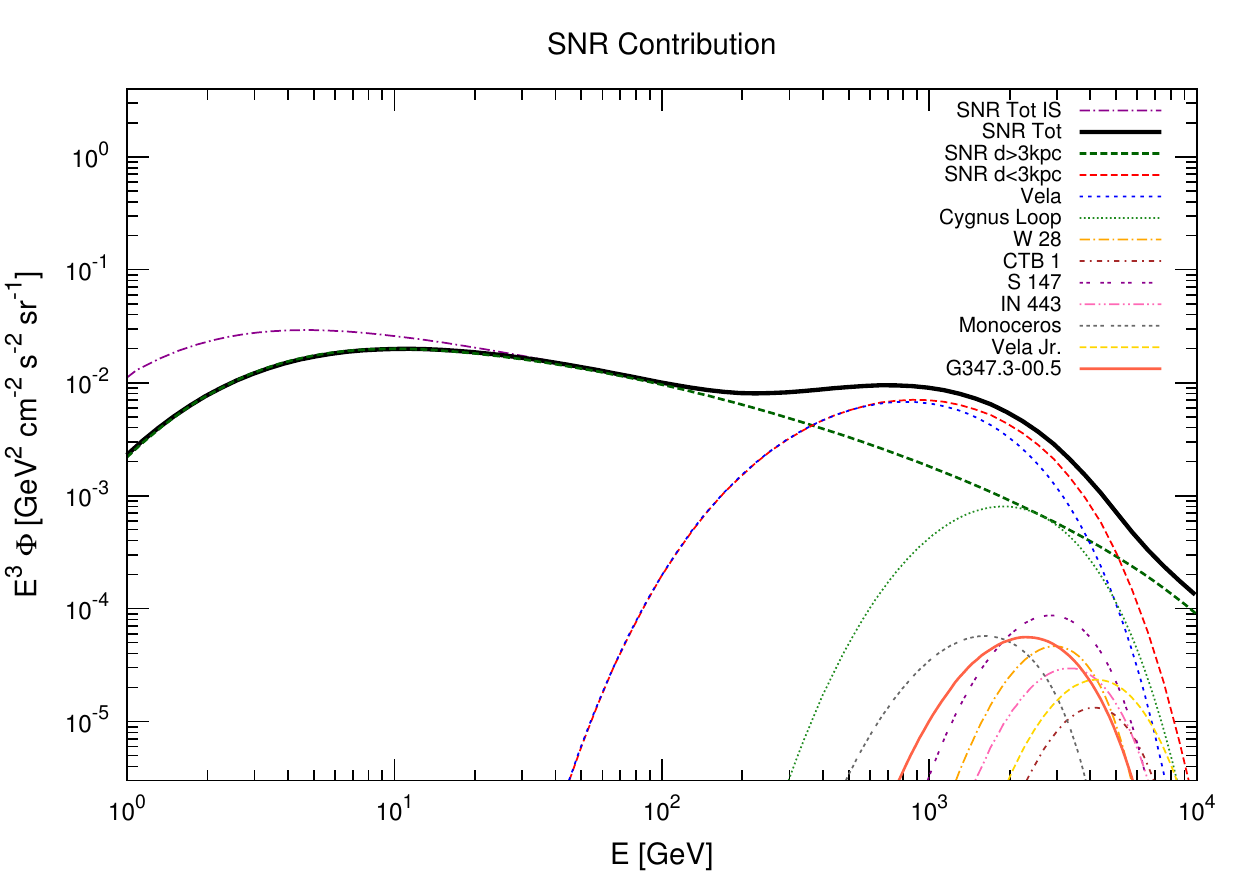}
\includegraphics[width=0.48\textwidth]{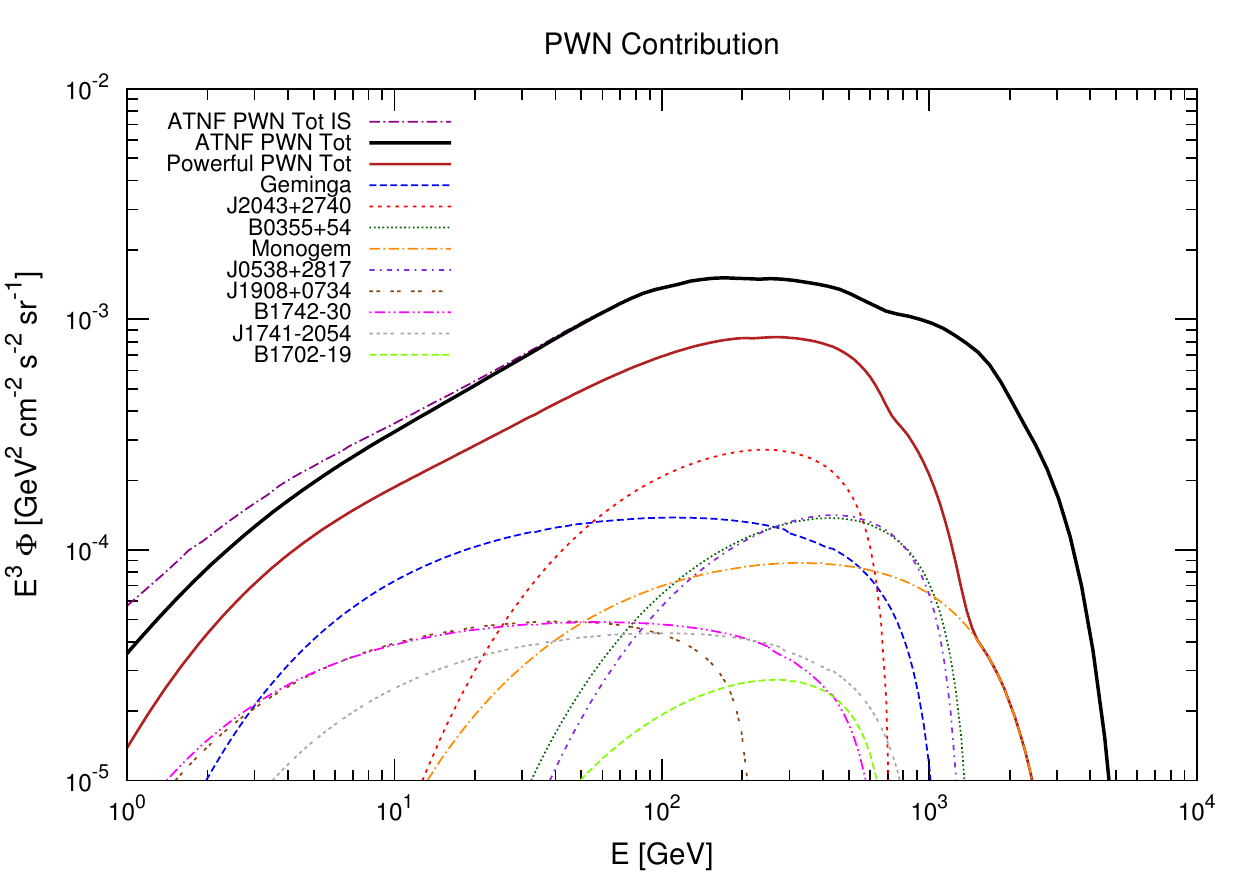}
\caption{{\sl Left}: Top of atmosphere electron flux (times $E^3$) from the nine most powerful and close  ($\leq$ 3 kpc) 
SNRs (from Tab.~\ref{tab:snrn}), shown together with their sum (red dashed line).
The green dashed line represents the electron flux from the far ($>$ 3 kpc) SNR population, 
while the solid black line is the sum of all the contributions. {\sl Right}: Positron flux (the same occurs for electrons) from the nine most powerful pulsars of the ATNF catalog, along with their sum (solid red line) 
and  the sum of the fluxes of all the pulsars of the catalogue (solid black line). Galactic propagation with the {\sl MED} model; solar modulation parameter $\phi=830$ MV.
In both panels, the dot-dashed (violet) line refers
to the interstellar flux.
}
\label{fig:nearsources} 
\end{figure}

\section{The propagation of electrons and positrons in the Galaxy}
\label{sec:propagation}
Once produced in their respective sources, electrons and positrons propagate throughout the Galaxy, 
where they diffuse on the magnetic field inhomogeneities. Most importantly, they lose their energy 
by electromagnetic interactions with the interstellar radiation field (ISRF) through inverse Compton (IC)
scattering, and by synchrotron emission on the galactic magnetic field (notably, bremsstrahlung, ionization 
and Coulomb interactions on the interstellar medium are negligible and can be safely neglected). 
The diffusion equation for the electron (positron) number density 
${\cal N}={\cal N}(E,\vec{x},t)\equiv dn/dE$ 
may be written as (see Ref.~\citep{2010AA...524A..51D} and references therein):
\begin{equation}
\partial_t {\cal N} - \vec{\nabla}\cdot 
\left\{ K(E)  \vec{\nabla}{\cal N} \right\} + 
\partial_E \left\{ \frac{dE}{dt} {\cal N} \right\} = {\cal Q}(E,\vec{x},t)\;.
\label{eq:prop}
\end{equation}
where we have neglected the effect of convection and reacceleration \citep{2009AA...501..821D}. 
In this equation, $K(E)$ is the energy-dependent diffusion coefficient while $dE/dt$
is the energy-loss term and ${\cal Q}(E,\vec{x},t)$ denotes the source term (discussed in the previous Section). 
The solution to Eq.~(\ref{eq:prop}) is found within a semi-analytical model in which the Galaxy is shaped as a cylinder, 
made of the stellar thin disk (with half-height of 100 pc), and a thick magnetic halo whose height $L$ varies from 1 to 15 kpc
\citep{2001ApJ...555..585M}. 
In our analysis we closely follow Ref.~\cite{2010AA...524A..51D}, to which we refer for any detail. 
We only remind here that we have included a full relativistic treatment of the IC energy losses, while
for the synchrotron emission we have set the magnetic field to 1 $\mu$G. 
The spatial diffusion coefficient 
$K(E)= \beta K_0 ({\cal R}/{1~\rm GV})^{\delta}$ is set to one of the three benchmark 
sets of parameters derived in Ref.~\cite{2004PhRvD..69f3501D} and compatible with the boron-to-carbon analysis \citep{2001ApJ...555..585M}. 
Namely, for the {\sl MIN/MED/MAX} models we fix $\delta=0.85/0.70/0.46$,
$K_0= 0.0016/0.0112/0.0765$ kpc$^2$/Myr and $L= 1/4/15$ kpc, respectively. 
Finally, for the solar modulation affecting low energy (about $<$ 10 GeV) charged CRs, we use the force field 
approximation \cite{1971JGR....76..221F, 1987AA...184..119P}, with a solar modulation parameter $\phi$ determined within our fitting procedure on the electron and positron data sets. Apart from the force field approximation, more complex models, in which solar modulation is assumed to depend on the  sign of the particles charge, have been employed (see, for example,  Refs. \cite{Maccione:2012cu,DellaTorre:2012zz}).

We show in Fig.~\ref{fig:nearsources}  the electron and positron fluxes produced from SNR and pulsars, 
 propagated according to the above prescriptions, for the {\sl MED} propagation parameters. 
The left panel shows the electron flux from the nine most-powerful among the near  ($\leq$ 3 kpc around the Solar System) SNRs, 
along with the sum of their single fluxes. The source parameters have been derived from Tab.~\ref{tab:snrn}, as explained in 
Sect. \ref{sec:SNR}.  We also plot the contribution from the average population of distant ($>$ 3 kpc)
SNR, 
whose spectral index $\gamma$ and normalization $Q_0$ have been fixed to 2.38 and $2.75\times 10^{50}$ GeV$^{-1}$, 
respectively, according to the results of the fit on the AMS-02 data explained in the next Sect.~\ref{sec:fit}. 
As expected, the far SNRs contribute predominantly to the electron flux up to about 100 GeV, above which the local sources dominate. 
We observe that Vela(XYZ) -- a near, young and strong radio-emitter SNR -- 
is the dominant contributor, exhibiting an electron flux much higher than the other SNRs. 
In the right panel, we plot the positron flux (which is the same as for electrons)  from the nine most powerful 
pulsars of the ATNF catalog, their sum  and  the sum of the fluxes of all the pulsars of the catalogue.
We display also the top of atmosphere (solid black line) and interstellar (dot-dashed violet line) total contribution for both SNR and PWN emission. Again, the source parameters are fixed according to the analysis performed in Sect.~\ref{sec:fit}: $\gamma_{\rm PWN}=1.90$ and
$\eta=0.032$. 
The two highest fluxes in the AMS-02 high-energy range are provided by Geminga and J2043-2740, but do not really 
dominate over the other ones. We also observe that the flux of the most powerful PWN 
is indeed lower (by a factor of two up to an order of magnitude) than the flux provided by the whole PWN in the ATNF catalog. 
As a consistency check, we point out that our positrons and electrons interstellar fluxes appear to be in good agreement with the determination of these fluxes given in Refs. \cite{2011A&A...534A..54S} and \cite{2013MNRAS.436.2127O} as the result of an analysis based on synchrotron observations, thus completely independent from any detail concerning  solar modulation.

\section{Fit to AMS-02 data: method and free parameters}
\label{sec:fit}
The AMS-02 Collaboration has recently published data about the positron fraction ($e^+/(e^+ + e^-)$) \cite{Aguilar:2013qda}
and presented preliminary results on the electron, positron and electron plus positron flux
\citep{electrons_AMS02,totalelectrons_AMS02}.
For the latter three quantities, for which a specific information on the experimental
uncertainty is not currently available, we assume an energy independent error of 8\%, comprehensive of statistical 
and systematic uncertainties.
We employ all the four observables ($e^++e^-, e^-, e^+, e^+/(e^+ + e^-)$) in order to explore whether a unique source-model can explain AMS-02 data. 

Our model is built up by the components described in Sect.~\ref{sec:sources}: $i$) electrons produced by near SNRs treated as individual sources;
$ii$) electrons from an average population of distant SNR; $iii$) electrons and positrons from PWN, considered as individual sources; $iv$)  secondary electrons and 
positrons produced by the spallation of p and He primary cosmic rays. 
For the electrons produced by the closest ($\leq$ 3 kpc) SNRs, we derive their source parameters according to the prescriptions given in Sect.~\ref{sec:SNR} and employ
the radio, distance and age data listed in Tab.~\ref{tab:snrn}. 
For the electrons arriving to the Earth from the population of far ($>$ 3 kpc) SNRs, we proceed as described in Sect.~\ref{sec:SNR}, leaving the spectral index 
$\gamma$ and the overall normalization $Q_0$ as free parameters. 
The ATNF catalog pulsars are included here by making the simplifying hypothesis that they all shine with a common 
spectral index $\gamma_{\rm PWN}$ and efficiency $\eta$, following the discussion outlined in Sect.~\ref{sec:pulsar}. 
Finally, the secondary positrons and electrons are computed from the observed primary p and He (see Sect.~\ref{sec:secondary}), and do not depend on any
free parameter. However, we allow the normalization to be adjusted (by an overall renormalization factor that we call here $\tilde{q}_{\rm sec}$)
in the fit to the AMS-02 data, in order to verify {\em a-posteriori} if the secondary positron production (determined by CR hadrons) is consistent with the measured lepton data. 
In summary, the free parameters of the model are:  $\gamma$, $Q_0$, $\gamma_{\rm PWN}$, $\eta$, $\tilde{q}_{\rm sec}$ and $\phi$, 
where the latter (the solar modulation potential) is let free in order to accommodate low energy data. We jointly fit all the four datasets together.

\begin{figure}[t]
	\centering
	\includegraphics[width=0.48\columnwidth]{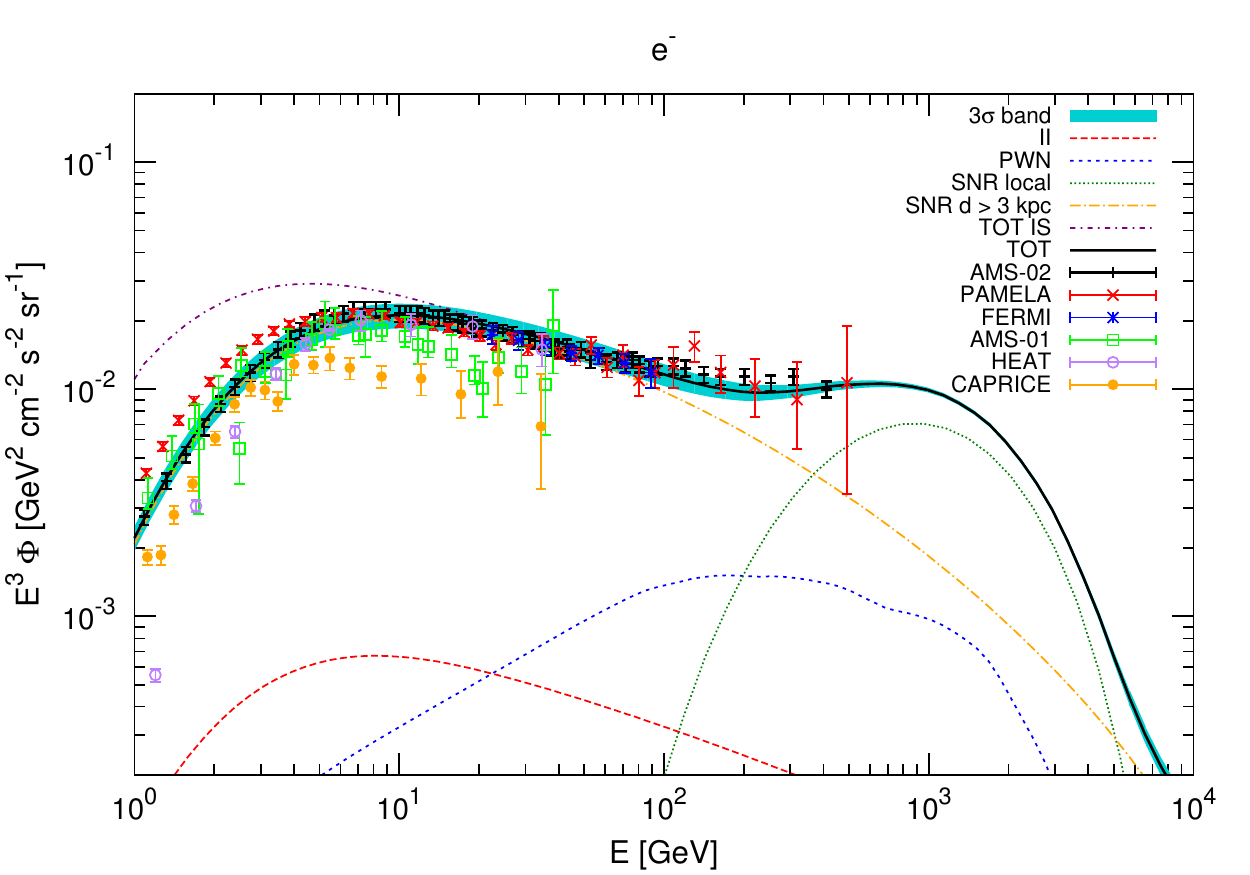}
	\includegraphics[width=0.48\columnwidth]{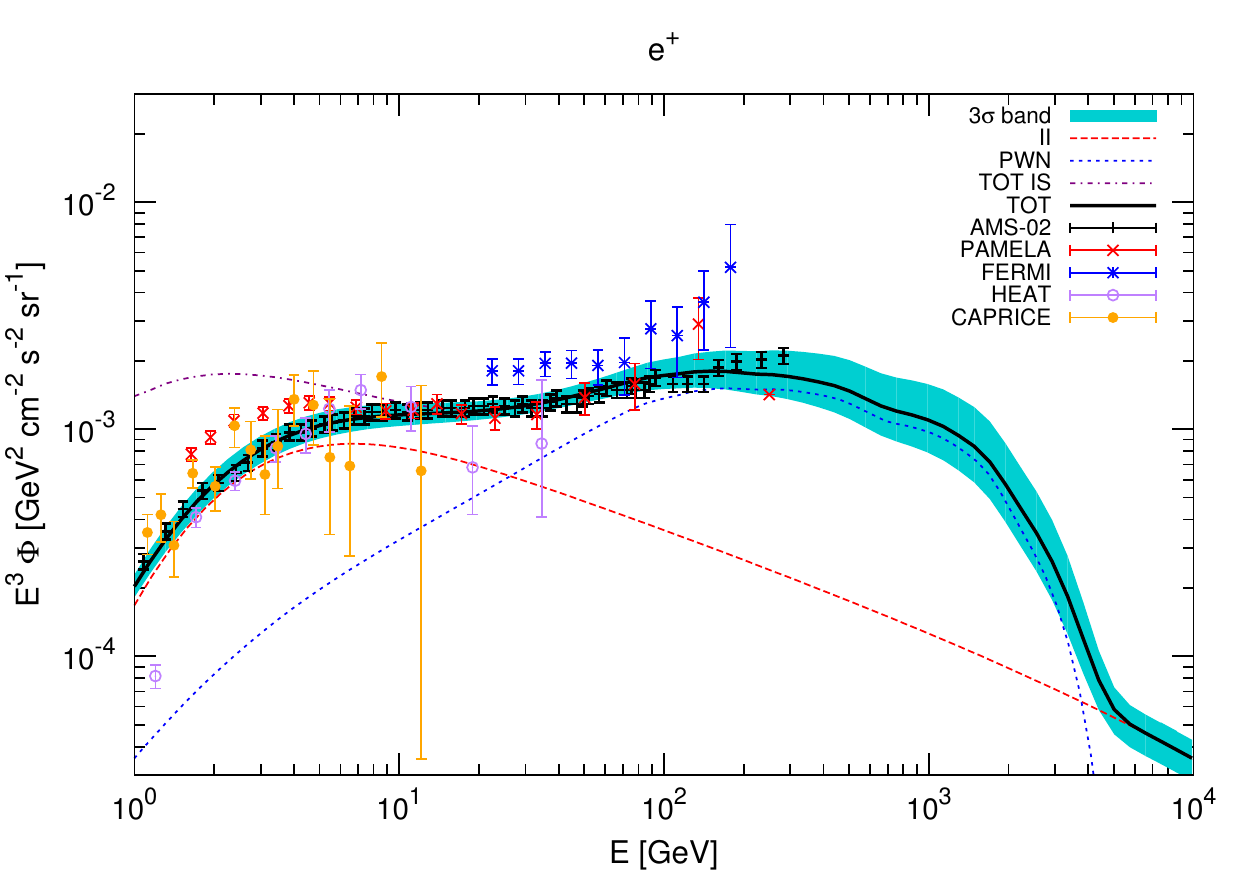}
	\includegraphics[width=0.48\columnwidth]{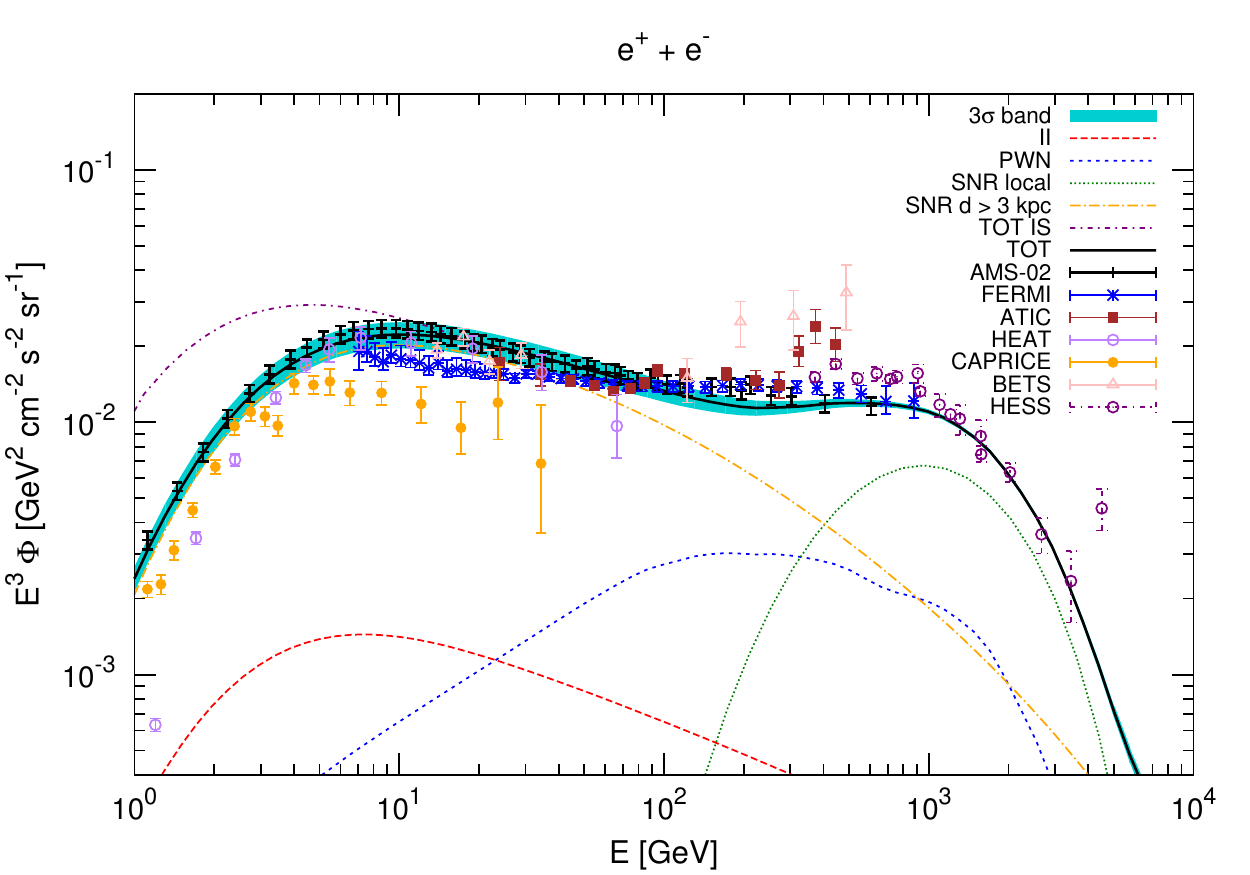}
    \includegraphics[width=0.48\columnwidth]{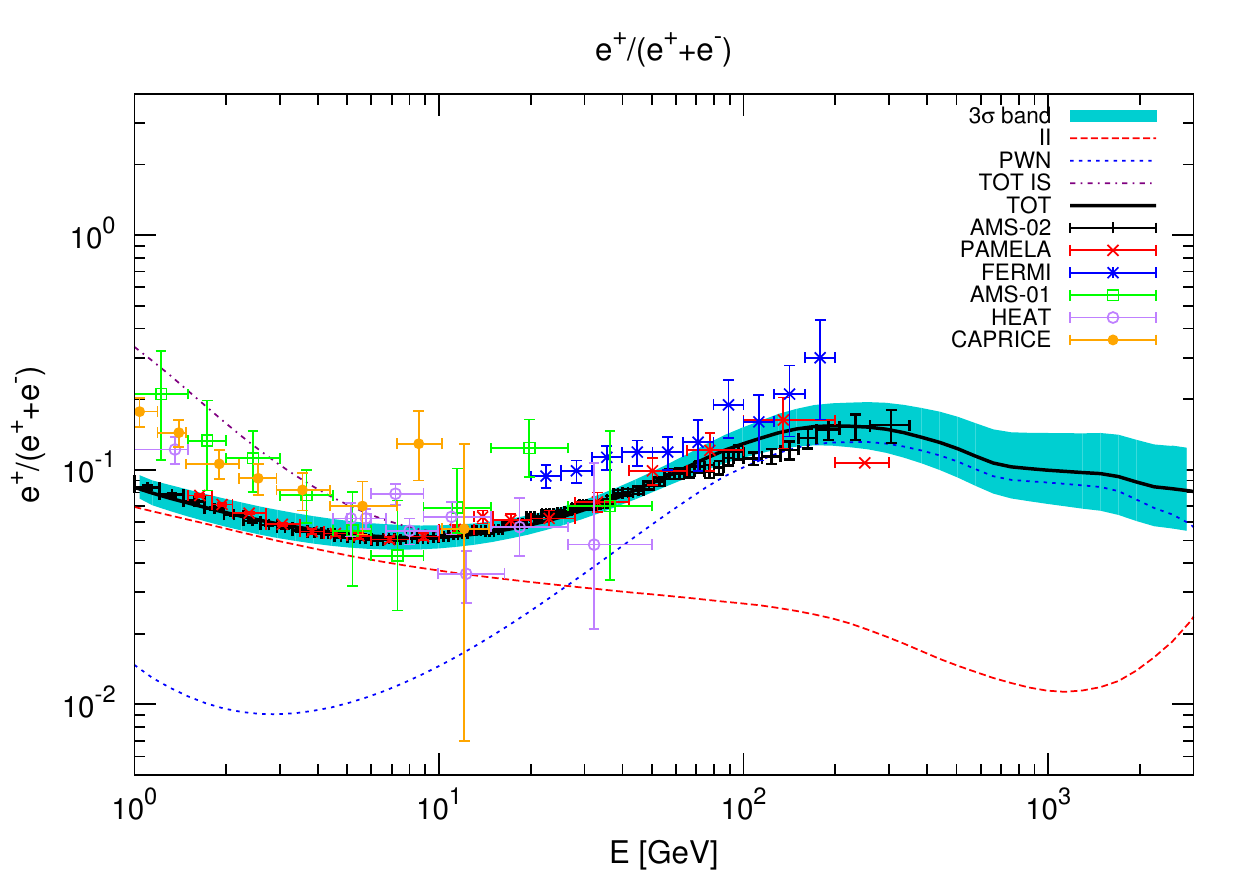} 
\caption{Results of our simultaneous fit on the AMS-02 data for the electron flux (top left), positron flux (top right), electron plus positron flux (bottom left)
and positron fraction (bottom right). The best fit model is represented by the solid black line, and is embedded in its 3$\sigma$ uncertainty band (cyan strip).
In each panel, the dot-dashed yellow line represents the electron flux from the far ($>$3 kpc) SNR population, the dotted green line the electrons from the local SNRs,
while the short dashed blue line describes the positron and electron flux from PWN and the long dashed red takes into account the secondary contribution to both  
electron and positron flux. The fit is performed on all the AMS-02 data simultaneously.
Together with our theoretical model, data from AMS-02 \cite{Aguilar:2013qda,totalelectrons_AMS02,electrons_AMS02}, 
Fermi-LAT \cite{2012PhRvL.108a1103A,2010PhRvD..82i2004A}, 
Pamela \cite{2009Natur.458..607A,2011PhRvL.106t1101A,2013arXiv1308.0133P}, 
Heat \cite{2004PhRvL..93x1102B,1998ApJ...498..779B,1997ApJ...482L.191B,2001ApJ...559..296D}, 
Caprice \cite{2000ApJ...532..653B,2001AdSpR..27..669B}, Bets \cite{2008AdSpR..42.1670Y,2001ApJ...559..973T} and
Hess experiments \cite{2008PhRvL.101z1104A,2009AA...508..561A} are reported.
Long-dashed lines report the corresponding interstellar fluxes, before solar modulation.}
\label{fig:fit_ele_pos} 
\end{figure}

In Fig.~\ref{fig:fit_ele_pos}, we show the result of the fit on all the four leptonic observables: the flux of electrons plus positrons, electrons, positrons and the positron 
fraction. The four panels report the total flux for each observable, together with the single subcomponents arising from the different categories of sources.
Fig.~\ref{fig:fit_ele_pos} also shows AMS-02 data and data from previous experiments. 
The best fit to each observable is shown as a solid line, embedded in its 3$\sigma$ uncertainty band.
The result of the analysis shows a quite remarkable agreement with AMS-02 data: this is confirmed
by the value of the best-fit chi-squared: $\chi^2/{\rm d.o.f.}=0.65$, for 236 data points and 6 degrees of freedom. The best fit-values of the 6 parameters are: $\eta=0.0320\pm0.0016$, $\gamma_{\rm PWN}=1.90\pm0.03$ for the PWN sources, 
$Q_0=(2.748\pm0.027)\times 10^{50}$ GeV$^{-1}$ and $\gamma= 2.382 \pm 0.004$ for the far SNRs,  
the renormalization of $e^+$ and $e^-$ secondary contribution is 
$\tilde{q}_{\rm sec} =1.080 \pm 0.026$,  and the Fisk 
potential turns out be $830\pm22$ MV.  The value of $Q_0$ is 
 similar to the one derived in Sec.~\ref{sec:SNR} for the 88 sources of the Green catalog with measured radio index, flux and distance.

The various electrons and positrons sources have different impact in the reconstruction
of the properties of the four set of observables. At high energies, local sources are the most relevant:
SNR for the electron flux and the $(e^+ + e^-)$ total flux, PWN for the positron flux and, in turn, the
positron fraction; at lower energies, far SNR are dominating the flux of electrons and
of $(e^+ + e^-)$ (this occurs for energies below about 100 GeV), while secondaries determine the positron flux and the positron fraction (for energies below 10-20 GeV).
It is therefore remarkable that a single model for all the source components, for both positron and electrons, fits simultaneously all the leptonic 
AMS-02 data, without any further ad-hoc adjustment. 
The best fit values found for the free parameters of SRN and PWN are in very good agreement with the ones quoted in Sections \ref{sec:SNR} and \ref{sec:pulsar}.   

Another quite interesting result concerns the positron flux interpretation. The secondary positron component adopted in our analysis, as discussed above, depends only on the p and He primary
fluxes (which we have determined by a separate, independent, fit on the recent AMS-02 data),
on the nuclear cross sections involved in the spallation process and on propagation in the Galaxy.  Therefore, this component does not require additional assumptions (like
it is in the case of the SNR and PWN contributions, which have free unknown parameters), and
is therefore somehow fixed once a specific propagation model is assumed. In order to check {\em a posteriori} the compatibility with the AMS-02 data, we have allowed the normalization parameter $\tilde{q}_{\rm sec}$ to freely vary: the fact that we find a best-fit value of  $\tilde{q}_{\rm sec}$ very close to one, for the {\sl MED} propagation parameters, is a confirmation of the good level of consistency in the analysis.
A further discussion of the secondary positron component is given in the next Section.

\subsection{The case for secondary positrons}
\label{sec:secondaries}

\begin{figure}[t]
\centering
\includegraphics[width=0.55\textwidth]{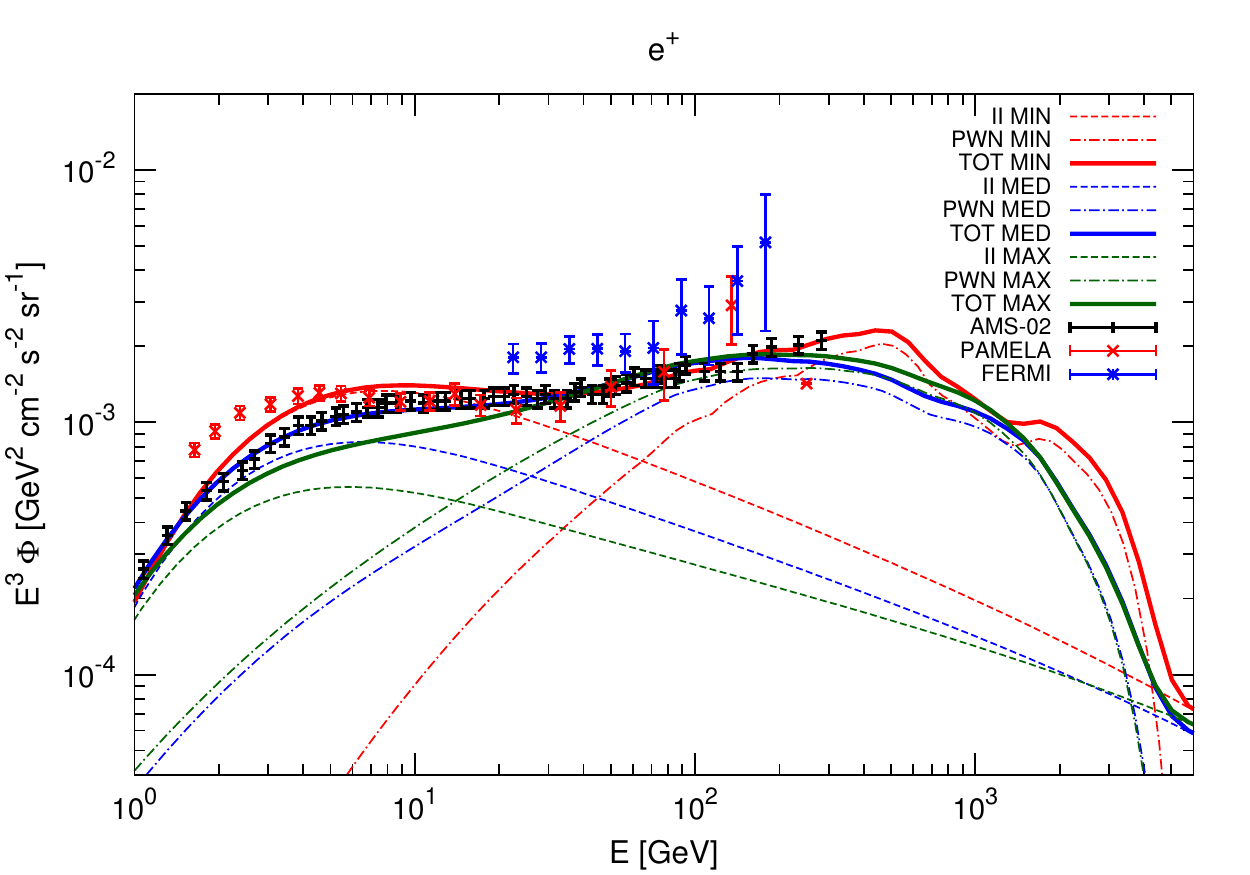}
\caption{The positron spectrum for the {\sl MIN} (red), {\sl MED} (blue) and {\sl MAX} (green) propagation models are displayed together with AMS-02, 
Fermi-LAT \cite{2012PhRvL.108a1103A,2010PhRvD..82i2004A} and Pamela \cite{2013arXiv1308.0133P} data. 
The theoretical contribution has been derived for the secondary (dashed), PWNs (dot-dashed) and total spectra (solid).}
\label{fig:secondary}
\end{figure}

The positron spectra is interpreted in terms of a secondary production at low energy and of a PWNs emission at higher ($>$10 GeV) energies. 
As already recalled, the secondary positrons depend on their progenitor p and He spectra, on the involved spallation cross sections, and on the propagation 
in the Galaxy. The uncertainties of the first two ingredients are definitely smaller than the ones induced by propagation. 
We study here the effect of the different {\sl MIN, MED} and {\sl MAX} models described in Sect.~\ref{sec:propagation} on the secondary and PWN positrons.
This theoretical emission is then compared with the $e^+$ spectrum measured by AMS-02 \cite{electrons_AMS02}.

We derive the secondary and PWNs production of positrons considering the {\sl MIN, MED, MAX} propagation models and fit the measured spectrum of positrons  with the Fisk potential $\phi$, the efficiency $\eta$ and the index $\gamma_{\rm PWN}$ for PWN as free parameters.
We have allowed the Fisk potential to vary in the range $(0.6,1.0)$ GV, in accordance
to results\footnote{\url{http://cosmicrays.oulu.fi/phi/Phi_mon.txt}} of combined analysis of proton and helium spectra correlated with neutron monitors data \cite{2005JGRA..11012108U,2011JGRA..116.2104U}, and compatible with our determination
for the AMS-02 data taking period derived in Sect. \ref{sec:secondary} with the fit on AMS-02 proton and helium fluxes.

The positron spectra  are displayed in Fig.~\ref{fig:secondary} for {\sl MIN}, {\sl MED} and {\sl MAX} models, and 
for the secondary, PWNs and total spectra. 
The best fit values for the Fisk potential are 0.6, 0.77 and 1.0 GV, for the PWNs efficiency 0.011, 0.032 and 0.087 
while for $\gamma_{\rm PWN}$ are 1.43, 1.90 and 2.08 for the {\sl MIN, MED, MAX} respectively.
Notice that in the case of {\sl MIN} and {\sl MAX} the Fisk potential best fit values are the minimal and maximal allowed in this analysis.
The best-fit chi-squared is for 56 data points and 3 degrees of freedom $\chi^2/{\rm d.o.f.}=\{2.43,0.66,4.62\}$ for the {\sl MIN, MED, MAX}.
We see that the {\sl MED} set of parameters predicts a positron spectrum fully 
 compatible with the data,  as previously derived in Sect.~\ref{sec:fit}.
On the other hand,  the {\sl MIN}  ({\sl MAX}) are not compatible with the data, mostly because of the low energy secondary positrons, which 
depend sensibly on galactic diffusion. We have checked that one  would need to renormalize the secondary component by
 a factor $\tilde{q}_{\rm sec}$ = 0.72 (1.78) for the {\sl MIN} ({\sl MAX}) cases,  in order to reproduce the $e^+$ AMS-02 measurements below 10 GeV. 
Remarkably, the {\sl MIN} model predicts an exceedingly high positron flux and indicates that a small halo size together with a very soft diffusion coefficient 
are strongly disfavored by low energy positron data.

\subsection{The case for pulsars cut-off energy}
\label{ecut}
\label{sec:secondaries}

\begin{figure}[t]
\centering
\includegraphics[width=0.48\textwidth]{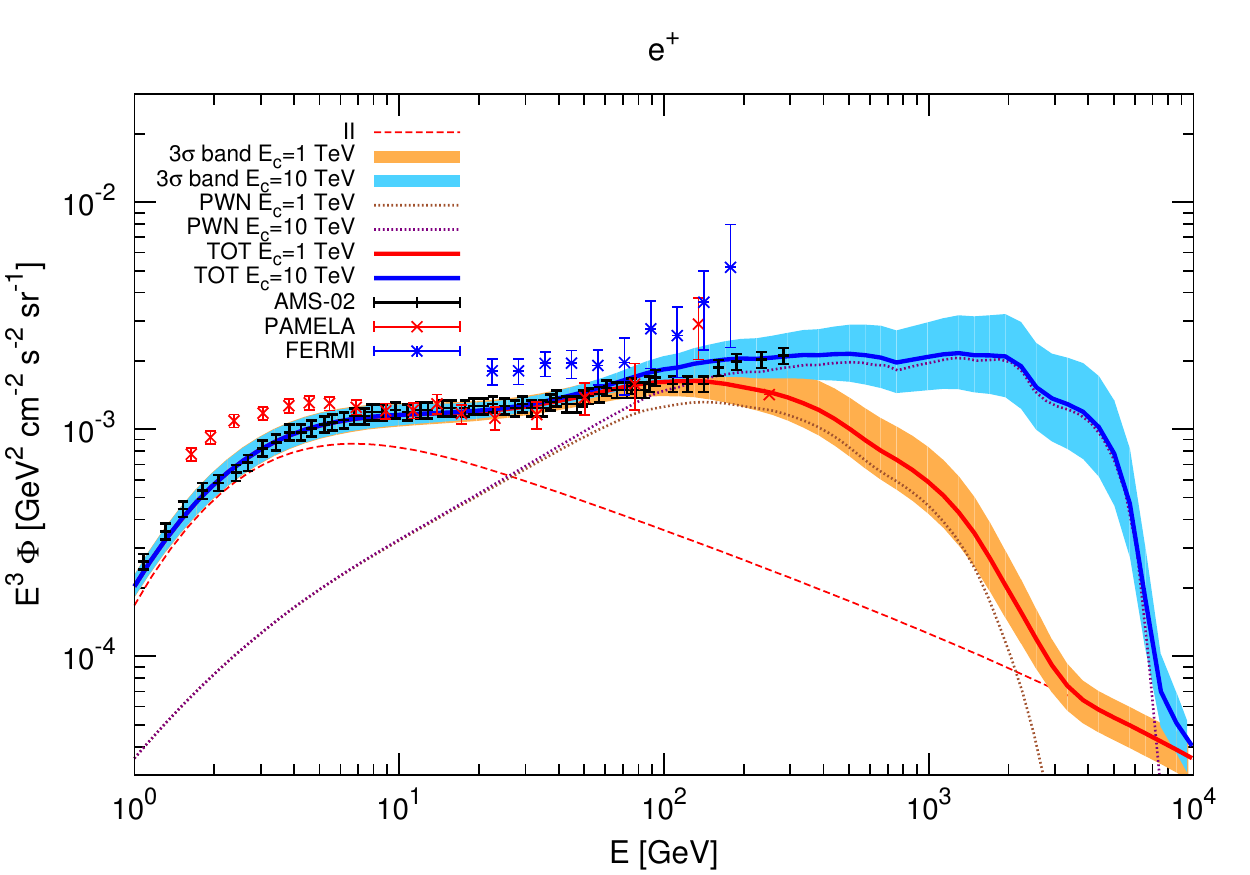}
\includegraphics[width=0.48\textwidth]{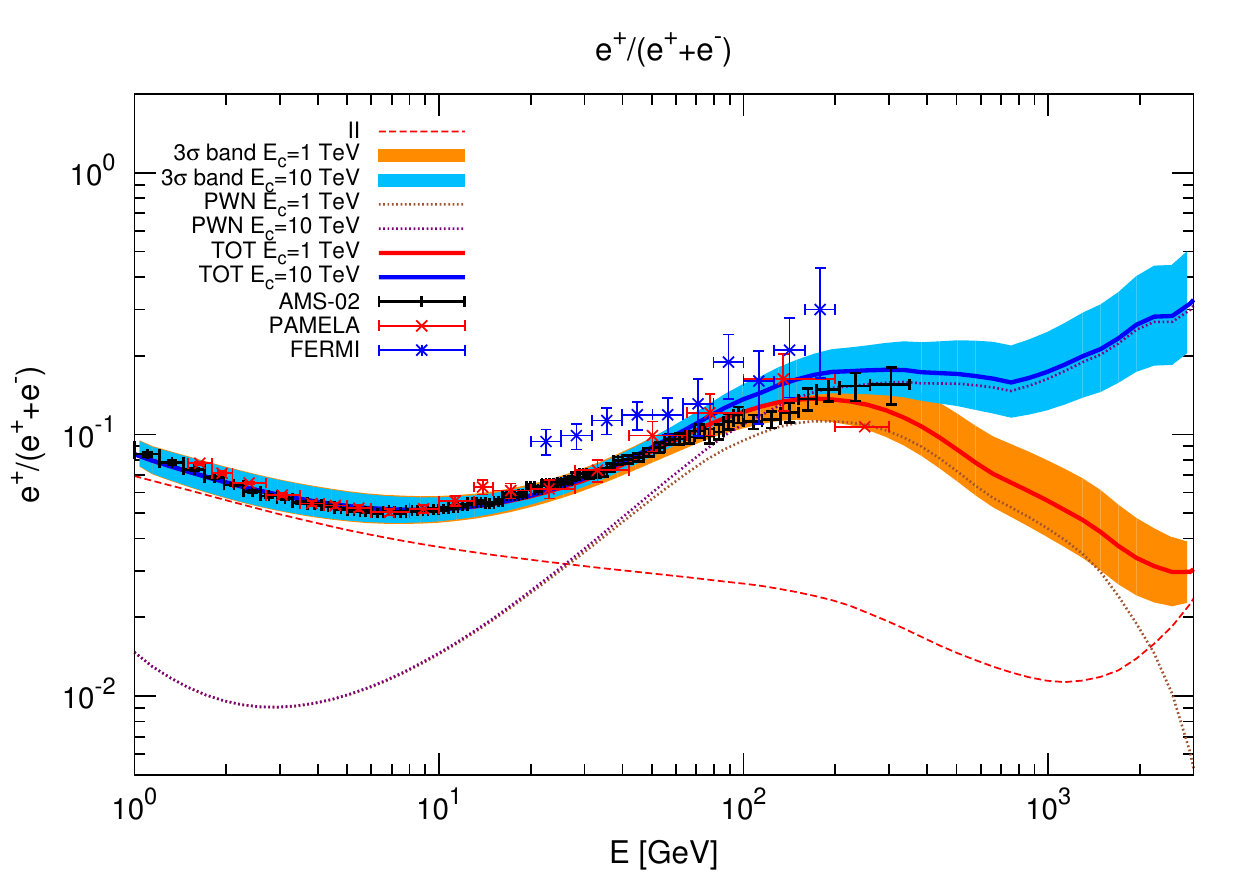}
\caption{Positron flux (left) and positron fraction (right) for two different extreme values
of the cut-off energy of pulsars: $E_c = 1$ TeV (lower red curve) and $E_c = 10$ TeV (upper blue curve). 
The curves show the best-fit agreement with the whole AMS-02 data set; the band around each curve represents the $3\sigma$ allowed range. 
AMS-02 \cite{Aguilar:2013qda,electrons_AMS02}, Fermi-LAT \cite{2012PhRvL.108a1103A,2010PhRvD..82i2004A} and Pamela \cite{2009Natur.458..607A,2013arXiv1308.0133P} data are displayed together with theoretical expectations.}
\label{fig:cutoff}
\end{figure}

The high-energy part of the positron data (positron flux and positron fraction) is
of special interest, since it might disclose relevant information on their source (including
a very intriguing dark matter origin). We have examined the impact of the uncertain cut-off
energy in pulsar emission. Fig.~\ref{fig:cutoff} shows the positron flux and the positron
fractions calculated under two extreme situations (which encompass the case adopted in all
other analysis in this paper, i.e. $E_c = 2$ TeV): the lower red curve shows the best-fit to the whole AMS-02 data 
when we assume $E_c = 1$ TeV; the upper blue curve is the best-fit obtained when $E_c = 10$ TeV. 
The band around each curve represents the $3\sigma$ allowed range. We notice that current data, which extend up to about 300 GeV, can be explained remarkably 
well for a wide interval of variation for $E_c$: they are not yet sensitive to the
drop expected from the exponential cut-off, and the expectation for the
positron flux and for the positron fraction above the current AMS-02 highest energy,
suggest a either a mild decrease (for $E_c$ close to 1 TeV) or even an almost constant value up to energies well above the TeV scale (for $E_c$ close to 10 TeV). An increase in positron fraction is not likely to be expected in the $300-1000$ TeV energy range. Notice that the
positron fraction at these energies depends also on the cut-off energy of SNR, discussed
in Sect.~\ref{sec:SNR}: however, the electron+positron flux measured up to energies
of few TeV by HESS, points toward a value around 2 TeV (as can be seen in Fig.~\ref{fig:fit_ele_pos}) and therefore it does not appear to be allowed to substantially vary,
and therefore alter the positron fraction shown in Fig.~\ref{fig:cutoff}. We also wish to comment
that the two extreme values adopted here for the pulsars cut-off energy are representative cases, adopted to encompass the possible maximal effect.
Expected values for $E_c$ should be more close to a few TeV, at most.

From Fig.~\ref{fig:cutoff} we can also observe that
a sharp drop in the positron observables just above the current AMS-02 highest energy 
could hardly be attributed to a pulsar origin: it would therefore represent an
interesting clue to a positron exotic origin, like dark matter annihilation or decay in
a hard production channel. On the contrary, if
pulsars are a major contributor at the current experimental energies, it would be 
difficult to have a clear signal of dark matter at higher energies, unless dark matter is very heavy.

\begin{figure}[t]
\centering
\includegraphics[width=0.48\textwidth]{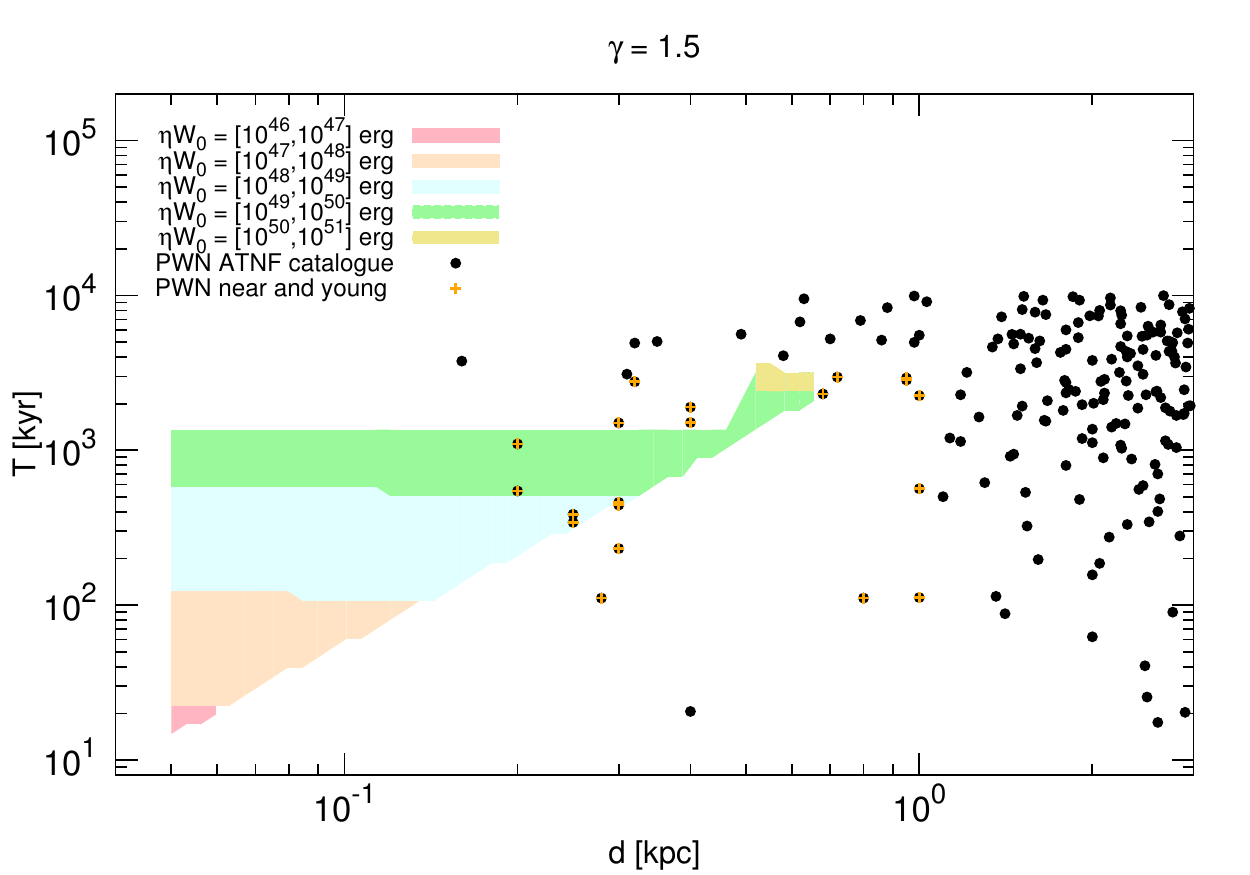}
\includegraphics[width=0.48\textwidth]{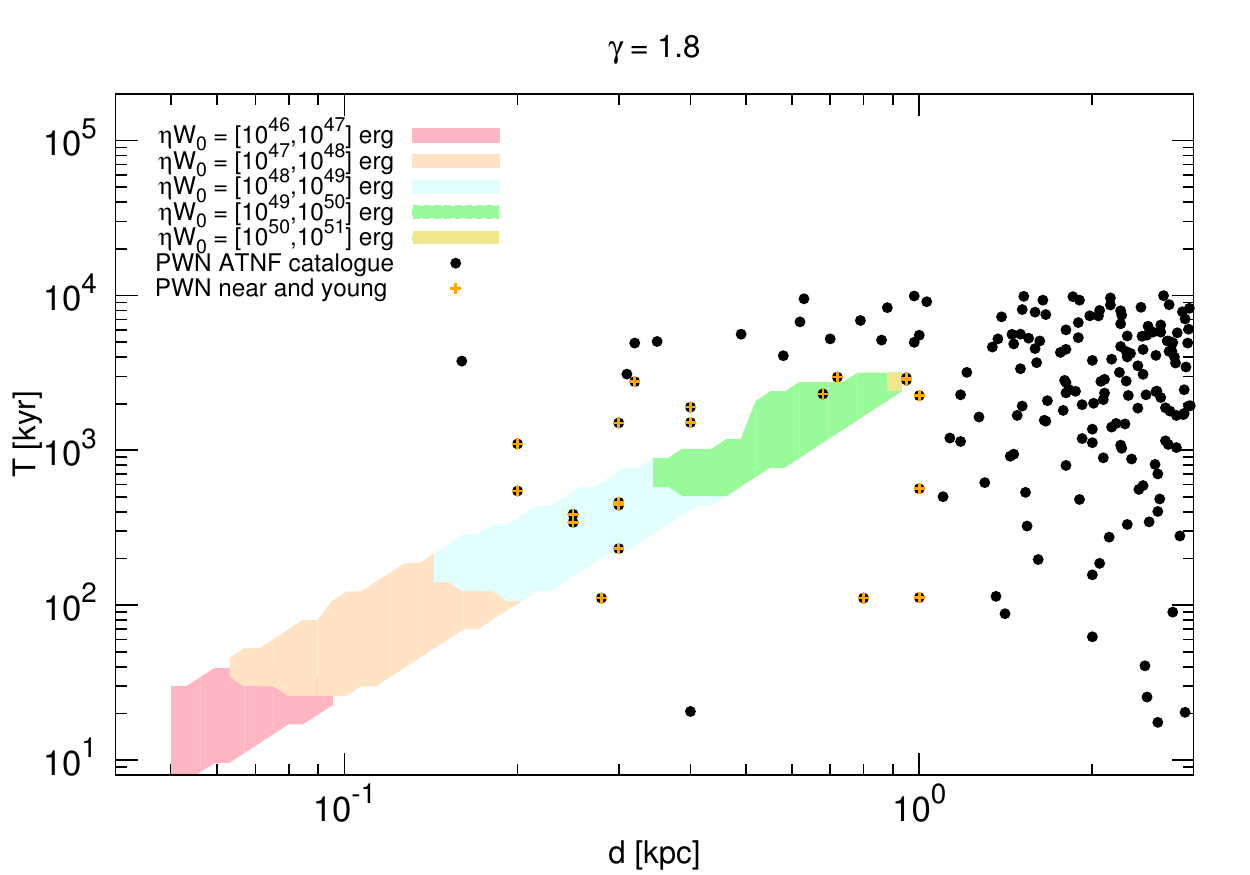}
\includegraphics[width=0.48\textwidth]{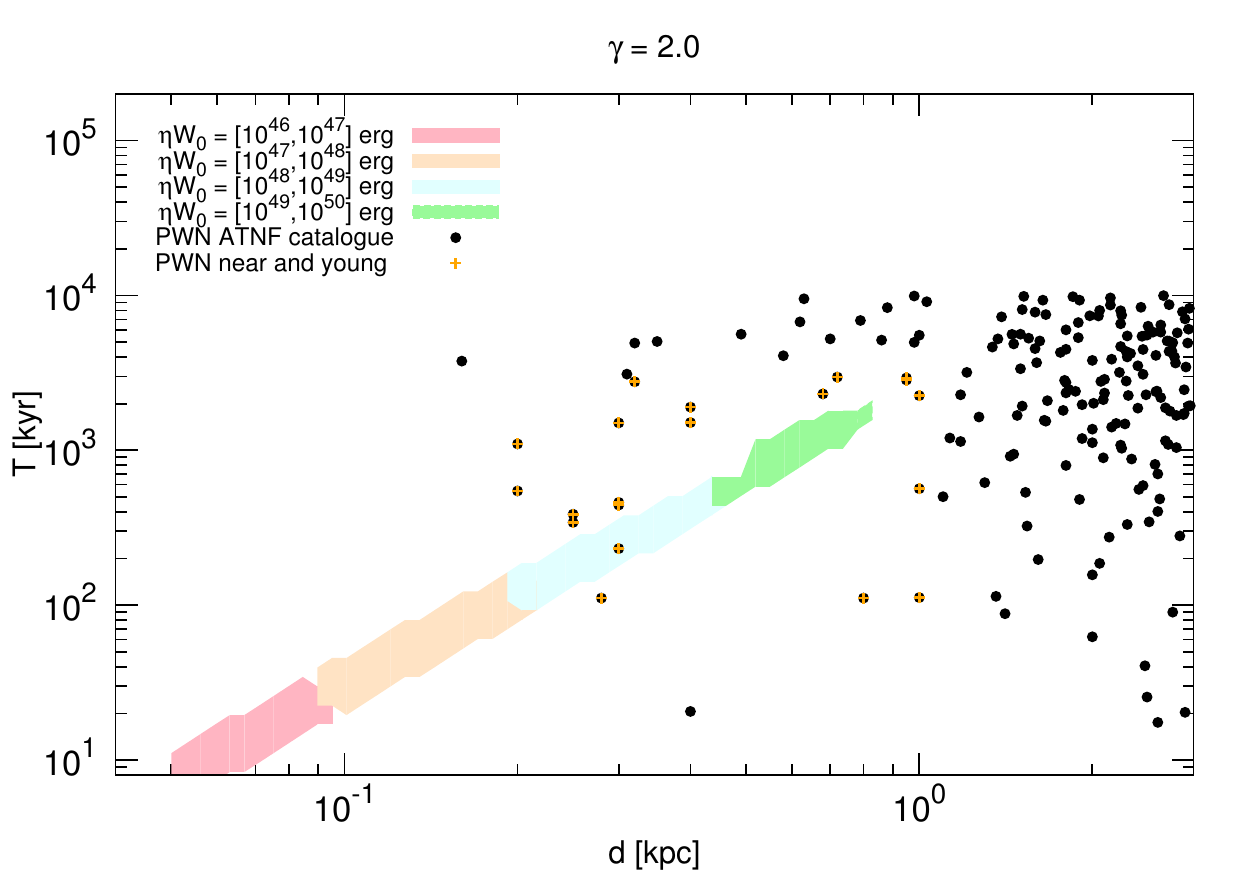}
\includegraphics[width=0.48\textwidth]{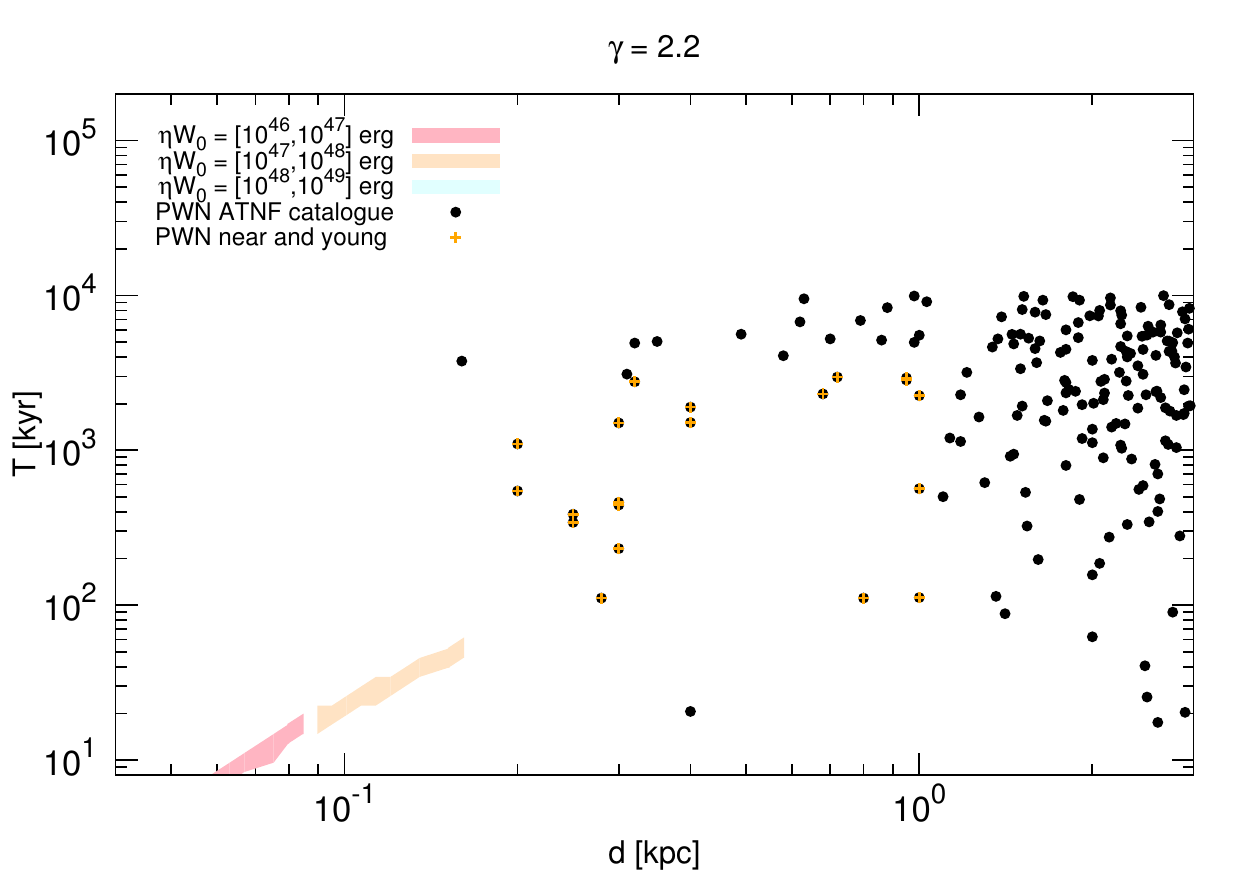}
\caption{``Single-source'' analysis. The colored areas correspond to the 3$\sigma$ allowed regions, in the single-PWN parameter space, compatible with the four observables measured by AMS-02.
The different panels refer to four representative values of the spectral index: $\gamma_{\rm PWN}=$ 1.5, 1.8, 2.0, 2.2. Bands of different colors correspond to various
 decades of values for the parameter $\eta W_0$, as reported in the insets
(from the lower to the upper band: $(10^{46},10^{47})$ erg, $(10^{47},10^{48})$ erg,
$(10^{48},10^{49})$ erg, $(10^{50},10^{51})$ erg, $(10^{51},10^{52})$ erg).
  The circles correspond to PWN listed in the ATNF catalogue (an orange mark differentiates PWN which are young and close, defined as PWN with $T<3000$ kyr and $d<1$ kpc). 
 }
\label{fig:planedt1}
\end{figure}

\section{High-energy window and local sources}
\label{sec:local}

In this Section we attempt additional analyses of the full set of AMS-02 observables, with a special
attention to the interpretation of the higher energy window. Data above about 10 GeV 
are of special interest, since they clearly show a rise in the positron fraction,
which is due to a positron production at high energies much larger than what is expected by secondary interactions only. The analysis of Sect.~\ref{sec:fit} shows that this can be ascribed to the positron
emission from local pulsars. In Sect.~\ref{sec:fit} we considered the whole integrated contribution
from the PWN reported in the ATNF catalog, where each pulsar was tuned to its catalog parameters as
far as the spin-down energy $W_0$ is concerned, while the spectral index $\gamma_{\rm PWN}$ and the efficiency $\eta$ of $e^\pm$ emission were allowed to vary, but they were assumed to be common to all the PWN in the catalog (the actual values of these two parameters were then determined by the fit).

In this Section we attempt a more detailed inspection of the PWN contribution, and
to this aim we carry out two different analyses, with somewhat opposite strategies.
In the first approach (called ``single-source'' analysis) we try to understand if a single, powerful, 
pulsar among the ones present in the ATNF catalog, is able alone to properly explain the high-energy part
of the AMS-02 data, still retaining, in the global analysis, a good agreement in the whole energy range: in fact, we include in our analysis all the electron and positron sources and analyze the whole
energy range for all the four AMS-02 observables, one of which is the single-source PWN. It is a global-fit to the AMS-02 data, where the
PWN contribution is ascribed to a single source. We therefore model a generic single
PWN contribution with free parameters in terms of distance $d$, age $T$, spectral index 
$\gamma_{\rm PWN}$ and energy released in $e^\pm$ (i.e., the quantity $\eta W_0$). We determine the allowed regions in the 4-dimensional parameter space ($d$, $T$, $\gamma_{\rm PWN}$ and $\eta W_0$) and then we check in the ATNF catalog if there are sources which are compatible with the requirements derived from this analysis. We will show that the ATNF catalog contains a few, close and relatively (but not extremely) young PWN which can potentially have the correct properties to explain the high-energy part of
the AMS-02 (and PAMELA, as well) data.

In the second analysis (called ``powerful-sources'' analysis) we take a different approach. We identify
in the ATNF catalog the 5 most powerful sources in terms of spin-down energy. For each of them,
we adopt the distance $d$ and age $T$ provided in the catalog, as well as the spin-down energy $W_0$.
We, instead, allow to be free parameters the efficiency $\eta$ and spectral index $\gamma_{\rm PWN}$
for each one of the five PWN (in total, we therefore have 10 free parameters). By a scan in the 10-dimensional parameter space, we derive statistical distributions of the allowed ranges of the efficiencies and spectral indexes of the five PWN. This allows to determine the preferred properties of the five pulsars and to discern if some of them are constrained to possess specific features 
or, at the opposite, if the data are somehow blind to PWN properties (all this, we recall, under the hypothesis that the 5 most powerful PWN are the dominant contributors). We will find that
in this context, Geminga is required to have a relatively soft emission spectrum and a relatively large efficiency, regardless of the properties of the other four pulsars, which instead are allowed to span a broad band
of values both for the efficiency and for the spectral index.

Finally, we extend the ``powerful-sources'' analisys to comprise an additional component, represented by
all the other PWN listed in ATNF catalogue (a sort of additional ``PWN background''): while
we assume for all of them the position, age and spin-down energy reported in the catalog, we attribute
to them a common spectral index and efficiency. We therefore repeat the analysis with a 12-dimensional
parameter space and derive the statistical distribution of the values which allow a good fit to the
AMS-02 data. In this case, we will obtain that the prominent role of Geminga is reduced (its efficiency can now be smaller than in the previous case) but the additional ``backgound'' pulsars are constrained to possess relatively
small efficiencies.

Let us move now to the discussion of the two types of analyses.

\subsection{``Single-source'' analysis}
\label{sec:singlesource}

In the ``single-source'' analysis we attempt to reproduce the full set of AMS-02 data
by invoking a single PWN contributing to the high-energy part of the positron flux.
While the SNR contribution and the secondaries are fixed at their best-fit configuration
determined in Sect.~\ref{sec:fit}, pulsar emission is attributed to a single source,
for which we vary the spectral index $\gamma_{\rm PWN}$ in the interval $[1.4,2.2]$, the distance $d$ in $[0.01,3]$ kpc, the age $T$ in $[1,20000]$ kyr, 
and the power emitted in the electron/positron channel $\eta W_0$ in $[10^{46}, 10^{51}]$ erg ($\eta$ representing the
efficiency of emission in this channel). We therefore determine the regions in this 4-dimensional parameter space which are able to reproduce the AMS-02 observables. 
The four panels
of Fig.~\ref{fig:planedt1} show the 3$\sigma$ allowed regions in the plane distance-age, at different
values of the spectral index ($\gamma_{\rm PWN}=$ 1.5, 1.8, 2.0, 2.2) and for
various decades of values for the parameter $\eta W_0$, depicted by bands of different
colors (from the lower to the upper band: $(10^{46},10^{47})$ erg, $(10^{47},10^{48})$ erg, $(10^{48},10^{49})$ erg, $(10^{50},10^{51})$ erg, $(10^{51},10^{52})$ erg).
In our 4-dimensional parameter space, the
best-fit configurations has a reduced-$\chi^2$ of 0.45 for 236
data points and 6 degrees of freedom, and the regions denote
the 3$\sigma$ allowed area. 
The figure shows that, regardless of the spectral index, only local (closer than about 1 kpc) and young (age below about 3000 kyr) sources are compatible with the AMS-02 data, and that
very soft spectra would require extremely young and close PWN.

We can now verify if in the ATNF catalog we can identify sources which have the right properties
to explain the AMS-02 leptonic data. This is obtained by reporting, in the same panels
of Fig.~\ref{sec:fit}, the position and age of all observed PWN with $d \le 3$ kpc and $T \le 10000$ kyr (an orange mark differentiates PWN which are young and close, defined as objects with $T<3000$ kyr and $d<1$ kpc). We can see that, depending on the spectral index and on the
emission power, we can identify 9 PWN which are potentially able to explain the AMS-02 data
with a high level of confidence 
These PWN are listed in Tab.~\ref{tab:table2}, together
with their catalog name and parameters. Tab.~\ref{tab:table2} also shows, for different allowed
spectral indexes, the allowed interval for the effective emission power (in the electron/positron channel) $\eta W_{0, \rm fit}$ determined by our fit (in units of $10^{49}$ erg). From
the information on the total emitted power $W_{0, \rm cat}$, we can infer information on
the efficiency $\eta_{\rm fit}$ required by these sources in order to reproduce the data: 
\begin{equation}
\eta_{\rm fit}=\frac{\eta W_{0,\rm fit}}{W_{0,\rm cat}}
\end{equation}  
The last columns of Tab.~\ref{tab:table2} reports the allowed intervals obtained for 
$\eta_{\rm fit}$: we notice that in most of the cases the required emitted power is too large (i.e. $\eta_{\rm fit}$ is too large, even much larger than 1) as compared to observations. However, and most notably, in a few cases the required values of the efficiency are quite
reasonable: this occurs for Geminga, for which efficiencies of the order of $0.3-0.4$ are obtained for a wide interval of spectral indexes (ranging from 1.5 to 1.9); for
B1742-30, where efficiencies of the
order of 0.6 are possible in the case
of hard spectra ($\gamma_{\rm PWM} \sim 1.6-1.7$); and for J1741-2054, with $\eta_{\rm fit} \sim 0.6$ for $\gamma_{\rm PWM} \sim 1.7-1.8$.
All other PWN instead appear quite disfavored. Although in Tab.~\ref{tab:table2} we emphasize (in boldface) all solutions with $\gamma_{\rm PWM}<2$, to account for possible uncertainties in the determination of the emitted power $W_{0,\rm cat}$, we nevertheless conclude that a ``single-source'' solution to the AMS-02 data is indeed possible, but only for a very limited number of PWN, namely Geminga \cite{Hooper:2008kg,Yuksel:2008rf,Cholis:2013psa}, B1742-30 or J1741-2054. 

\begin{figure}[t]
\centering
\includegraphics[width=0.48\textwidth]{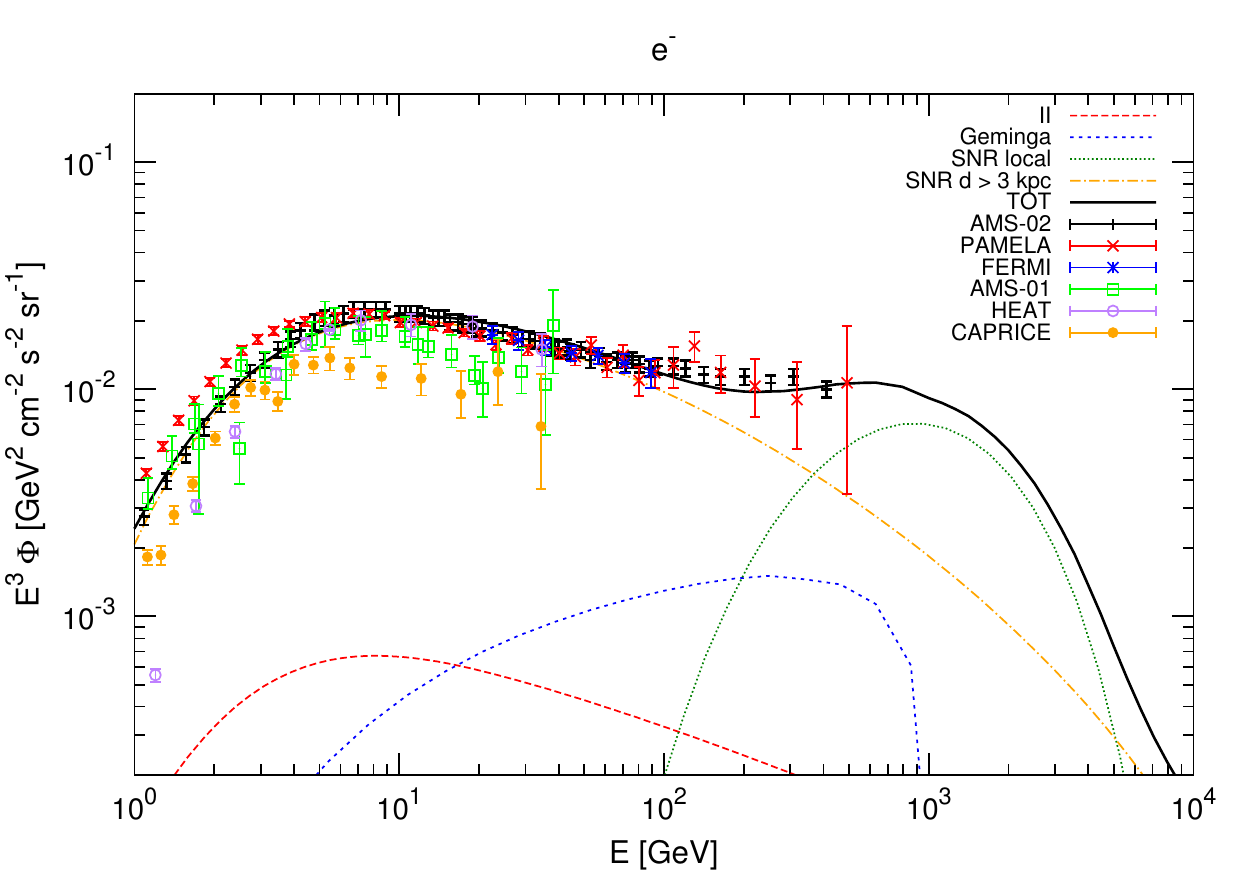}
\includegraphics[width=0.48\textwidth]{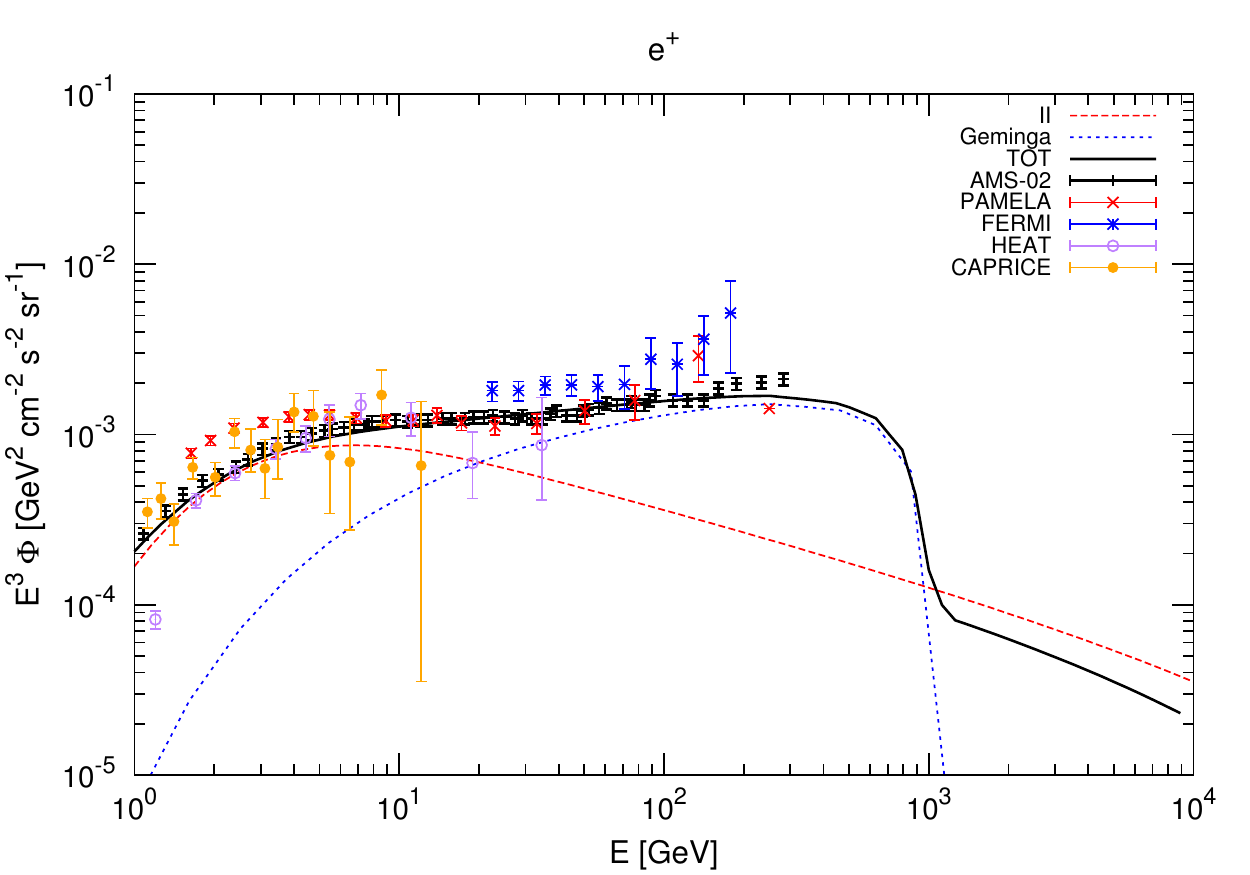}
\includegraphics[width=0.48\textwidth]{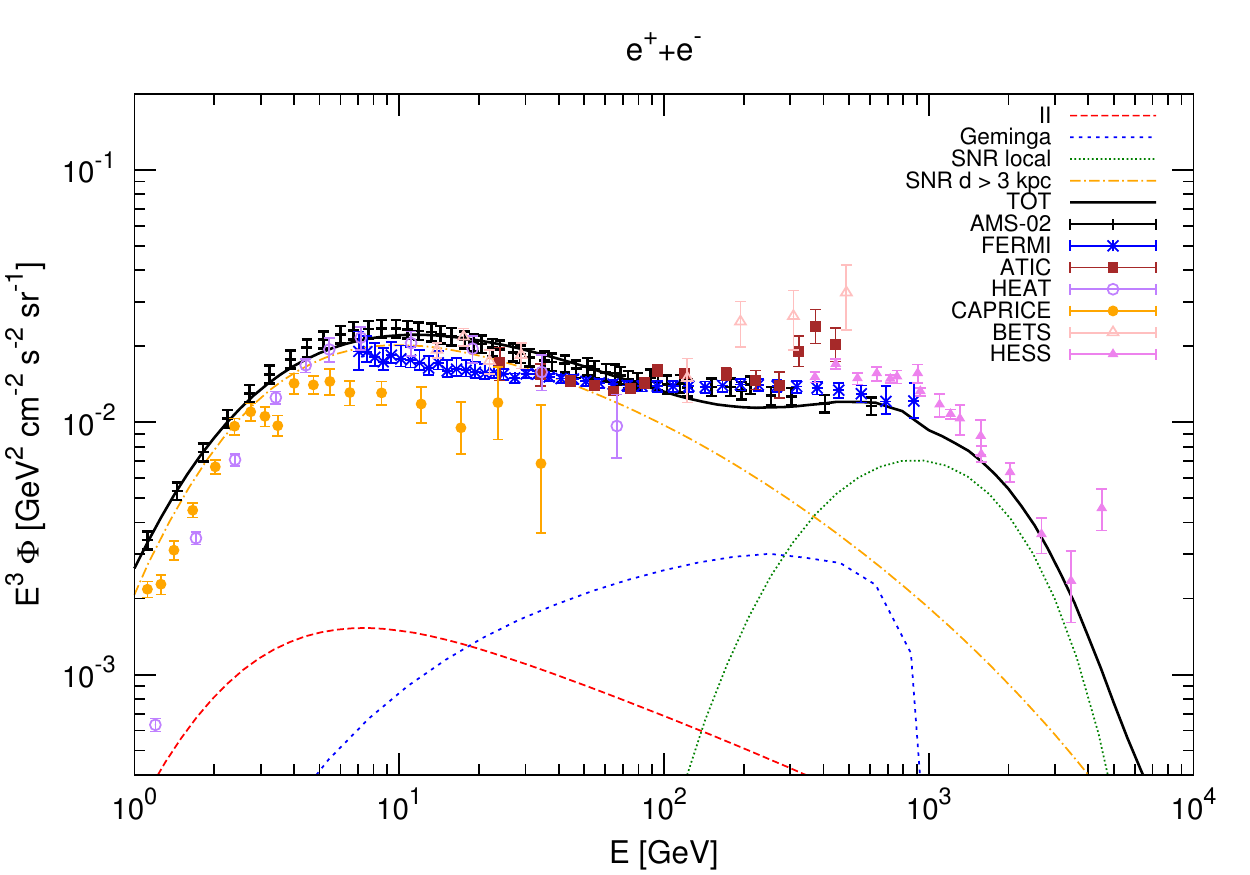}
\includegraphics[width=0.48\textwidth]{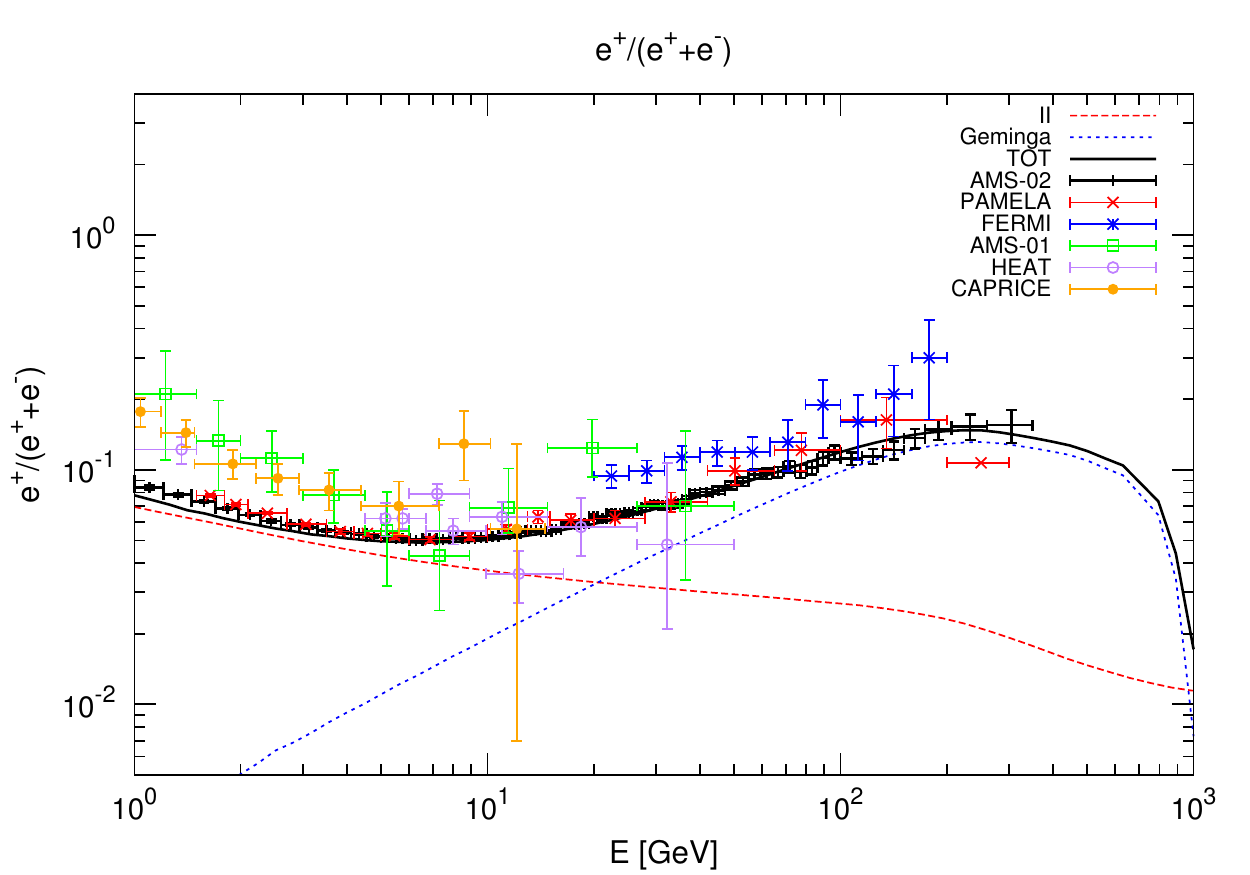}
\caption{``Single-source'' analysis. Results of our simultaneous fit on the AMS-02 data for the electron flux (top left), positron flux (top right), electron plus positron flux (bottom left)
and positron fraction (bottom right) when the pulsar contribution is fully supplied by
Geminga alone. The fit is performed on all the AMS-02 data simultaneously and the
derived Geminga parameters are those reported in Tab.~\ref{tab:table3}.
The colors and styles of the lines are the same as in Fig.~\ref{fig:fit_ele_pos}.
Together with our theoretical model, data from AMS-02 \cite{Aguilar:2013qda,totalelectrons_AMS02,electrons_AMS02}, 
Fermi-LAT \cite{2012PhRvL.108a1103A,2010PhRvD..82i2004A}, 
Pamela \cite{2009Natur.458..607A,2011PhRvL.106t1101A,2013arXiv1308.0133P}, 
Heat \cite{2004PhRvL..93x1102B,1998ApJ...498..779B,1997ApJ...482L.191B,2001ApJ...559..296D}, 
Caprice \cite{2000ApJ...532..653B,2001AdSpR..27..669B}, Bets \cite{2008AdSpR..42.1670Y,2001ApJ...559..973T} and
Hess experiments \cite{2008PhRvL.101z1104A,2009AA...508..561A} are reported.}
\label{fig:bestfit}
\end{figure}

For those three emitters, plus J1918+1541 which is the only remaining candidate which admits
solution with $\gamma_{\rm PWM} < 2$, we show in Tab.~\ref{tab:table3} their best-fit
solutions: it is remarkable (although expected from the above analysis) that the best-fit
values for the distance and age of the sources reported in Tab. \ref{tab:table3} are quite close to the corresponding values in the ATNF
catalog. From Tab.~\ref{tab:table3} we can conclude that in the case of a single-source
contributor, Geminga appears to be best option, with a derived spectral index $\gamma_{\rm PWN} = 1.74$ and efficiency $\eta=0.27$. The electron, positron, electron+positron fluxes
and the positron flux obtained with the Geminga solution are shown in Fig.~\ref{fig:bestfit}.

\begin{table}[th!]
\center
\begin{tabular}{|c|c|c|c|c|c|c|c|c|}
\hline
Name & $d_{\rm{cat}}$ & $T_{\rm{cat}}$& $W_{0,\rm{cat}}$ & $\gamma_{\rm PWN}$ & $\eta W_{0,\rm{fit}}$ & $\eta_{\rm{fit}}$	  \\
\hline
B1742-30   & 0.20 &  546 &  0.829 	& 1.4 & (0.85,1.2)  &	{\bf (1.0,1.5)}     \\
           &      &      &	     	& 1.5 & (0.61,1.00) &	{\bf (0.79,1.2)}	\\
           &      &      &	    	& 1.6 & (0.52,0.85) &	{\bf (0.63,1.0)}	\\
           &      &      & 	    	& 1.7 & (0.52,0.61) &   {\bf (0.63,0.74)}	\\
\hline
B1749-28   & 0.20 &  1100 &	0.700	& 1.4 & (2.3,3.2)	  &	(3.3,4.6)	 	    \\
           &      &       &	       	& 1.5 & (1.9,2.7)	  &	(2.71,3.86)	 	    \\
           &      &       & 	   	& 1.6 & (1.4,1.9)	  &	(2.0,2.7)     	    \\
\hline
Geminga    & 0.25 &  342  &	1.25 	& 1.5 & (0.44,0.61) &	{\bf (0.35,0.41)}	\\
           &      &       &	     	& 1.6 & (0.32,0.52) &	{\bf (0.26,0.42)}   \\
           &      &       &	      	& 1.7 & (0.27,0.44) &	{\bf (0.22,0.35)}	\\
           &      &       &     	& 1.8 & (0.27,0.37) &	{\bf (0.22,4.57)}	\\
           &      &       &	      	& 1.9 & (0.32,0.37) &	{\bf (0.26,0.30)}	\\
\hline
J1741-2054 & 0.25 &  386  &	0.470  	& 1.5 & (0.44,0.61) &	{\bf (0.94,1.1)}	\\
           &      &       &	      	& 1.6 & (0.32,0.52) &	{\bf (0.68,1.1)}	\\
           &      &       &	      	& 1.7 & (0.27,0.44) &	{ \bf (0.57,0.94)}	\\
           &      &       & 	    & 1.8 & (0.27,0.52) &	{\bf (0.57,1.1)}	\\
\hline
B0959-54   & 0.30 &  443  &	0.044 	& 1.5 & (0.72,0.85) &	(16,19)	 	        \\
           &      &       &	     	& 1.6 & (0.44,0.85) &	(10,19)	   	        \\
           &      &       &	      	& 1.7 & (0.44,0.72) &	(10,16)	   	        \\
           &      &       & 	    & 1.8 & (0.44,0.61) &	(10,14)	  	        \\
           &      &       & 	    & 1.9 & (0.37,0.52) &	(8.4,12)	  	    \\
\hline
B0940-55   & 0.30 &  461  &	0.217  	& 1.5 & (0.72,0.85) &	(3.3,3.9)	 	    \\
           &      &       &	      	& 1.6 & (0.44,0.85) &	(2.0,3.9)	  	    \\
           &      &       &	     	& 1.7 & (0.44,0.72) &	(2.2,3.3)	 	    \\
           &      &       & 	  	& 1.8 & (0.44,0.61) &	(2.0,2.8)    \\
           &      &       & 	   	& 1.9 & (0.44,0.52) &	(2.0,2.4)    \\
\hline
B0834+0   & 0.72 	&  2970  	&	0.364 		& 1.6 	& (8.4,10)	&	(23,28)	  	\\
          &        &         	&	     		& 1.7 	& (7.2,8.5)	&	(20,4.23)	    	\\
\hline
J1918+1541  & 0.68  &  2310  	&	3.39 			& 1.6 	& (6.1,10)	&	{\bf (1.8,2.9)}	  	\\
            &       &         	&	     			& 1.7 	& (5.1,8.5)	&	{\bf (1.5,2.5)}	   \\
            &       &         	&	     			& 1.8 	& (4.4,6.1)	&	{\bf (1.3,1.8)}	    \\
            &       &         	&	     			& 1.9 	& (6.1,7.2)	&	{\bf (1.8,2.2)}	   \\
\hline
B1822-09  & 0.30  &  232 		& 	0.0849  	& 1.8 	& (0.19,0.27)	&	(2.2,3.2)  		\\
          &       &      		& 	       		& 1.9 	& (0.23,0.32)	&	(2.7,3.8)	  	\\
          &     	&      		& 	       		& 2.0 	& (0.23,0.32)	&	(2.7,3.8)	   	\\
          &     	&      		& 	       		& 2.1 	& (0.32,0.37)	&	(3.8,4.4)	   	\\
\hline
\end{tabular}
\caption{``Single-source'' analysis.
List of the pulsars reported in the ATNF catalogue whose distance and age lie inside the regions of parameter space compatible with 
AMS-02 measurements, identified by our single-source analysis (for a few representative values of the spectral index $\gamma_{\rm PWN}$, these pulsars
are those shown in Fig.~\ref{fig:planedt1} which fall inside the reconstructed regions).
The columns report the pulsar catalog name, the distance $d_{\rm cat}$ (in kpc), age $T_{\rm cat}$ (in kyr) and total emitted power $W_{0,\rm cat}$ (in units of $10^{49}$ erg) reported
in the ATNF catalog, the spectral index $\gamma_{\rm PWN}$, the range of the emissivity $\eta W_{0,\rm fit}$ for which the source is able 
to reproduce the AMS-02 observables and finally the reconstructed value of the pulsar efficiency which is required to match the emissivity $W_{0,\rm cat}$ (in bold, we emphasize those cases
where this effective efficiency parameter is smaller than 2).
}
\label{tab:table2}
\end{table}
\clearpage

We wish to emphasize that, with the results of this analysis, we are not claiming that we have unambiguously identified the source of the high-energy positron flux (equivalently good solutions have
been obtained in Sect.~\ref{sec:fit}, where all pulsars in the ATNF catalog are contributing,
and others will be found in the next Section with the ``'powerful sources'' analysis). Instead, we attempted to investigate if a solution in terms of a single emitter
is  possible and if the ATNF catalog contains viable candidates, which indeed occurs.
A verification that the sources reported in Tab.~\ref{tab:table3} have the spectral index and efficiency quoted in the table would require additional observational data. 
At the same time, we wish to comment that the use of catalog sources might be biased
from incompleteness of the catalog. The ATNF catalog might not (and very likely, does not)
contain all local PWN, since for a fraction of them the electromagnetic emission may not
be resolved. This might occur for the SNR in the Green catalog, as well. Nevertheless, it is remarkable that current data can be properly and fully explained in terms of known sources. In the case of the ``single-source analysis'' discussed in this Section, the results of Fig. \ref{fig:planedt1} can also be interpreted as bringing information on the age-distance parameters required for any unknown single, powerful PWN to explain the AMS-02 leptonic 
data: any putative source likely needs to be closer than 1 kpc and younger than about 3000 kyr,
with specific correlations with its spectral index and emitted power, as reported in the panels of Fig. \ref{fig:planedt1}.

\begin{table}
\center
\begin{tabular}  {|c|c|c|c|c|c||c|c|c||c|}
\hline
Name &	$\gamma_{\rm fit}$ 	&	$d_{\rm fit}$ &	$T_{\rm fit}$	&    	$\eta W_{0,\rm fit}$	&  $\chi^2/{\rm dof}$  &	$d_{\rm cat}$  &		$T_{\rm cat}$ &	$W_{0,\rm cat}$ & $\eta_{\rm fit} $ \\
\hline
Geminga    & 1.74 & 0.24 & 344.6 & 0.341 & 0.68 &	0.25 & 342 & 1.25 & 0.27 \\
J1741-2054 & 1.68 & 0.25 & 378.0 &	 0.413 & 0.62 &	0.25 & 386 & 0.47 & 0.88 \\	
B1742-30   & 1.52 & 0.19 & 539.1 &	 0.770 & 0.54 &	0.2	 & 546 & 0.83 & 0.92 \\
\hline 
J1918+1541 & 1.65 & 0.64 & 2355  &	 6.48  & 0.92 & 0.68 & 2310	 & 3.4 & 1.90 \\
\hline
\end{tabular}
\caption{``Single-source'' analysis.  For the four PWN identified in the ``single-source'' analysis as those which
can provide the best agreement to the AMS-02 data (the ones that in Tab.~\ref{tab:table2} exhibit an effective efficiency smaller than 2), we report here the best-fit values obtained
for their spectral index $\gamma_{\rm fit}$, distance $d_{\rm fit}$ (in kpc), 
age $T_{\rm fit}$ (in kyr) and emitted power $\eta W_{0,\rm fit}$ (in units of $10^{49}$ erg), and for
comparison the catalog values of distance $d_{\rm cat}$, age $T_{\rm cat}$
and power $W_{0,\rm cat}$. From $\eta W_{0,\rm fit}$ and $W_{0,\rm cat}$ we derive the
efficiency $\eta_{\rm fit}$.
}
\label{tab:table3}
\end{table}

\subsection{``Powerful-sources'' analisys}
\label{sec:powsource}

In this Section we adopt a somehow complementary approach to the ``single-source analysis''
discussed in Sect. \ref{sec:singlesource}: we identify, in the ATNF catalog, a limited number of PWN
which are potentially able to sizably contribute to the local positron flux at high-energies, and we use
them in the global analysis of the full set of AMS-02. For definiteness, we adopt the 5 ``most powerful'' sources, and for each of them we allow a free spectral index $\gamma_{\rm PWN}$ and a free
efficiency factor $\eta$, which are then determined by fitting the AMS-02 data. All the other leptonic components (SNR and secondaries) are taken at their best-fit configuration of Sect. \ref{sec:fit}, for definiteness. We label this analysis "powerful-sources'' analysis.

The criterion used to identify the 5 ``most powerful'' sources relies on a ranking-algorithm based
on the contribution of the PWN to the high-energy part of the positron flux. Since pulsars contribution
becomes dominant above about 10 GeV, we have subdivided the energy range $(10-550)$ GeV into 4 equally spaced in log-scale bins, and for each bin we have calculated the integrated positron flux for all the PWN
present in the ATNF catalog, by adopting for them a common spectral index and efficiency 
(taken at
the best-fit values of Sect.  \ref{sec:fit}). By using the calculated fluxes $\Phi_i^a$ ($i=1,\dots,4$ counts the energy bins, $a$ counts the PWN in the catalog) we have created a rank $R_i^a$ of the sources $a$ in each bin $i$ ($R_i^a = 1,2,\dots$ for the most-powerful, second most-powerful, and so on).  The 5 ``most-powerful'' sources are then identified as those who possess the highest ranking $\overline{R^a} = \sum_i R_i^a$ (i.e. the smaller value of $\overline{R^a}$). These pulsars are listed in Table \ref{tab:pwn}.

\begin{figure}[t]
\centering
\includegraphics[width=0.48\textwidth]{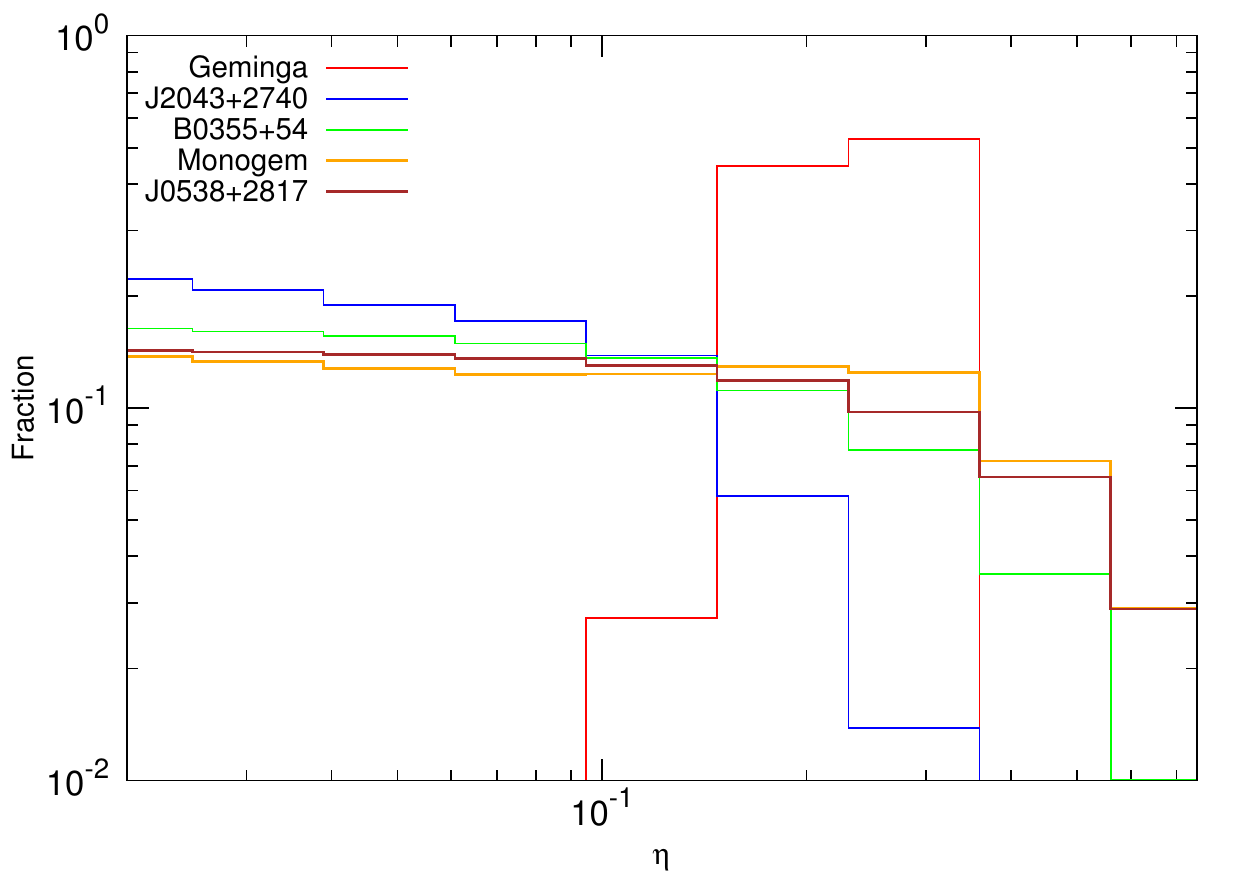}
\includegraphics[width=0.50\textwidth]{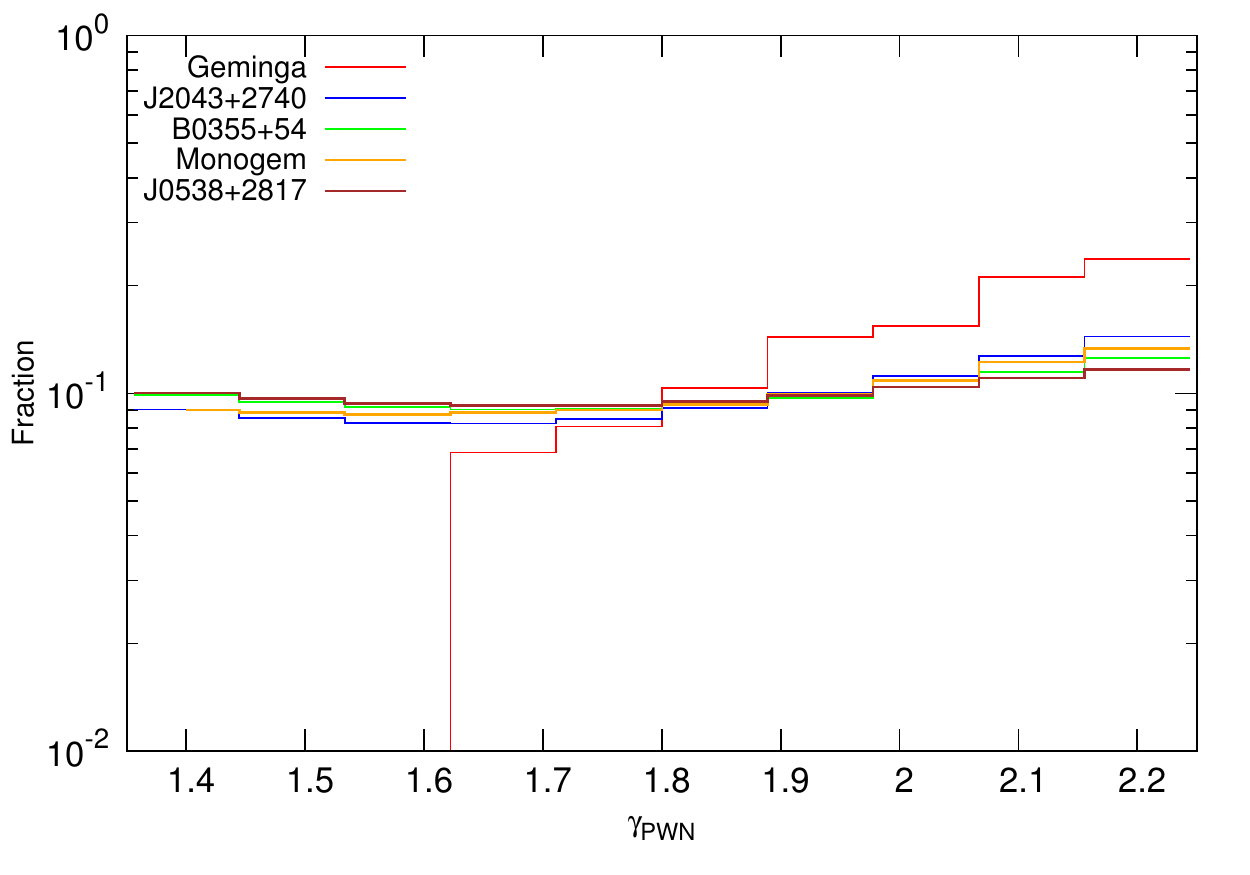}
\caption{``Powerful-source analysis''. Frequency distribution of the values of the efficiency $\eta$ (left) and the spectral index $\gamma$ (right) of the five ``most powerful'' sources. The distributions refer to the PWN configurations which lie inside the 3$\sigma$ allowed region around the best-fit configuration on the AMS-02 full data set.}
\label{fig:etagamma}
\end{figure}

Now that we have identified the PWN to be used in the analysis, we allow for them a variation of the spectral index $\gamma_{\rm PWN}$ and efficiency $\eta$ parameters, while assuming their distance $d$, age $T$ and total emitted power $W_0$ at their catalog values. The analysis therefore relies on 10 free parameters, which are varied independently. By fitting the whole set of AMS-02
data, we can identify the best-fit configuration and the corresponding 3$\sigma$ allowed region in this
10-dimensional parameter space. For those configurations falling in the 3$\sigma$ allowed region,
Fig. \ref{fig:etagamma} shows the frequency distribution of
the two parameters for each of the 5 ``'most-powerful'' sources. We can notice that there is
a preferred trend for Geminga: the efficiency is required to be larger than about 0.1, with a peak value
around $0.2-0.3$ (not far from the best-fit value 0.27 obtained in the ``single-source'' analysis) and that
its spectral index is lower-bounded arounded 1.6, with a small (but not significant) preference toward
softer spectra (we can notice that in the case Geminga is the only, largely dominant, contributor
the ``single-source'' analysis has determined a best-fit value of 1.74). The other four ``most-powerful'' PWN are much less constrained: they have a mild preference for efficiencies lower than 0.1 and no clear
preference for the value of the spectral index.

\begin{table}
\center
\begin{tabular}{|c|c|c|c|c|}
\hline
ATNF	&   Association & $d [\rm{kpc}]$&	$T [\rm{kyr}]$	&	$W_0 [10^{49}\,\rm{erg}]$	\\
\hline
J0633+1746 	& 	Geminga	&	0.25  	&	343 			&	1.26		\\
J2043+2740 	& 	 		&	1.13  	&	1204  		&	26.0		\\
B0355+54  	&			&  	1  		&	567  		&	4.73		\\
B0656+14  	&	Monogem	&	0.28  	&	112  		&	0.178	\\
J0538+2817 	& 			&	1.3 		&	622  		&	6.18		\\
\hline
\end{tabular}
\caption{``Powerful-sources analysis''. List of the 5 pulsars identified as ``most-powerful''
with the criteria defined in the text, as used in the analysis.
The columns report the ATNF-catalog name, the association name,  the distance $d$ $[\rm{kpc}]$,
 the age $T$ $[\rm{kyr}]$ and the emitted power $W_0$ (in units of $10^{49}$ erg), as reported in the
catalog.
\label{tab:pwn}}
\end{table}

\begin{figure}[t]
\centering
\includegraphics[width=0.48\textwidth]{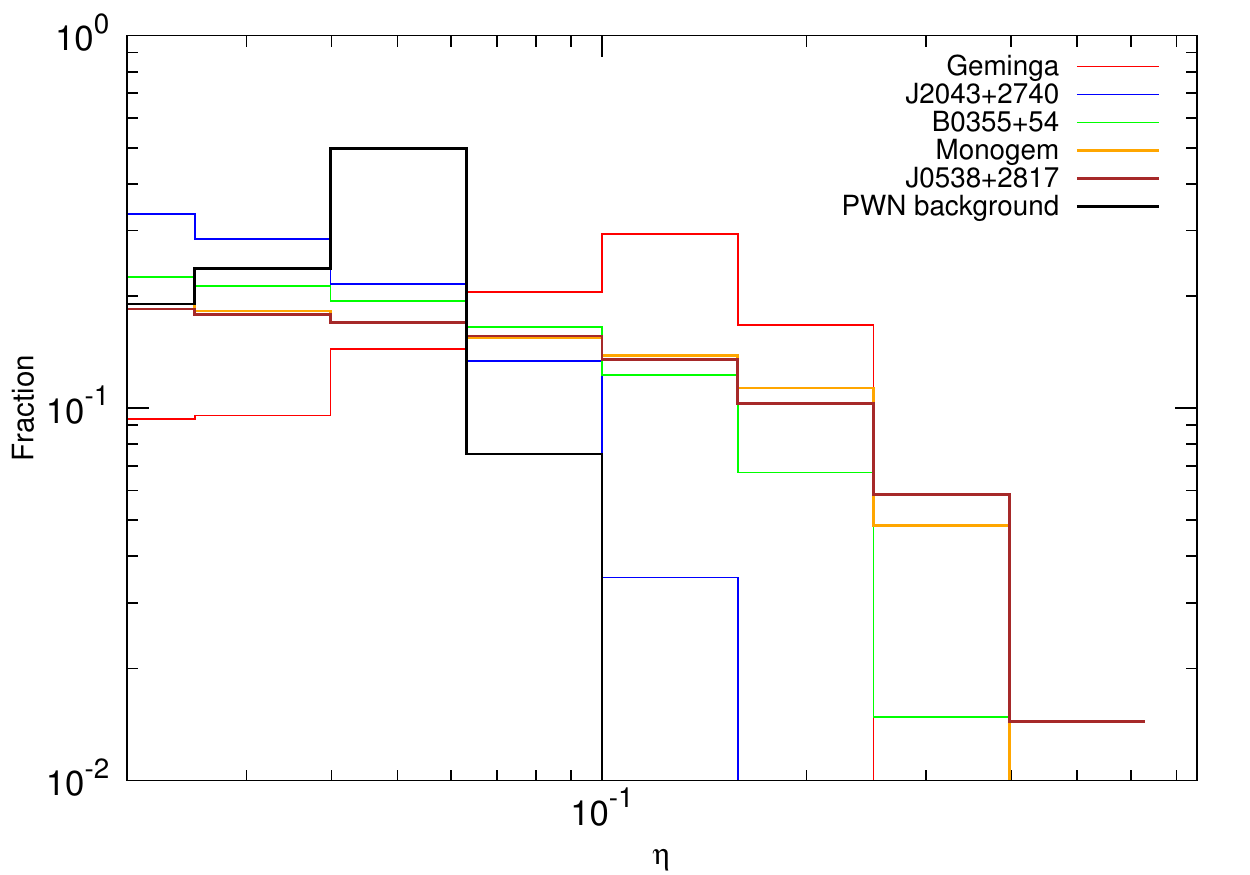}
\includegraphics[width=0.48\textwidth]{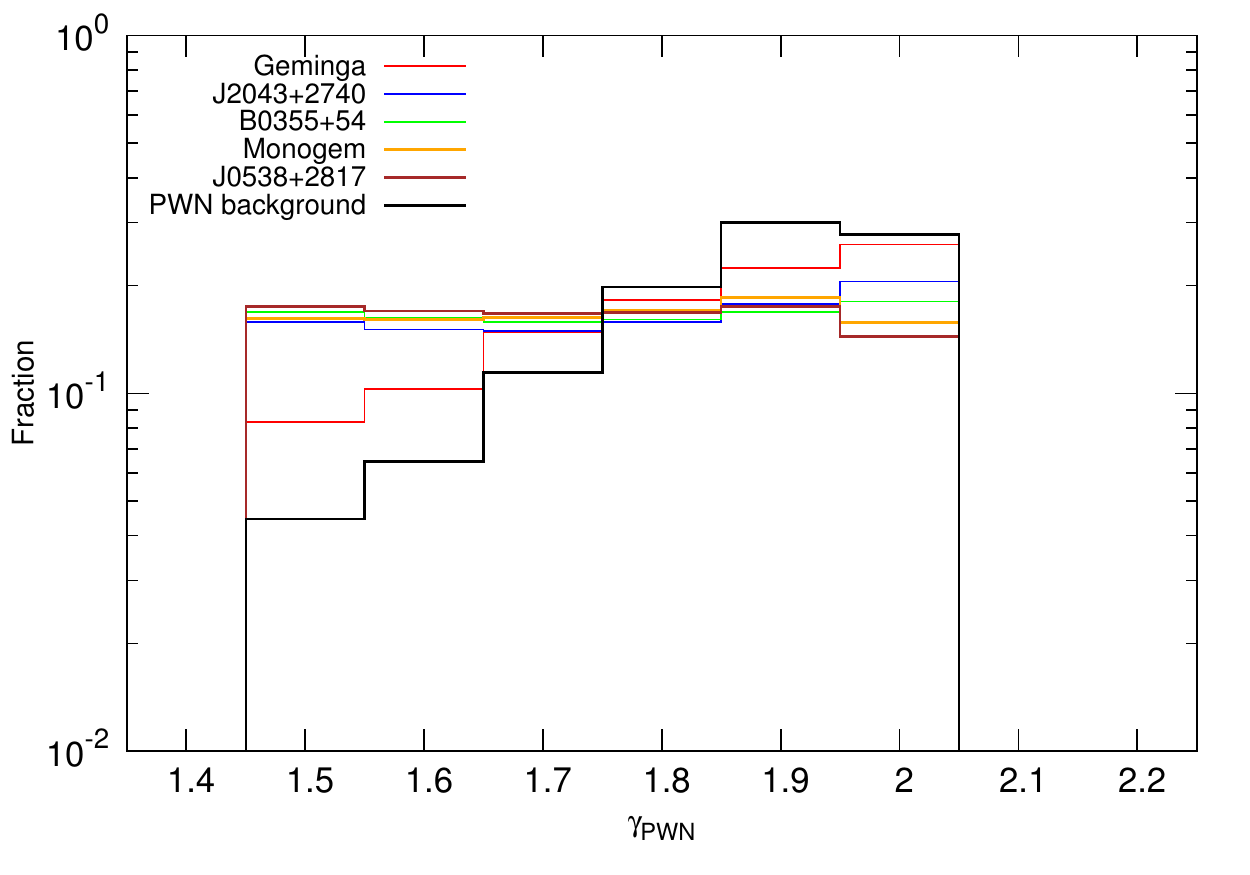}
\caption{``Powerful-source analysis''. The same as in Fig. \ref{fig:etagamma}, with the difference that,
in addition
to the 5 ``most-powerful'' pulsars, an aggregate contribution from all the additional PWN in the ATNF catalog is added (considered as a ``PWN background''). All the additional pulsars are assumed to have
a common efficiency and a common spectral index.
\label{fig:etagamman} 
}
\end{figure}

In order to understand the role of the additional PWN present in the ATNF catalog, we have performed an extended version of the ``power-source'' analysis where, in addition to the 5 sources defined above, we have added the contribution of 
all the remaining pulsars in the catalog (a sort of ``PWN background'', just to fix a denomination): for each of them, we adopt a common spectral index and efficiency, while the other 3 parameters ($d$, $T$, and $W_0$) are taken at the value reported in the catalog. The analysis now deals with 12 parameters.
We have again performed a fit on the whole set of AMS-02 data, identified the 3$\sigma$ allowed
regions around the best-fit configuration on the 12-dimensional parameter space:
Fig. \ref{fig:etagamman} shows the frequency distribution of the values of $\gamma_{\rm PWN}$
and $\eta$ for each of the 5 ``most-powerful'' pulsars, as well as the frequency distribution
of the spectral index and efficiency of the pulsars contributing to the ``PWN background''. The
presence of the additional pulsars makes the role of Geminga less relevant, as can be seen by the fact that
now the allowed interval for the efficiency is widely distributed, contrary to the previous case: while the
most probable value is still around $0.1 - 0.2$, much lower efficiencies are now accepted, while only
efficiencies in excess of 0.1 were accepted with a negligible contribution of the ``PWN background''.
The additional pulsars have a tendency toward low efficiencies, around 0.05. The spectral features
do not exhibit strong preferences, except for Geminga and for the ``PWN background'', where a mild
tendency toward soft spectra appears, as shown in the right panel of  Fig. \ref{fig:etagamman}.

\begin{figure}[t]
\centering
\includegraphics[width=0.48\textwidth]{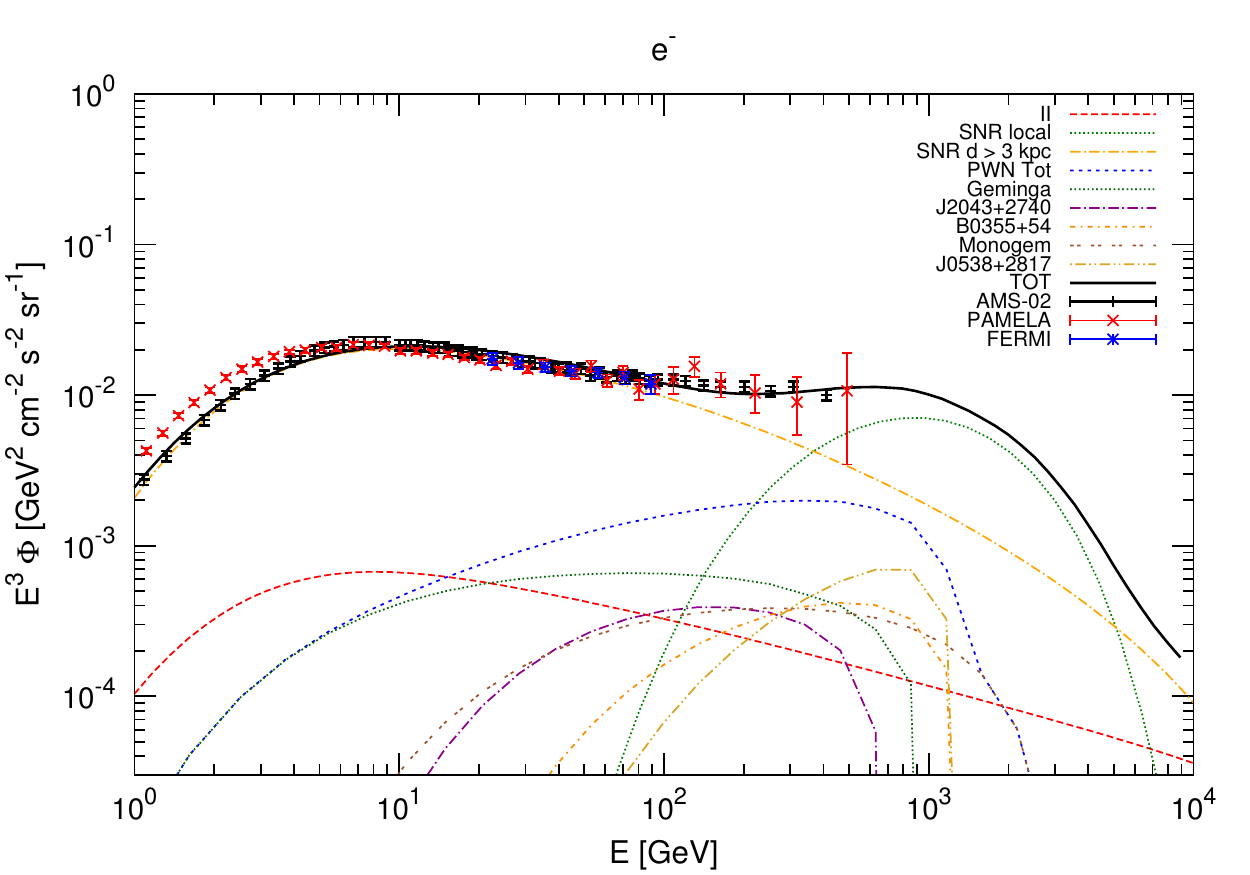}
\includegraphics[width=0.48\textwidth]{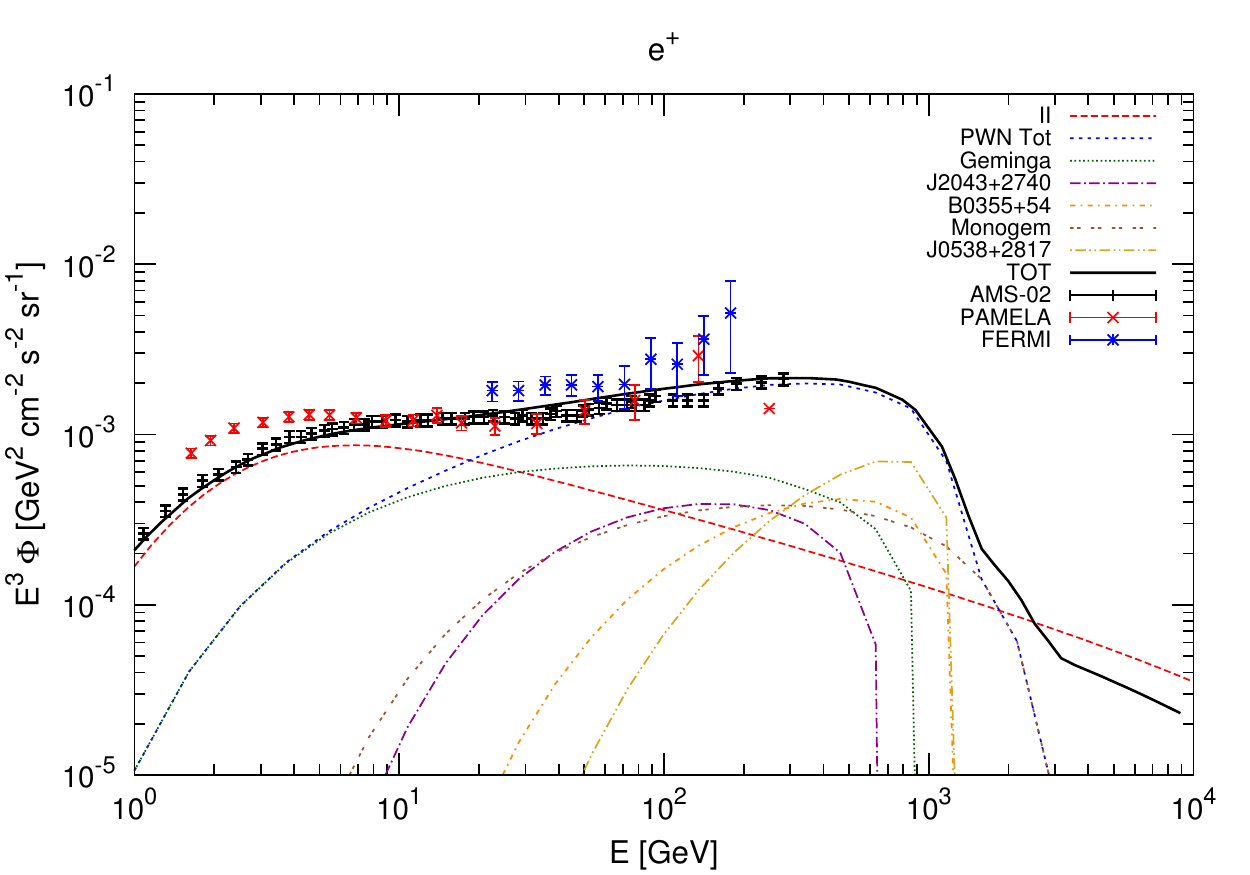}
\includegraphics[width=0.48\textwidth]{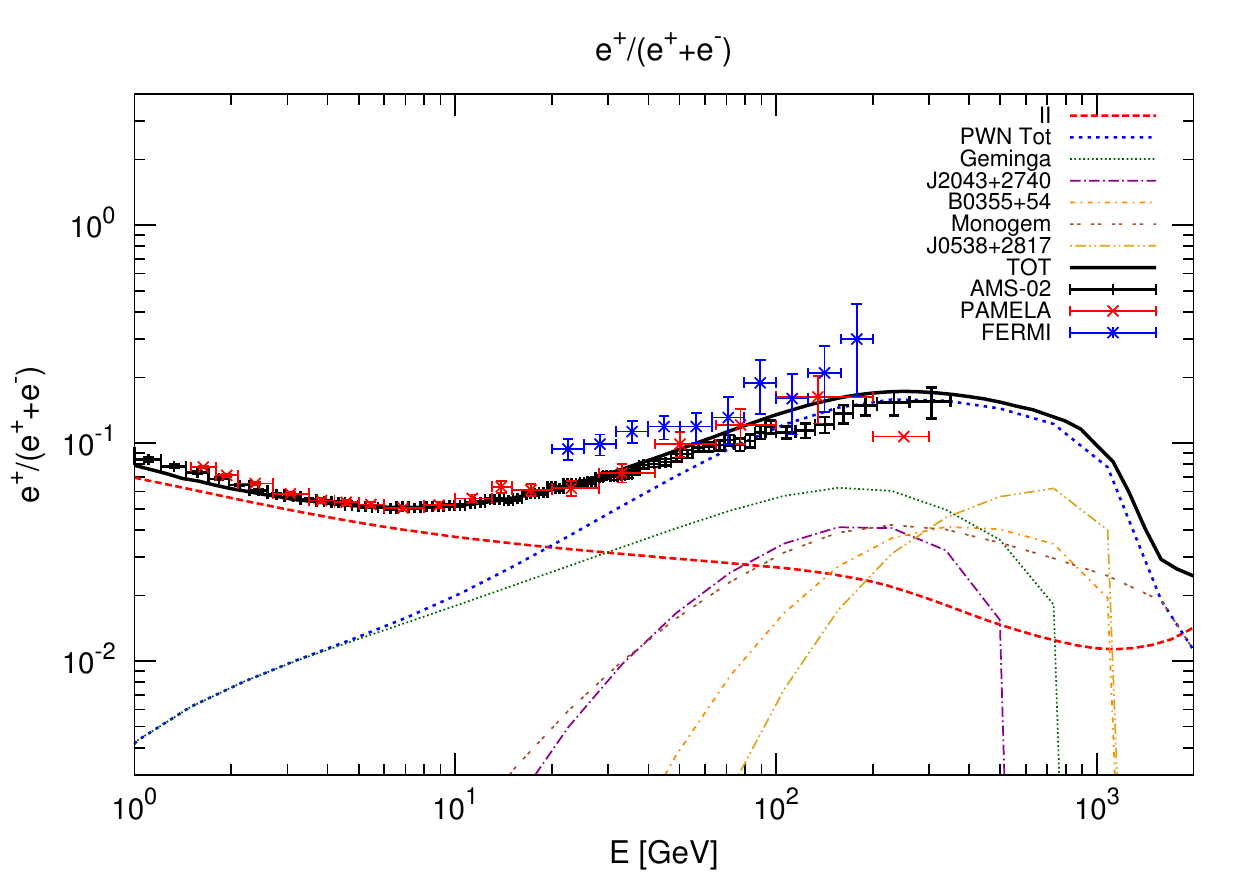}
\includegraphics[width=0.48\textwidth]{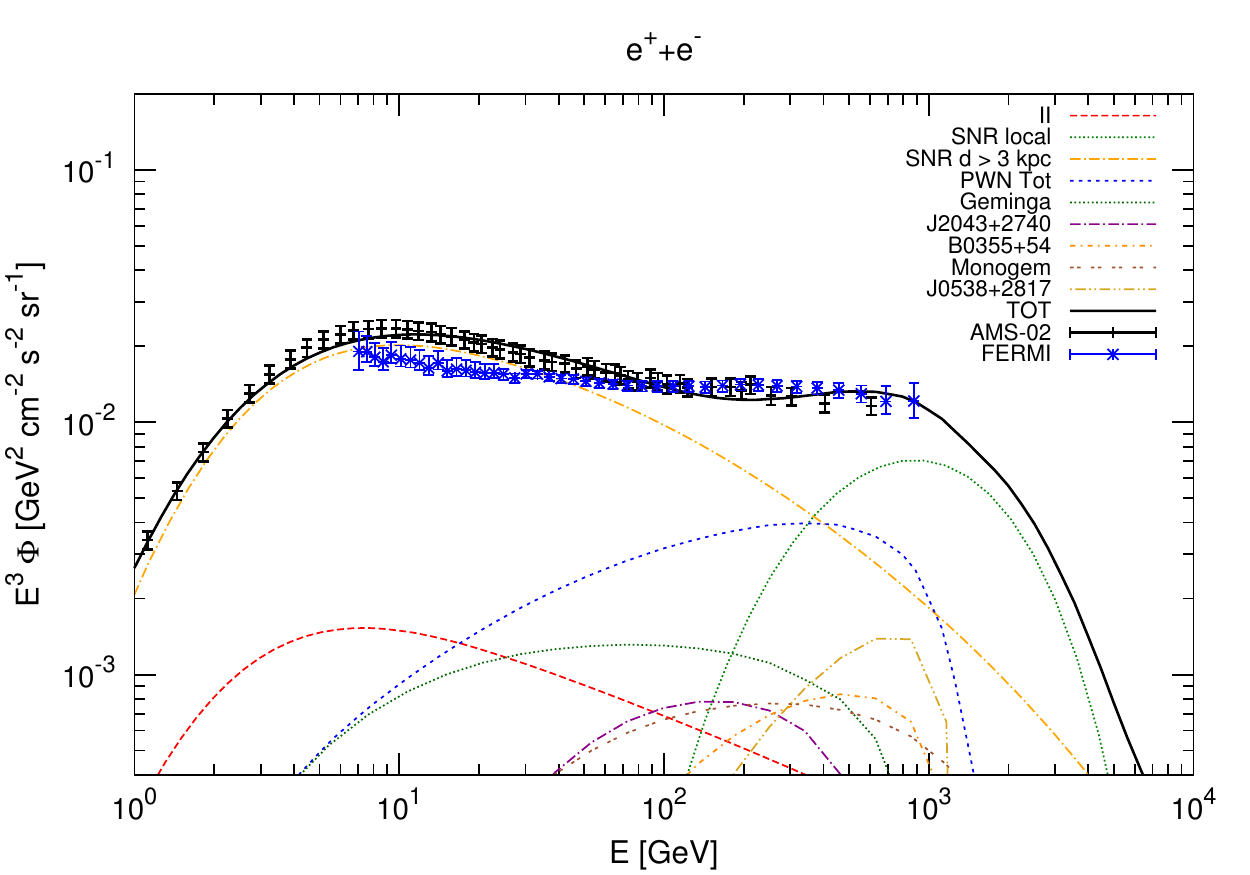}
\caption{``Powerful-source'' analysis. Results of our simultaneous fit on the AMS-02 data for the electron flux (top left), positron flux (top right), electron plus positron flux (bottom left)
and positron fraction (bottom right) when the pulsar contribution is supplied by
the 5 ``most-powerful'' pulsars in the ATNF catalog, listed in Table \ref{tab:pwn}. The fit is performed on all the AMS-02 data simultaneously and the result shown
in the figure refers to best-fit configuration for the
5+1 efficiences and the 5+1 spectra indexes.
The colors and styles of the lines are the same as in Fig.~\ref{fig:fit_ele_pos}.
Together with our theoretical model, data from AMS-02 \cite{Aguilar:2013qda,totalelectrons_AMS02,electrons_AMS02}, 
Fermi-LAT \cite{2012PhRvL.108a1103A,2010PhRvD..82i2004A}, 
Pamela \cite{2009Natur.458..607A,2011PhRvL.106t1101A,2013arXiv1308.0133P}, 
Heat \cite{2004PhRvL..93x1102B,1998ApJ...498..779B,1997ApJ...482L.191B,2001ApJ...559..296D}, 
Caprice \cite{2000ApJ...532..653B,2001AdSpR..27..669B}, Bets \cite{2008AdSpR..42.1670Y,2001ApJ...559..973T} and
Hess experiments \cite{2008PhRvL.101z1104A,2009AA...508..561A} are reported.
}
\label{fig:powerful}
\end{figure}

The kind of agreement which can be obtained with the ``powerful-sources'' approach can be appreciated in Fig. \ref{fig:powerful}, where we show the best-fit configuration of the analysis for the 5 ``most-powerful'' pulsars for the combined analysis of the electron flux (top left), positron flux (top right), electron plus positron flux (bottom left) and positron fraction (bottom right). All four data sets
are reproduced quite well, with a similar level of agreement already obtained with the other approaches
discusses above. The best-fit configuration corresponds to a reduced-$\chi^2$ of 0.41 for 236 data points and 12 free parameters. Also in this case, the agreement is remarkably good.

\section{Conclusions}
\label{sec:concl}

In this paper we have performed a combined analysis of the recent AMS-02 data on the
electron flux, positron flux, electrons plus positrons flux and positron fraction, in a theoretical  framework that self-consistently accounts for all the astrophysical components able to contribute to the
leptonic fluxes in their whole energy range. 

The primary electron contribution is modeled through the sum of an average flux produced by distant sources and the fluxes arising from the local supernova remnants in the Green catalog. 
The secondary electron and positron fluxes originate from interactions on the interstellar medium of primary cosmic rays, for which we have derived a novel determination by using AMS-02 proton
and helium data. Finally, the pulsar wind nebulae contribution to the positron (and electron) fluxes
at high energies relies on a modeling of the sources reported in the ATNF catalog (where
information on age, distance and total emitted power of the pulsars is available), under a number
of  assumptions on the way the different pulsars might contribute to the local
fluxes. We have in fact specifically performed three different type of analysis: we have studied the average contribution from the whole catalog; we have investigated if the ATNF catalog contains a
single, dominant, pulsar which alone can allow agreement with the data; finally we have
examined the possibility that a few powerful sources in the ATNF catalog may concurrently
contribute to the local observed fluxes.

For all three different types of analysis, we obtain a remarkable agreement between our modelings and the whole set of AMS-02 data. The supernova remnants and the secondary contribution are able to
properly explain the electron data and the low-energy part of the positron spectra, and to some extent they also point toward
a disfavoring of small cosmic-rays confinement volumes. The high-energy part of the positron
flux, which has received great attention because of its implications not only for the astrophysics
of sources but also for dark matter studies, finds a remarkable solution in terms of pulsars present
in the ATNF catalog. We find that AMS-02 data can be properly explained either in the case of an average contribution from the whole catalog, or for the situation where a single and close pulsar is the dominant contributor, or even in the case where a few and powerful dominant pulsars in the catalog are mostly contributing. For all cases, we have identified the required ranges of the relevant parameters (spectral index and efficiency of the emission) for the contributing pulsars, once the other parameters (age, distance, total emitted power) are fixed at their values reported in the ATNF catalog.

We can therefore conclude that the whole set of AMS-02 leptonic data admits a self-consistent interpretation in terms of astrophysical contributions.  Alternative solutions, like e.g. a dark matter production of electrons and positrons, are indeed a viable alternative or complementary possibility: however, a self-consistent solution in terms of purely astrophysical sources can be properly met.


\acknowledgments

We thank Pasquale Serpico for interesting and useful discussions on the physics of supernova remnants and pulsar wind nebulae and Timur Delahaye for discussions on the study of electron and positron cosmic rays. This
work is supported by the research grant {\sl Theoretical Astroparticle Physics} number 2012CPPYP7 under the program PRIN 2012  funded by the Ministero dell'Istruzione, Universit\`a e della Ricerca (MIUR), by the research grant {\sl TAsP (Theoretical Astroparticle Physics)}
funded by the Istituto Nazionale di Fisica Nucleare (INFN), by the  {\sl Strategic Research Grant: Origin and Detection of Galactic and Extragalactic Cosmic Rays} funded by Torino University and Compagnia di San Paolo, by the Spanish MINECO under grants FPA2011-22975 and MULTIDARK CSD2009-00064 (Consolider-Ingenio 2010 Programme), by Prometeo/2009/091 (Generalitat Valenciana), and by the EU ITN UNILHC PITN-GA-2009-237920. This work is also supported in part by the European Research Council ({\sc Erc}) under the EU Seventh Framework Programme (FP7/2007-2013) / {\sc Erc} Starting Grant (agreement n. 278234 - `{\sc NewDark}' project).
At LAPTh this activity was supported by the Labex grant ENIGMASS.
R.L. is supported by a Juan de la Cierva contract (MINECO). 
A.V. acknowledges the hospitality of the Institut d'Astrophysique de Paris (IAP). 

\bibliography{positrons2014}

\providecommand{\href}[2]{#2}\begingroup\raggedright\begin{thebibliography}{100}

\bibitem{2009Natur.458..607A}
O.~{Adriani}, G.~C. {Barbarino}, G.~A. {Bazilevskaya}, et~al., {\it {An
  anomalous positron abundance in cosmic rays with energies 1.5-100GeV}},  {\em
  \nat} {\bf 458} (Apr., 2009) 607--609,
  [\href{http://xxx.lanl.gov/abs/0810.4995}{{\tt arXiv:0810.4995}}].

\bibitem{2013arXiv1308.0133P}
{PAMELA Collaboration}, {\it {The cosmic-ray positron energy spectrum measured
  by PAMELA}},  {\em ArXiv e-prints} (Aug., 2013)
  [\href{http://xxx.lanl.gov/abs/1308.0133}{{\tt arXiv:1308.0133}}].

\bibitem{2011PhRvL.106t1101A}
O.~{Adriani}, G.~C. {Barbarino}, G.~A. {Bazilevskaya}, et~al., {\it {Cosmic-Ray
  Electron Flux Measured by the PAMELA Experiment between 1 and 625 GeV}},
  {\em Phys.Rev.Lett.} {\bf 106} (May, 2011) 201101,
  [\href{http://xxx.lanl.gov/abs/1103.2880}{{\tt arXiv:1103.2880}}].

\bibitem{2012PhRvL.108a1103A}
M.~{Ackermann}, M.~{Ajello}, {Allafort}, et~al., {\it {Measurement of Separate
  Cosmic-Ray Electron and Positron Spectra with the Fermi Large Area
  Telescope}},  {\em Phys.Rev.Lett.} {\bf 108} (Jan., 2012) 011103,
  [\href{http://xxx.lanl.gov/abs/1109.0521}{{\tt arXiv:1109.0521}}].

\bibitem{2010PhRvD..82i2004A}
M.~{Ackermann}, M.~{Ajello}, W.~B. {Atwood}, L.~{Baldini}, et~al., {\it {Fermi
  LAT observations of cosmic-ray electrons from 7 GeV to 1 TeV}},  {\em \prd}
  {\bf 82} (Nov., 2010) 092004, [\href{http://xxx.lanl.gov/abs/1008.3999}{{\tt
  arXiv:1008.3999}}].

\bibitem{Aguilar:2013qda}
{\bf AMS} Collaboration, M.~Aguilar et~al., {\it {First Result from the Alpha
  Magnetic Spectrometer on the International Space Station: Precision
  Measurement of the Positron Fraction in Primary Cosmic Rays of 0.5--350
  GeV}},  {\em Phys.Rev.Lett.} {\bf 110} (2013), no.~14 141102.

\bibitem{electrons_AMS02}
{S. Shael} and {the Ams-02 Collaboration}, {\it {Precision measurements of the
  electron spectrum and the positron spectrum with AMS}},  {\em Talk at the
  33rd ICRC Conference} (2013).

\bibitem{totalelectrons_AMS02}
{B. Bertucci} and {the Ams-02 Collaboration}, {\it {Precision measurement of
  the electron plus positron spectrum with AMS}},  {\em Talk at the 33rd ICRC
  Conference} (2013).

\bibitem{2012APh....39....2S}
P.~D. {Serpico}, {\it {Astrophysical models for the origin of the positron
  ''excess''}},  {\em Astroparticle Physics} {\bf 39} (Dec., 2012) 2--11,
  [\href{http://xxx.lanl.gov/abs/1108.4827}{{\tt arXiv:1108.4827}}].

\bibitem{Hooper:2008kg}
D.~Hooper, P.~Blasi, and P.~D. Serpico, {\it {Pulsars as the Sources of High
  Energy Cosmic Ray Positrons}},  {\em JCAP} {\bf 0901} (2009) 025,
  [\href{http://xxx.lanl.gov/abs/0810.1527}{{\tt arXiv:0810.1527}}].

\bibitem{Profumo:2008ms}
S.~Profumo, {\it {Dissecting cosmic-ray electron-positron data with Occam's
  Razor: the role of known Pulsars}},  {\em Central Eur.J.Phys.} {\bf 10}
  (2011) 1--31, [\href{http://xxx.lanl.gov/abs/0812.4457}{{\tt
  arXiv:0812.4457}}].

\bibitem{2009AA...501..821D}
T.~{Delahaye}, R.~{Lineros}, F.~{Donato}, N.~{Fornengo}, J.~{Lavalle},
  P.~{Salati}, and R.~{Taillet}, {\it {Galactic secondary positron flux at the
  Earth}},  {\em \aap} {\bf 501} (July, 2009) 821--833,
  [\href{http://xxx.lanl.gov/abs/0809.5268}{{\tt arXiv:0809.5268}}].

\bibitem{2009PhRvL.103h1104M}
P.~{Mertsch} and S.~{Sarkar}, {\it {Testing Astrophysical Models for the PAMELA
  Positron Excess with Cosmic Ray Nuclei}},  {\em Physical Review Letters} {\bf
  103} (Aug., 2009) 081104, [\href{http://xxx.lanl.gov/abs/0905.3152}{{\tt
  arXiv:0905.3152}}].

\bibitem{2010AA...524A..51D}
T.~{Delahaye}, J.~{Lavalle}, R.~{Lineros}, F.~{Donato}, and N.~{panel}, {\it
  {Galactic electrons and positrons at the Earth: new estimate of the primary
  and secondary fluxes}},  {\em \aap} {\bf 524} (Dec., 2010) A51,
  [\href{http://xxx.lanl.gov/abs/1002.1910}{{\tt arXiv:1002.1910}}].

\bibitem{2013arXiv1312.3483D}
S.~{Della Torre}, M.~{Gervasi}, P.~{Rancoita}, D.~{Rozza}, and A.~{Treves},
  {\it {Possible Contribution to Electron and Positron Fluxes from Pulsars and
  their Nebulae}},  {\em ArXiv e-prints} (Dec., 2013)
  [\href{http://xxx.lanl.gov/abs/1312.3483}{{\tt arXiv:1312.3483}}].

\bibitem{2013APh....49...23E}
A.~D. {Erlykin} and A.~W. {Wolfendale}, {\it {Cosmic ray positrons from a
  local, middle-aged supernova remnant}},  {\em Astroparticle Physics} {\bf 49}
  (Sept., 2013) 23--27, [\href{http://xxx.lanl.gov/abs/1308.4878}{{\tt
  arXiv:1308.4878}}].

\bibitem{Gaggero:2013nfa}
D.~Gaggero, D.~Grasso, L.~Maccione, G.~Di~Bernardo, and C.~Evoli, {\it {PAMELA
  positron and electron spectra are reproduced by 3-dimensional cosmic-ray
  modeling}},  \href{http://xxx.lanl.gov/abs/1311.5575}{{\tt arXiv:1311.5575}}.

\bibitem{Gaggero:2013oga}
D.~Gaggero and L.~Maccione, {\it {Model independent interpretation of recent CR
  lepton data after AMS-02}},  {\em JCAP} {\bf 1312} (2013) 011,
  [\href{http://xxx.lanl.gov/abs/1307.0271}{{\tt arXiv:1307.0271}}].

\bibitem{Gaggero:2013rya}
D.~Gaggero, L.~Maccione, G.~Di~Bernardo, C.~Evoli, and D.~Grasso, {\it
  {Three-Dimensional Model of Cosmic-Ray Lepton Propagation Reproduces Data
  from the Alpha Magnetic Spectrometer on the International Space Station}},
  {\em Phys.Rev.Lett.} {\bf 111} (2013) 021102,
  [\href{http://xxx.lanl.gov/abs/1304.6718}{{\tt arXiv:1304.6718}}].

\bibitem{Grasso:2013nyx}
D.~Grasso, G.~Di~Bernardo, C.~Evoli, D.~Gaggero, and L.~Maccione, {\it
  {Galactic electron and positron properties from cosmic ray and radio
  observations}},  \href{http://xxx.lanl.gov/abs/1306.6885}{{\tt
  arXiv:1306.6885}}.

\bibitem{Yuan:2013eba}
Q.~Yuan and X.-J. Bi, {\it {Reconcile the AMS-02 positron fraction and
  Fermi-LAT/HESS total $e^{\pm}$ spectra by the primary electron spectrum
  hardening}},  {\em Phys.Lett.} {\bf B727} (2013) 1--7,
  [\href{http://xxx.lanl.gov/abs/1304.2687}{{\tt arXiv:1304.2687}}].

\bibitem{Cholis:2013lwa}
I.~Cholis and D.~Hooper, {\it {Constraining the origin of the rising cosmic ray
  positron fraction with the boron-to-carbon ratio}},
  \href{http://xxx.lanl.gov/abs/1312.2952}{{\tt arXiv:1312.2952}}.

\bibitem{2013arXiv1305.0084I}
M.~{Ibe}, S.~{Matsumoto}, S.~{Shirai}, and T.~T. {Yanagida}, {\it {AMS-02
  Positrons from Decaying Wino in the Pure Gravity Mediation Model}},  {\em
  ArXiv e-prints} (May, 2013) [\href{http://xxx.lanl.gov/abs/1305.0084}{{\tt
  arXiv:1305.0084}}].

\bibitem{2013arXiv1307.6204B}
P.~S. {Bhupal Dev}, D.~K. {Ghosh}, N.~{Okada}, and I.~{Saha}, {\it {Neutrino
  Mass and Dark Matter in light of recent AMS-02 results}},  {\em ArXiv
  e-prints} (July, 2013) [\href{http://xxx.lanl.gov/abs/1307.6204}{{\tt
  arXiv:1307.6204}}].

\bibitem{2013arXiv1312.7841S}
V.~C. {Spanos}, {\it {The Price of a Dark Matter Annihilation Interpretation of
  AMS-02 Data}},  {\em ArXiv e-prints} (Dec., 2013)
  [\href{http://xxx.lanl.gov/abs/1312.7841}{{\tt arXiv:1312.7841}}].

\bibitem{2013arXiv1309.2570I}
A.~{Ibarra}, A.~S. {Lamperstorfer}, and J.~{Silk}, {\it {Dark matter
  annihilations and decays after the AMS-02 positron measurements}},  {\em
  ArXiv e-prints} (Sept., 2013) [\href{http://xxx.lanl.gov/abs/1309.2570}{{\tt
  arXiv:1309.2570}}].

\bibitem{2013PhRvD..88j3509D}
K.~R. {Dienes}, J.~{Kumar}, and B.~{Thomas}, {\it {Dynamical Dark Matter and
  the positron excess in light of AMS results}},  {\em \prd} {\bf 88} (Nov.,
  2013) 103509, [\href{http://xxx.lanl.gov/abs/1306.2959}{{\tt
  arXiv:1306.2959}}].

\bibitem{2014PhLB..728...58H}
A.~{Hektor}, M.~{Raidal}, A.~{Strumia}, and E.~{Tempel}, {\it {The cosmic-ray
  positron excess from a local Dark Matter over-density}},  {\em Physics
  Letters B} {\bf 728} (Jan., 2014) 58--62,
  [\href{http://xxx.lanl.gov/abs/1307.2561}{{\tt arXiv:1307.2561}}].

\bibitem{Cirelli:2008pk}
M.~Cirelli, M.~Kadastik, M.~Raidal, and A.~Strumia, {\it {Model-independent
  implications of the e+-, anti-proton cosmic ray spectra on properties of Dark
  Matter}},  {\em Nuclear Physics B} {\bf B813} (2009) 1--21,
  [\href{http://xxx.lanl.gov/abs/0809.2409}{{\tt arXiv:0809.2409}}].

\bibitem{Zhao:2014nsa}
Y.~Zhao and K.~M. Zurek, {\it {Indirect Detection Signatures for the Origin of
  Asymmetric Dark Matter}},  \href{http://xxx.lanl.gov/abs/1401.7664}{{\tt
  arXiv:1401.7664}}.

\bibitem{Baek:2014awa}
S.~Baek, H.~Okada, and T.~Toma, {\it {Radiative Lepton Model and Dark Matter
  with Global $U(1)'$ Symmetry}},
  \href{http://xxx.lanl.gov/abs/1401.6921}{{\tt arXiv:1401.6921}}.

\bibitem{Kopp:2014tsa}
J.~Kopp, L.~Michaels, and J.~Smirnov, {\it {Loopy Constraints on Leptophilic
  Dark Matter and Internal Bremsstrahlung}},
  \href{http://xxx.lanl.gov/abs/1401.6457}{{\tt arXiv:1401.6457}}.

\bibitem{Hryczuk:2014hpa}
A.~Hryczuk, I.~Cholis, R.~Iengo, M.~Tavakoli, and P.~Ullio, {\it {Indirect
  Detection Analysis: Wino Dark Matter Case Study}},
  \href{http://xxx.lanl.gov/abs/1401.6212}{{\tt arXiv:1401.6212}}.

\bibitem{Choi:2013oaa}
K.-Y. Choi, B.~Kyae, and C.~S. Shin, {\it {Decaying WIMP dark matter for AMS-02
  cosmic positron excess}},  \href{http://xxx.lanl.gov/abs/1307.6568}{{\tt
  arXiv:1307.6568}}.

\bibitem{Ibarra:2013cra}
A.~Ibarra, D.~Tran, and C.~Weniger, {\it {Indirect Searches for Decaying Dark
  Matter}},  {\em Int.J.Mod.Phys.} {\bf A28} (2013), no.~27 1330040,
  [\href{http://xxx.lanl.gov/abs/1307.6434}{{\tt arXiv:1307.6434}}].

\bibitem{Bergstrom:2013jra}
L.~Bergstrom, T.~Bringmann, I.~Cholis, D.~Hooper, and C.~Weniger, {\it {New
  limits on dark matter annihilation from AMS cosmic ray positron data}},  {\em
  Phys.Rev.Lett.} {\bf 111} (2013) 171101,
  [\href{http://xxx.lanl.gov/abs/1306.3983}{{\tt arXiv:1306.3983}}].

\bibitem{Jin:2013nta}
H.-B. Jin, Y.-L. Wu, and Y.-F. Zhou, {\it {Implications of the first AMS-02
  measurement for dark matter annihilation and decay}},  {\em JCAP} {\bf 1311}
  (2013) 026, [\href{http://xxx.lanl.gov/abs/1304.1997}{{\tt
  arXiv:1304.1997}}].

\bibitem{DeSimone:2013fia}
A.~De~Simone, A.~Riotto, and W.~Xue, {\it {Interpretation of AMS-02 Results:
  Correlations among Dark Matter Signals}},  {\em JCAP} {\bf 1305} (2013) 003,
  [\href{http://xxx.lanl.gov/abs/1304.1336}{{\tt arXiv:1304.1336}}].

\bibitem{Feng:2013zca}
L.~Feng, R.-Z. Yang, H.-N. He, T.-K. Dong, Y.-Z. Fan, et~al., {\it {AMS-02
  positron excess: new bounds on dark matter models and hint for primary
  electron spectrum hardening}},  {\em Phys.Lett.} {\bf B728} (2014) 250--255,
  [\href{http://xxx.lanl.gov/abs/1303.0530}{{\tt arXiv:1303.0530}}].

\bibitem{Masina:2013yea}
I.~Masina and F.~Sannino, {\it {Hints of a Charge Asymmetry in the Electron and
  Positron Cosmic-Ray Excesses}},  {\em Phys.Rev.} {\bf D87} (2013) 123003,
  [\href{http://xxx.lanl.gov/abs/1304.2800}{{\tt arXiv:1304.2800}}].

\bibitem{2007ApJ...661..879E}
D.~C. {Ellison}, D.~J. {Patnaude}, P.~{Slane}, P.~{Blasi}, and S.~{Gabici},
  {\it {Particle Acceleration in Supernova Remnants and the Production of
  Thermal and Nonthermal Radiation}},  {\em \apj} {\bf 661} (June, 2007)
  879--891, [\href{http://xxx.lanl.gov/abs/astro-ph/0702674}{{\tt
  astro-ph/0702674}}].

\bibitem{2009AA...499..191T}
V.~{Tatischeff}, {\it {Radio emission and nonlinear diffusive shock
  acceleration of cosmic rays in the supernova SN 1993J}},  {\em \aap} {\bf
  499} (May, 2009) 191--213, [\href{http://xxx.lanl.gov/abs/0903.2944}{{\tt
  arXiv:0903.2944}}].

\bibitem{2013MNRAS.434.2748C}
P.~{Cristofari}, S.~{Gabici}, S.~{Casanova}, R.~{Terrier}, and E.~{Parizot},
  {\it {Acceleration of cosmic rays and gamma-ray emission from supernova
  remnants in the Galaxy}},  {\em \mnras} {\bf 434} (Oct., 2013) 2748--2760,
  [\href{http://xxx.lanl.gov/abs/1302.2150}{{\tt arXiv:1302.2150}}].

\bibitem{Blasi:2013rva}
P.~Blasi, {\it {The Origin of Galactic Cosmic Rays}},
  \href{http://xxx.lanl.gov/abs/1311.7346}{{\tt arXiv:1311.7346}}.

\bibitem{2012SSRv..173..369H}
E.~A. {Helder}, J.~{Vink}, A.~M. {Bykov}, Y.~{Ohira}, J.~C. {Raymond}, and
  R.~{Terrier}, {\it {Observational Signatures of Particle Acceleration in
  Supernova Remnants}},  {\em \ssr} {\bf 173} (Nov., 2012) 369--431,
  [\href{http://xxx.lanl.gov/abs/1206.1593}{{\tt arXiv:1206.1593}}].

\bibitem{2008PhRvL.101z1104A}
F.~{Aharonian}, A.~G. {Akhperjanian}, et~al., {\it {Energy Spectrum of
  Cosmic-Ray Electrons at TeV Energies}},  {\em Phys.Rev.Lett.} {\bf 101}
  (Dec., 2008) 261104, [\href{http://xxx.lanl.gov/abs/0811.3894}{{\tt
  arXiv:0811.3894}}].

\bibitem{2009ApJ...692.1500A}
F.~{Aharonian}, A.~G. {Akhperjanian}, U.~B. {de Almeida}, et~al., {\it
  {Discovery of Gamma-Ray Emission From the Shell-Type Supernova Remnant RCW 86
  With Hess}},  {\em \apj} {\bf 692} (Feb., 2009) 1500--1505,
  [\href{http://xxx.lanl.gov/abs/0810.2689}{{\tt arXiv:0810.2689}}].

\bibitem{2010ApJ...714..163A}
{\bf VERITAS} Collaboration, V.~A. {Acciari} et~al., {\it {Observations of the
  Shell-type Supernova Remnant Cassiopeia A at TeV Energies with VERITAS}},
  {\em \apj} {\bf 714} (May, 2010) 163--169,
  [\href{http://xxx.lanl.gov/abs/1002.2974}{{\tt arXiv:1002.2974}}].

\bibitem{Aharonian:2001mz}
F.~{Aharonian}, A.~{Akhperjanian}, J.~{Barrio}, K.~{Bernl{\"o}hr},
  H.~{B{\"o}rst}, H.~{Bojahr}, O.~{Bolz}, J.~{Contreras}, J.~{Cortina},
  S.~{Denninghoff}, V.~{Fonseca}, J.~{Gonzalez}, N.~{G{\"o}tting},
  G.~{Heinzelmann}, G.~{Hermann}, A.~{Heusler}, W.~{Hofmann}, D.~{Horns},
  A.~{Ibarra}, C.~{Iserlohe}, I.~{Jung}, R.~{Kankanyan}, M.~{Kestel},
  J.~{Kettler}, A.~{Kohnle}, A.~{Konopelko}, H.~{Kornmeyer}, D.~{Kranich},
  H.~{Krawczynski}, H.~{Lampeitl}, M.~{Lopez}, E.~{Lorenz}, F.~{Lucarelli},
  N.~{Magnussen}, O.~{Mang}, H.~{Meyer}, R.~{Mirzoyan}, A.~{Moralejo},
  E.~{Ona}, L.~{Padilla}, M.~{Panter}, R.~{Plaga}, A.~{Plyasheshnikov},
  J.~{Prahl}, G.~{P{\"u}hlhofer}, G.~{Rauterberg}, A.~{R{\"o}hring},
  W.~{Rhode}, G.~P. {Rowell}, V.~{Sahakian}, M.~{Samorski}, M.~{Schilling},
  F.~{Schr{\"o}der}, M.~{Siems}, W.~{Stamm}, M.~{Tluczykont}, H.~J. {V{\"o}lk},
  C.~A. {Wiedner}, and W.~{Wittek}, {\it {Evidence for TeV gamma ray emission
  from Cassiopeia A}},  {\em \aap} {\bf 370} (Apr., 2001) 112--120,
  [\href{http://xxx.lanl.gov/abs/astro-ph/0102391}{{\tt astro-ph/0102391}}].

\bibitem{Aharonian:2006ws}
F.~{Aharonian}, A.~G. {Akhperjanian}, A.~R. {Bazer-Bachi}, M.~{Beilicke},
  W.~{Benbow}, D.~{Berge}, K.~{Bernl{\"o}hr}, C.~{Boisson}, O.~{Bolz},
  V.~{Borrel}, I.~{Braun}, E.~{Brion}, A.~M. {Brown}, R.~{B{\"u}hler},
  I.~{B{\"u}sching}, S.~{Carrigan}, P.~M. {Chadwick}, L.-M. {Chounet},
  G.~{Coignet}, R.~{Cornils}, L.~{Costamante}, B.~{Degrange}, H.~J.
  {Dickinson}, A.~{Djannati-Ata{\"i}}, L.~{O'C.~Drury}, G.~{Dubus},
  K.~{Egberts}, D.~{Emmanoulopoulos}, P.~{Espigat}, F.~{Feinstein},
  E.~{Ferrero}, A.~{Fiasson}, G.~{Fontaine}, S.~{Funk}, S.~{Funk},
  M.~{F{\"u}{\ss}ling}, Y.~A. {Gallant}, B.~{Giebels}, J.~F. {Glicenstein},
  B.~{Gl{\"u}ck}, P.~{Goret}, C.~{Hadjichristidis}, D.~{Hauser}, M.~{Hauser},
  G.~{Heinzelmann}, G.~{Henri}, G.~{Hermann}, J.~A. {Hinton}, A.~{Hoffmann},
  W.~{Hofmann}, M.~{Holleran}, S.~{Hoppe}, D.~{Horns}, A.~{Jacholkowska}, O.~C.
  {de Jager}, E.~{Kendziorra}, M.~{Kerschhaggl}, B.~{Kh{\'e}lifi}, N.~{Komin},
  A.~{Konopelko}, K.~{Kosack}, G.~{Lamanna}, I.~J. {Latham}, R.~{Le Gallou},
  A.~{Lemi{\`e}re}, M.~{Lemoine-Goumard}, T.~{Lohse}, J.~M. {Martin},
  O.~{Martineau-Huynh}, A.~{Marcowith}, C.~{Masterson}, G.~{Maurin}, T.~J.~L.
  {McComb}, E.~{Moulin}, M.~{de Naurois}, D.~{Nedbal}, S.~J. {Nolan},
  A.~{Noutsos}, J.-P. {Olive}, K.~J. {Orford}, J.~L. {Osborne}, M.~{Panter},
  G.~{Pelletier}, S.~{Pita}, G.~{P{\"u}hlhofer}, M.~{Punch}, S.~{Ranchon},
  B.~C. {Raubenheimer}, M.~{Raue}, S.~M. {Rayner}, A.~{Reimer}, O.~{Reimer},
  J.~{Ripken}, L.~{Rob}, L.~{Rolland}, S.~{Rosier-Lees}, G.~{Rowell},
  V.~{Sahakian}, A.~{Santangelo}, L.~{Saug{\'e}}, S.~{Schlenker},
  R.~{Schlickeiser}, R.~{Schr{\"o}der}, U.~{Schwanke}, S.~{Schwarzburg},
  S.~{Schwemmer}, A.~{Shalchi}, H.~{Sol}, D.~{Spangler}, F.~{Spanier},
  R.~{Steenkamp}, C.~{Stegmann}, G.~{Superina}, P.~H. {Tam}, J.-P. {Tavernet},
  R.~{Terrier}, M.~{Tluczykont}, C.~{van Eldik}, G.~{Vasileiadis}, C.~{Venter},
  J.~P. {Vialle}, P.~{Vincent}, H.~J. {V{\"o}lk}, S.~J. {Wagner}, and
  M.~{Ward}, {\it {Primary particle acceleration above 100 TeV in the
  shell-type supernova remnant <ASTROBJ>RX J1713.7-3946</ASTROBJ> with deep
  HESS observations}},  {\em \aap} {\bf 464} (Mar., 2007) 235--243,
  [\href{http://xxx.lanl.gov/abs/astro-ph/0611813}{{\tt astro-ph/0611813}}].

\bibitem{2013PASJ...65...69S}
T.~{Shimizu}, K.~{Masai}, and K.~{Koyama}, {\it {Non-Thermal Radio and
  Gamma-Ray Emissions from a Supernova Remnant by Blast Wave Breaking Out of
  the Circumstellar Matter}},  {\em \pasj} {\bf 65} (June, 2013) 69,
  [\href{http://xxx.lanl.gov/abs/1303.6049}{{\tt arXiv:1303.6049}}].

\bibitem{2014AA...561A.139S}
I.~{Sushch} and B.~{Hnatyk}, {\it {Modelling of the radio emission from the
  Vela supernova remnant}},  {\em \aap} {\bf 561} (Jan., 2014) A139,
  [\href{http://xxx.lanl.gov/abs/1312.0777}{{\tt arXiv:1312.0777}}].

\bibitem{2009BASI...37...45G}
D.~A. {Green}, {\it {A revised Galactic supernova remnant catalogue}},  {\em
  Bulletin of the Astronomical Society of India} {\bf 37} (Mar., 2009) 45--61,
  [\href{http://xxx.lanl.gov/abs/0905.3699}{{\tt arXiv:0905.3699}}].

\bibitem{2000MNRAS.314...65L}
S.~G. {Lucek} and A.~R. {Bell}, {\it {Non-linear amplification of a magnetic
  field driven by cosmic ray streaming}},  {\em \mnras} {\bf 314} (May, 2000)
  65--74.

\bibitem{2006ApJ...648L..33V}
J.~{Vink}, J.~{Bleeker}, K.~{van der Heyden}, A.~{Bykov}, A.~{Bamba}, and
  R.~{Yamazaki}, {\it {The X-Ray Synchrotron Emission of RCW 86 and the
  Implications for Its Age}},  {\em \apjl} {\bf 648} (Sept., 2006) L33--L37,
  [\href{http://xxx.lanl.gov/abs/astro-ph/0607307}{{\tt astro-ph/0607307}}].

\bibitem{2010ApJ...718..348A}
A.~A. {Abdo}, M.~{Ackermann}, M.~{Ajello}, {Allafort}, et~al., {\it {Fermi
  Large Area Telescope Observations of the Supernova Remnant W28 (G6.4-0.1)}},
  {\em \apj} {\bf 718} (July, 2010) 348--356.

\bibitem{2013MNRAS.431..415B}
A.~R. {Bell}, K.~M. {Schure}, B.~{Reville}, and G.~{Giacinti}, {\it {Cosmic-ray
  acceleration and escape from supernova remnants}},  {\em \mnras} {\bf 431}
  (May, 2013) 415--429, [\href{http://xxx.lanl.gov/abs/1301.7264}{{\tt
  arXiv:1301.7264}}].

\bibitem{2013MNRAS.435.1174S}
K.~M. {Schure} and A.~R. {Bell}, {\it {Cosmic ray acceleration in young
  supernova remnants}},  {\em \mnras} {\bf 435} (Oct., 2013) 1174--1185,
  [\href{http://xxx.lanl.gov/abs/1307.6575}{{\tt arXiv:1307.6575}}].

\bibitem{2009MNRAS.392..240M}
G.~{Morlino}, E.~{Amato}, and P.~{Blasi}, {\it {Gamma-ray emission from SNR RX
  J1713.7-3946 and the origin of galactic cosmic rays}},  {\em \mnras} {\bf
  392} (Jan., 2009) 240--250, [\href{http://xxx.lanl.gov/abs/0810.0094}{{\tt
  arXiv:0810.0094}}].

\bibitem{2013IJMPS..23...82H}
J.~L. {Han}, W.~{Reich}, X.~H. {Sun}, X.~Y. {Gao}, L.~{Xiao}, W.~B. {Shi},
  P.~{Reich}, and R.~{Wielebinski}, {\it {The Sino-German {$\lambda$}6 CM
  Polarization Survey of the Galactic Plane: a Summary}},  {\em International
  Journal of Modern Physics Conference Series} {\bf 23} (2013) 82--91,
  [\href{http://xxx.lanl.gov/abs/1202.1875}{{\tt arXiv:1202.1875}}].

\bibitem{1998MNRAS.297L..17M}
P.~{Madau}, M.~{della Valle}, and N.~{Panagia}, {\it {On the evolution of the
  cosmic supernova rates}},  {\em \mnras} {\bf 297} (June, 1998) L17,
  [\href{http://xxx.lanl.gov/abs/astro-ph/9803284}{{\tt astro-ph/9803284}}].

\bibitem{2006Natur.439...45D}
R.~{Diehl}, H.~{Halloin}, K.~{Kretschmer}, G.~G. {Lichti},
  V.~{Sch{\"o}nfelder}, A.~W. {Strong}, A.~{von Kienlin}, W.~{Wang}, P.~{Jean},
  J.~{Kn{\"o}dlseder}, J.-P. {Roques}, G.~{Weidenspointner}, S.~{Schanne},
  D.~H. {Hartmann}, C.~{Winkler}, and C.~{Wunderer}, {\it {Radioactive
  $^{26}$Al from massive stars in the Galaxy}},  {\em \nat} {\bf 439} (Jan.,
  2006) 45--47, [\href{http://xxx.lanl.gov/abs/astro-ph/0601015}{{\tt
  astro-ph/0601015}}].

\bibitem{2004IAUS..218..105L}
D.~R. {Lorimer}, {\it {The Galactic Population and Birth Rate of Radio
  Pulsars}},  in {\em Young Neutron Stars and Their Environments} (F.~{Camilo}
  and B.~M. {Gaensler}, eds.), vol.~218 of {\em IAU Symposium}, p.~105, 2004.
\newblock \href{http://xxx.lanl.gov/abs/astro-ph/0308501}{{\tt
  astro-ph/0308501}}.

\bibitem{1976MNRAS.174..267C}
D.~H. {Clark} and J.~L. {Caswell}, {\it {A study of galactic supernova
  remnants, based on Molonglo-Parkes observational data}},  {\em \mnras} {\bf
  174} (Feb., 1976) 267--305.

\bibitem{2002AJ....124.2145V}
P.~F. {Vel{\'a}zquez}, G.~M. {Dubner}, W.~M. {Goss}, and A.~J. {Green}, {\it
  {Investigation of the Large-scale Neutral Hydrogen near the Supernova Remnant
  W28}},  {\em \aj} {\bf 124} (Oct., 2002) 2145--2151,
  [\href{http://xxx.lanl.gov/abs/astro-ph/0207530}{{\tt astro-ph/0207530}}].

\bibitem{2011AA...536A..83S}
X.~H. {Sun}, P.~{Reich}, W.~{Reich}, L.~{Xiao}, X.~Y. {Gao}, and J.~L. {Han},
  {\it {A Sino-German {$\lambda$}6 cm polarization survey of the Galactic
  plane. VII. Small supernova remnants}},  {\em \aap} {\bf 536} (Dec., 2011)
  A83, [\href{http://xxx.lanl.gov/abs/1110.1106}{{\tt arXiv:1110.1106}}].

\bibitem{1989AA...209..361F}
E.~{Furst}, E.~{Hummel}, W.~{Reich}, Y.~{Sofue}, W.~{Sieber}, K.~{Reif}, and
  R.-J. {Dettmar}, {\it {A study of the composite supernova remnant G
  18.95-1.1}},  {\em \aap} {\bf 209} (Jan., 1989) 361--368.

\bibitem{1997AA...319..655F}
E.~{Fuerst}, W.~{Reich}, and B.~{Aschenbach}, {\it {New radio and soft X-ray
  observations of the supernova remnant G 18.95-1.1.}},  {\em \aap} {\bf 319}
  (Mar., 1997) 655--663.

\bibitem{2004ApJ...603..152H}
I.~M. {Harrus}, P.~O. {Slane}, J.~P. {Hughes}, and P.~P. {Plucinsky}, {\it {An
  X-Ray Study of the Supernova Remnant G18.95-1.1}},  {\em \apj} {\bf 603}
  (Mar., 2004) 152--158, [\href{http://xxx.lanl.gov/abs/astro-ph/0311410}{{\tt
  astro-ph/0311410}}].

\bibitem{1989MNRAS.238..737G}
D.~A. {Green}, {\it {Sensitive OH observations towards 16 supernova remnants}},
   {\em \mnras} {\bf 238} (May, 1989) 737--751.

\bibitem{1997AJ....113.1379G}
E.~B. {Giacani}, G.~M. {Dubner}, N.~E. {Kassim}, D.~A. {Frail}, W.~M. {Goss},
  P.~F. {Winkler}, and B.~F. {Williams}, {\it {New Radio and Optical Study of
  the Supernova Remnant W44}},  {\em \aj} {\bf 113} (Apr., 1997) 1379.

\bibitem{1975AA....45..239C}
J.~L. {Caswell}, J.~D. {Murray}, R.~S. {Roger}, D.~J. {Cole}, and D.~J.
  {Cooke}, {\it {Neutral hydrogen absorption measurements yielding kinematic
  distances for 42 continuum sources in the galactic plane}},  {\em \aap} {\bf
  45} (Dec., 1975) 239--258.

\bibitem{1991ApJ...372L..99W}
A.~{Wolszczan}, J.~M. {Cordes}, and R.~J. {Dewey}, {\it {Discovery of a young,
  267 millisecond pulsar in the supernova remnant W44}},  {\em \apjl} {\bf 372}
  (May, 1991) L99--L102.

\bibitem{1994ApJ...430..757R}
J.~{Rho}, R.~{Petre}, E.~M. {Schlegel}, and J.~J. {Hester}, {\it {An X-ray and
  optical study of the supernova remnant W44}},  {\em \apj} {\bf 430} (Aug.,
  1994) 757--773.

\bibitem{2009AA...503..827X}
L.~{Xiao}, W.~{Reich}, E.~{F{\"u}rst}, and J.~L. {Han}, {\it {Radio properties
  of the low surface brightness SNR G65.2+5.7}},  {\em \aap} {\bf 503} (Sept.,
  2009) 827--836, [\href{http://xxx.lanl.gov/abs/0904.3170}{{\tt
  arXiv:0904.3170}}].

\bibitem{1996ApJ...458..257G}
P.~W. {Gorham}, P.~S. {Ray}, S.~B. {Anderson}, S.~R. {Kulkarni}, and T.~A.
  {Prince}, {\it {A Pulsar Survey of 18 Supernova Remnants}},  {\em \apj} {\bf
  458} (Feb., 1996) 257.

\bibitem{2002AA...388..355}
F.~{Mavromatakis}, P.~{Boumis}, J.~{Papamastorakis}, and J.~{Ventura}, {\it
  {Deep optical observations of G 65.3+5.7}},  {\em \aap} {\bf 388} (June,
  2002) 355--362, [\href{http://xxx.lanl.gov/abs/astro-ph/0204079}{{\tt
  astro-ph/0204079}}].

\bibitem{2004ApJ...607..855K}
R.~{Kothes}, T.~L. {Landecker}, and M.~{Wolleben}, {\it {H I Absorption of
  Polarized Emission: A New Technique for Determining Kinematic Distances to
  Galactic Supernova Remnants}},  {\em \apj} {\bf 607} (June, 2004) 855--864.

\bibitem{2008ApJ...687..516K}
R.~{Kothes}, T.~L. {Landecker}, W.~{Reich}, S.~{Safi-Harb}, and
  Z.~{Arzoumanian}, {\it {DA 495: An Aging Pulsar Wind Nebula}},  {\em \apj}
  {\bf 687} (Nov., 2008) 516--531,
  [\href{http://xxx.lanl.gov/abs/0807.0811}{{\tt arXiv:0807.0811}}].

\bibitem{2005AA...440..171C}
G.~{Castelletti} and G.~{Dubner}, {\it {A multi-frequency study of the spectral
  index distribution in the SNR CTB 80}},  {\em \aap} {\bf 440} (Sept., 2005)
  171--177, [\href{http://xxx.lanl.gov/abs/astro-ph/0507716}{{\tt
  astro-ph/0507716}}].

\bibitem{2006AA...457.1081K}
R.~{Kothes}, K.~{Fedotov}, T.~J. {Foster}, and B.~{Uyan{\i}ker}, {\it {A
  catalogue of Galactic supernova remnants from the Canadian Galactic plane
  survey. I. Flux densities, spectra, and polarization characteristics}},  {\em
  \aap} {\bf 457} (Oct., 2006) 1081--1093.

\bibitem{2000ASPC..202..509S}
R.~G. {Strom} and B.~W. {Stappers}, {\it {Supernova Remnant CTB 80 and PSR
  1951+32}},  in {\em IAU Colloq. 177: Pulsar Astronomy - 2000 and Beyond}
  (M.~{Kramer}, N.~{Wex}, and R.~{Wielebinski}, eds.), vol.~202 of {\em
  Astronomical Society of the Pacific Conference Series}, p.~509, 2000.

\bibitem{2003AJ....126.2114C}
G.~{Castelletti}, G.~{Dubner}, K.~{Golap}, W.~M. {Goss}, P.~F. {Vel{\'a}zquez},
  M.~{Holdaway}, and A.~P. {Rao}, {\it {New High-Resolution Radio Observations
  of the Supernova Remnant CTB 80}},  {\em \aj} {\bf 126} (Nov., 2003)
  2114--2124, [\href{http://xxx.lanl.gov/abs/astro-ph/0310655}{{\tt
  astro-ph/0310655}}].

\bibitem{2006AA...447..937S}
X.~H. {Sun}, W.~{Reich}, J.~L. {Han}, P.~{Reich}, and R.~{Wielebinski}, {\it
  {New {$\lambda$}6 cm observations of the Cygnus Loop}},  {\em \aap} {\bf 447}
  (Mar., 2006) 937--947, [\href{http://xxx.lanl.gov/abs/astro-ph/0510509}{{\tt
  astro-ph/0510509}}].

\bibitem{2005AJ....129.2268B}
W.~P. {Blair}, R.~{Sankrit}, and J.~C. {Raymond}, {\it {Hubble Space Telescope
  Imaging of the Primary Shock Front in the Cygnus Loop Supernova Remnant}},
  {\em \aj} {\bf 129} (May, 2005) 2268--2280.

\bibitem{2009ApJ...692..335B}
W.~P. {Blair}, R.~{Sankrit}, S.~I. {Torres}, P.~{Chayer}, and C.~W. {Danforth},
  {\it {Far Ultraviolet Spectroscopic Explorer Observations of KPD 2055+3111, a
  Star behind the Cygnus Loop}},  {\em \apj} {\bf 692} (Feb., 2009) 335--345.

\bibitem{2008AA...490..197L}
Y.~{Ladouceur} and S.~{Pineault}, {\it {New perspectives on the supernova
  remnant G78.2+2.1}},  {\em \aap} {\bf 490} (Oct., 2008) 197--211.

\bibitem{2004AA...427L..21B}
A.~M. {Bykov}, A.~M. {Krassilchtchikov}, Y.~A. {Uvarov}, H.~{Bloemen}, R.~A.
  {Chevalier}, M.~Y. {Gustov}, W.~{Hermsen}, F.~{Lebrun}, T.~A. {Lozinskaya},
  G.~{Rauw}, T.~V. {Smirnova}, S.~J. {Sturner}, J.-P. {Swings}, R.~{Terrier},
  and I.~N. {Toptygin}, {\it {Hard X-ray emission clumps in the
  {$\gamma$}-Cygni supernova remnant: An INTEGRAL-ISGRI view}},  {\em \aap}
  {\bf 427} (Dec., 2004) L21--L24,
  [\href{http://xxx.lanl.gov/abs/astro-ph/0410688}{{\tt astro-ph/0410688}}].

\bibitem{2003AA...408..237M}
F.~{Mavromatakis}, {\it {Deep optical observations of the supernova remnant G
  78.2+2.1}},  {\em \aap} {\bf 408} (Sept., 2003) 237--243.

\bibitem{2004AA...415.1051M}
F.~{Mavromatakis}, B.~{Aschenbach}, P.~{Boumis}, and J.~{Papamastorakis}, {\it
  {Multi-wavelength study of the $\{$G 82.2+5.3$\}$ supernova remnant}},  {\em
  \aap} {\bf 415} (Mar., 2004) 1051--1063,
  [\href{http://xxx.lanl.gov/abs/astro-ph/0311530}{{\tt astro-ph/0311530}}].

\bibitem{1981RMxAA...5...93R}
M.~{Rosado} and J.~{Gonzalez}, {\it {The Radial Velocity Field of the Optical
  Filaments Associated with the Supernova Remnant W63}},  {\em \rmxaa} {\bf 5}
  (June, 1981) 93.

\bibitem{2003AA...408..961R}
W.~{Reich}, X.~{Zhang}, and E.~{F{\"u}rst}, {\it {35 cm observations of a
  sample of large supernova remnants}},  {\em \aap} {\bf 408} (Sept., 2003)
  961--969.

\bibitem{2006ApJ...637..283B}
D.-Y. {Byun}, B.-C. {Koo}, K.~{Tatematsu}, and K.~{Sunada}, {\it {Interaction
  between the Supernova Remnant HB 21 and Molecular Clouds}},  {\em \apj} {\bf
  637} (Jan., 2006) 283--295.

\bibitem{2006ApJ...647..350L}
J.~S. {Lazendic} and P.~O. {Slane}, {\it {Enhanced Abundances in Three
  Large-Diameter Mixed-Morphology Supernova Remnants}},  {\em \apj} {\bf 647}
  (Aug., 2006) 350--366, [\href{http://xxx.lanl.gov/abs/astro-ph/0505498}{{\tt
  astro-ph/0505498}}].

\bibitem{1999ApJ...527..866L}
T.~L. {Landecker}, D.~{Routledge}, S.~P. {Reynolds}, R.~J. {Smegal}, K.~J.
  {Borkowski}, and F.~D. {Seward}, {\it {DA 530: A Supernova Remnant in a
  Stellar Wind Bubble}},  {\em \apj} {\bf 527} (Dec., 1999) 866--878.

\bibitem{2003ApJ...598.1005F}
T.~{Foster} and D.~{Routledge}, {\it {A New Distance Technique for Galactic
  Plane Objects}},  {\em \apj} {\bf 598} (Dec., 2003) 1005--1016.

\bibitem{2002ApJ...565.1022U}
B.~{Uyaniker}, R.~{Kothes}, and C.~M. {Brunt}, {\it {The Supernova Remnant CTB
  104A: Magnetic Field Structure and Interaction with the Environment}},  {\em
  \apj} {\bf 565} (Feb., 2002) 1022--1034,
  [\href{http://xxx.lanl.gov/abs/astro-ph/0110001}{{\tt astro-ph/0110001}}].

\bibitem{1982AA...105..176M}
F.~{Mantovani}, M.~{Nanni}, C.~J. {Salter}, and P.~{Tomasi}, {\it {Further
  observations of radio sources from the BG survey. I - The non-thermal sources
  near L equals 94 deg}},  {\em \aap} {\bf 105} (Jan., 1982) 176--183.

\bibitem{2002ApJ...576..169K}
R.~{Kothes}, B.~{Uyaniker}, and A.~{Yar}, {\it {The Distance to Supernova
  Remnant CTB 109 Deduced from Its Environment}},  {\em \apj} {\bf 576} (Sept.,
  2002) 169--175, [\href{http://xxx.lanl.gov/abs/astro-ph/0205034}{{\tt
  astro-ph/0205034}}].

\bibitem{1998AA...330..175P}
A.~N. {Parmar}, T.~{Oosterbroek}, F.~{Favata}, S.~{Pightling}, M.~J. {Coe},
  S.~{Mereghetti}, and G.~L. {Israel}, {\it {A BeppoSAX observation of the
  X-ray pulsar 1E2259+586hfill and the supernova remnant (CTB109)}},  {\em
  \aap} {\bf 330} (Feb., 1998) 175--180,
  [\href{http://xxx.lanl.gov/abs/astro-ph/9709248}{{\tt astro-ph/9709248}}].

\bibitem{1981ApJ...246L.127H}
V.~A. {Hughes}, R.~H. {Harten}, and S.~{van den Bergh}, {\it {A new supernova
  remnant G109.2-1.0}},  {\em \apjl} {\bf 246} (June, 1981) L127--L131.

\bibitem{2005AA...444..871K}
R.~{Kothes}, B.~{Uyan{\i}ker}, and R.~I. {Reid}, {\it {Two new Perseus arm
  supernova remnants discovered in the Canadian Galactic Plane Survey}},  {\em
  \aap} {\bf 444} (Dec., 2005) 871--881.

\bibitem{2004ApJ...616..247Y}
A.~{Yar-Uyaniker}, B.~{Uyaniker}, and R.~{Kothes}, {\it {Distance of Three
  Supernova Remnants from H I Line Observations in a Complex Region:
  G114.3+0.3, G116.5+1.1, and CTB 1 (G116.9+0.2)}},  {\em \apj} {\bf 616}
  (Nov., 2004) 247--256, [\href{http://xxx.lanl.gov/abs/astro-ph/0408386}{{\tt
  astro-ph/0408386}}].

\bibitem{1997AA...324.1152P}
S.~{Pineault}, T.~L. {Landecker}, C.~M. {Swerdlyk}, and W.~{Reich}, {\it {The
  supernova remnant CTA1 (G 119.5+10.3): a study. of the breakout phenomenon}},
   {\em \aap} {\bf 324} (Aug., 1997) 1152--1164.

\bibitem{1993AJ....105.1060P}
S.~{Pineault}, T.~L. {Landecker}, B.~{Madore}, and S.~{Gaumont-Guay}, {\it {The
  supernova remnant CTA1 and the surrounding interstellar medium}},  {\em \aj}
  {\bf 105} (Mar., 1993) 1060--1073.

\bibitem{2006AA...451..251L}
D.~{Leahy} and W.~{Tian}, {\it {Radio observations and spectrum of the SNR
  G127.1+0.5 and its central source 0125+628}},  {\em \aap} {\bf 451} (May,
  2006) 251--257, [\href{http://xxx.lanl.gov/abs/astro-ph/0601487}{{\tt
  astro-ph/0601487}}].

\bibitem{1989AA...219..303J}
G.~{Joncas}, R.~S. {Roger}, and P.~E. {Dewdney}, {\it {New radio observations
  of two supernova remnants in Cassiopeia - G 126.2 + 1.6 and G 127.1 + 0.5}},
  {\em \aap} {\bf 219} (July, 1989) 303--307.

\bibitem{2008ApJS..174..379F}
R.~{Fesen}, G.~{Rudie}, A.~{Hurford}, and A.~{Soto}, {\it {Optical Imaging and
  Spectroscopy of the Galactic Supernova Remnant 3C 58 (G130.7+3.1)}},  {\em
  \apjs} {\bf 174} (Feb., 2008) 379--395.

\bibitem{2008AA...486..273S}
Y.~A. {Shibanov}, N.~{Lundqvist}, P.~{Lundqvist}, J.~{Sollerman}, and
  D.~{Zyuzin}, {\it {Optical identification of the 3C 58 pulsar wind nebula}},
  {\em \aap} {\bf 486} (July, 2008) 273--282,
  [\href{http://xxx.lanl.gov/abs/0802.2386}{{\tt arXiv:0802.2386}}].

\bibitem{2008AA...487..601S}
W.~B. {Shi}, J.~L. {Han}, X.~Y. {Gao}, X.~H. {Sun}, L.~{Xiao}, P.~{Reich}, and
  W.~{Reich}, {\it {The radio spectrum and magnetic field structure of SNR
  HB3}},  {\em \aap} {\bf 487} (Aug., 2008) 601--604,
  [\href{http://xxx.lanl.gov/abs/0806.1647}{{\tt arXiv:0806.1647}}].

\bibitem{2006Sci...311...54X}
Y.~{Xu}, M.~J. {Reid}, X.~W. {Zheng}, and K.~M. {Menten}, {\it {The Distance to
  the Perseus Spiral Arm in the Milky Way}},  {\em Science} {\bf 311} (Jan.,
  2006) 54--57, [\href{http://xxx.lanl.gov/abs/astro-ph/0512223}{{\tt
  astro-ph/0512223}}].

\bibitem{1992AA...256..214R}
W.~{Reich}, E.~{Fuerst}, and E.~M. {Arnal}, {\it {Radio observations of the
  bright X-ray supernova remnant G156.2+5.7}},  {\em \aap} {\bf 256} (Mar.,
  1992) 214--224.

\bibitem{2009PASJ...61S.155K}
S.~{Katsuda}, R.~{Petre}, U.~{Hwang}, H.~{Yamaguchi}, K.~{Mori}, and
  H.~{Tsunemi}, {\it {Suzaku Observations of Thermal and Non-Thermal X-Ray
  Emission from the Middle-Aged Supernova Remnant G156.2+5.7}},  {\em \pasj}
  {\bf 61} (Jan., 2009) 155, [\href{http://xxx.lanl.gov/abs/0902.1782}{{\tt
  arXiv:0902.1782}}].

\bibitem{2007AA...470..969X}
J.~W. {Xu}, J.~L. {Han}, X.~H. {Sun}, W.~{Reich}, L.~{Xiao}, P.~{Reich}, and
  R.~{Wielebinski}, {\it {Polarization observations of SNR G156.2+5.7 at
  {$\lambda$}6 cm}},  {\em \aap} {\bf 470} (Aug., 2007) 969--975,
  [\href{http://xxx.lanl.gov/abs/0705.2806}{{\tt arXiv:0705.2806}}].

\bibitem{1999PASJ...51...13Y}
S.~{Yamauchi}, K.~{Koyama}, H.~{Tomida}, J.~{Yokogawa}, and K.~{Tamura}, {\it
  {ASCA Observations of the Supernova Remnant G156.2+5.7}},  {\em \pasj} {\bf
  51} (Feb., 1999) 13--22.

\bibitem{2007AA...461.1013L}
D.~A. {Leahy} and W.~W. {Tian}, {\it {Radio spectrum and distance of the SNR
  HB9}},  {\em \aap} {\bf 461} (Jan., 2007) 1013--1018,
  [\href{http://xxx.lanl.gov/abs/astro-ph/0606598}{{\tt astro-ph/0606598}}].

\bibitem{2008AA...482..783X}
L.~{Xiao}, E.~{F{\"u}rst}, W.~{Reich}, and J.~L. {Han}, {\it {Radio spectral
  properties and the magnetic field of the SNR S147}},  {\em \aap} {\bf 482}
  (May, 2008) 783--792, [\href{http://xxx.lanl.gov/abs/0801.4803}{{\tt
  arXiv:0801.4803}}].

\bibitem{1996ApJ...468L..55A}
S.~B. {Anderson}, B.~J. {Cadwell}, B.~A. {Jacoby}, A.~{Wolszczan}, R.~S.
  {Foster}, and M.~{Kramer}, {\it {A 143 Millisecond Radio Pulsar in the
  Supernova Remnant S147}},  {\em \apjl} {\bf 468} (Sept., 1996) L55.

\bibitem{2007ApJ...654..487N}
C.-Y. {Ng}, R.~W. {Romani}, W.~F. {Brisken}, S.~{Chatterjee}, and M.~{Kramer},
  {\it {The Origin and Motion of PSR J0538+2817 in S147}},  {\em \apj} {\bf
  654} (Jan., 2007) 487--493,
  [\href{http://xxx.lanl.gov/abs/astro-ph/0611068}{{\tt astro-ph/0611068}}].

\bibitem{2003ApJ...593L..31K}
M.~{Kramer}, A.~G. {Lyne}, G.~{Hobbs}, O.~{L{\"o}hmer}, P.~{Carr}, C.~{Jordan},
  and A.~{Wolszczan}, {\it {The Proper Motion, Age, and Initial Spin Period of
  PSR J0538+2817 in S147}},  {\em \apjl} {\bf 593} (Aug., 2003) L31--L34,
  [\href{http://xxx.lanl.gov/abs/astro-ph/0306628}{{\tt astro-ph/0306628}}].

\bibitem{2006AA...457..899A}
F.~{Aharonian}, A.~G. {Akhperjanian}, A.~R. {Bazer-Bachi}, et~al., {\it
  {Observations of the Crab nebula with HESS}},  {\em \aap} {\bf 457} (Oct.,
  2006) 899--915, [\href{http://xxx.lanl.gov/abs/astro-ph/0607333}{{\tt
  astro-ph/0607333}}].

\bibitem{2003AA...408..545W}
B.~Y. {Welsh} and S.~{Sallmen}, {\it {High-velocity NaI and CaII absorption
  components observed towards the IC 443 SNR}},  {\em \aap} {\bf 408} (Sept.,
  2003) 545--551.

\bibitem{2008AJ....135..796L}
J.-J. {Lee}, B.-C. {Koo}, M.~S. {Yun}, S.~{Stanimirovi{\'c}}, C.~{Heiles}, and
  M.~{Heyer}, {\it {A 21 cm Spectral and Continuum Study of IC 443 Using the
  Very Large Array and the Arecibo Telescope}},  {\em \aj} {\bf 135} (Mar.,
  2008) 796--808.

\bibitem{1982AA...109..145G}
D.~A. {Graham}, C.~G.~T. {Haslam}, C.~J. {Salter}, and W.~E. {Wilson}, {\it {A
  continuum study of galactic radio sources in the constellation of
  Monoceros}},  {\em \aap} {\bf 109} (May, 1982) 145--154.

\bibitem{2009AN....330..741B}
V.~{Borka Jovanovi{\'c}} and D.~{Uro{\v s}evi{\'c}}, {\it {The Monoceros radio
  loop: Temperature, brightness, spectral index, and distance}},  {\em
  Astronomische Nachrichten} {\bf 330} (July, 2009) 741,
  [\href{http://xxx.lanl.gov/abs/0904.2261}{{\tt arXiv:0904.2261}}].

\bibitem{2009MNRAS.394.2127B}
C.~{Bonatto} and E.~{Bica}, {\it {Probing the age and structure of the nearby
  very young open clusters NGC2244 and 2239}},  {\em \mnras} {\bf 394} (Apr.,
  2009) 2127--2140, [\href{http://xxx.lanl.gov/abs/0901.0833}{{\tt
  arXiv:0901.0833}}].

\bibitem{2001AA...372..516W}
B.~Y. {Welsh}, D.~M. {Sfeir}, S.~{Sallmen}, and R.~{Lallement}, {\it {Far
  Ultraviolet Spectroscopic Explorer observations of high-velocity gas
  associated with the Monoceros Loop SNR}},  {\em \aap} {\bf 372} (June, 2001)
  516--526.

\bibitem{2006AA...459..535C}
G.~{Castelletti}, G.~{Dubner}, K.~{Golap}, and W.~M. {Goss}, {\it {New VLA
  observations of the SNR Puppis A: the radio properties and the correlation
  with the X-ray emission}},  {\em \aap} {\bf 459} (Nov., 2006) 535--544,
  [\href{http://xxx.lanl.gov/abs/astro-ph/0607200}{{\tt astro-ph/0607200}}].

\bibitem{1988AAS...75..363D}
G.~M. {Dubner} and E.~M. {Arnal}, {\it {Neutral hydrogen and carbon monoxide
  observations towards the SNR Puppis A}},  {\em \aaps} {\bf 75} (Nov., 1988)
  363--369.

\bibitem{1995AJ....110..318R}
E.~M. {Reynoso}, G.~M. {Dubner}, W.~M. {Goss}, and E.~M. {Arnal}, {\it {VLA
  Observations of Neutral Hydrogen in the Direction of Puppis A}},  {\em \aj}
  {\bf 110} (July, 1995) 318.

\bibitem{1988srim.conf...65W}
P.~F. {Winkler}, J.~H. {Tuttle}, R.~P. {Kirshner}, and M.~J. {Irwin}, {\it
  {Kinematics of Oxygen-Rich Filaments in Puppis a}},  in {\em IAU Colloq. 101:
  Supernova Remnants and the Interstellar Medium} (R.~S. {Roger} and T.~L.
  {Landecker}, eds.), p.~65, 1988.

\bibitem{2001AA...372..636A}
H.~{Alvarez}, J.~{Aparici}, J.~{May}, and P.~{Reich}, {\it {The radio spectral
  index of the Vela supernova remnant}},  {\em \aap} {\bf 372} (June, 2001)
  636--643.

\bibitem{1999ApJ...515L..25C}
A.~N. {Cha}, K.~R. {Sembach}, and A.~C. {Danks}, {\it {The Distance to the VELA
  Supernova Remnant}},  {\em \apjl} {\bf 515} (Apr., 1999) L25--L28,
  [\href{http://xxx.lanl.gov/abs/astro-ph/9902230}{{\tt astro-ph/9902230}}].

\bibitem{2001ApJ...561..930C}
P.~A. {Caraveo}, A.~{De Luca}, R.~P. {Mignani}, and G.~F. {Bignami}, {\it {The
  Distance to the Vela Pulsar Gauged with Hubble Space Telescope Parallax
  Observations}},  {\em \apj} {\bf 561} (Nov., 2001) 930--937,
  [\href{http://xxx.lanl.gov/abs/astro-ph/0107282}{{\tt astro-ph/0107282}}].

\bibitem{1993ApJS...88..529T}
J.~H. {Taylor}, R.~N. {Manchester}, and A.~G. {Lyne}, {\it {Catalog of 558
  pulsars}},  {\em \apjs} {\bf 88} (Oct., 1993) 529--568.

\bibitem{2008ApJ...676.1064M}
M.~{Miceli}, F.~{Bocchino}, and F.~{Reale}, {\it {Physical and Chemical
  Inhomogeneities inside the Vela SNR Shell: Indications of Ejecta Shrapnels}},
   {\em \apj} {\bf 676} (Apr., 2008) 1064--1072,
  [\href{http://xxx.lanl.gov/abs/0712.3017}{{\tt arXiv:0712.3017}}].

\bibitem{2008ApJ...678L..35K}
S.~{Katsuda}, H.~{Tsunemi}, and K.~{Mori}, {\it {The Slow X-Ray Expansion of
  the Northwestern Rim of the Supernova Remnant RX J0852.0-4622}},  {\em \apjl}
  {\bf 678} (May, 2008) L35--L38,
  [\href{http://xxx.lanl.gov/abs/0803.3266}{{\tt arXiv:0803.3266}}].

\bibitem{2005MNRAS.356..969R}
M.~P. {Redman} and J.~{Meaburn}, {\it {A possible association of a young pulsar
  (PSR J0855-4644) with the young Vela supernova remnant RX J0852.0-4622}},
  {\em \mnras} {\bf 356} (Jan., 2005) 969--973.

\bibitem{1998Natur.396..141A}
B.~{Aschenbach}, {\it {Discovery of a young nearby supernova remnant}},  {\em
  \nat} {\bf 396} (Nov., 1998) 141--142.

\bibitem{1998Natur.396..142I}
A.~F. {Iyudin}, V.~{Sch{\"o}nfelder}, K.~{Bennett}, H.~{Bloemen}, R.~{Diehl},
  W.~{Hermsen}, G.~G. {Lichti}, R.~D. {van der Meulen}, J.~{Ryan}, and
  C.~{Winkler}, {\it {Emission from $^{44}$Ti associated with a previously
  unknown Galactic supernova}},  {\em \nat} {\bf 396} (Nov., 1998) 142--144.

\bibitem{1997MNRAS.287..722D}
A.~R. {Duncan}, R.~T. {Stewart}, R.~F. {Haynes}, and K.~L. {Jones}, {\it
  {Supernova remnant candidates from the Parkes 2.4-GHz survey}},  {\em \mnras}
  {\bf 287} (June, 1997) 722--738.

\bibitem{2007MNRAS.374.1441S}
M.~{Stupar}, Q.~A. {Parker}, and M.~D. {Filipovi{\'c}}, {\it {G315.1+2.7: a new
  Galactic supernova remnant from the AAO/UKST H{$\alpha$} survey}},  {\em
  \mnras} {\bf 374} (Feb., 2007) 1441--1448,
  [\href{http://xxx.lanl.gov/abs/astro-ph/0612771}{{\tt astro-ph/0612771}}].

\bibitem{2001ApJ...546..447D}
J.~R. {Dickel}, R.~G. {Strom}, and D.~K. {Milne}, {\it {The Radio Structure of
  the Supernova Remnant G315.4-2.3 (MSH 14-63)}},  {\em \apj} {\bf 546} (Jan.,
  2001) 447--454, [\href{http://xxx.lanl.gov/abs/astro-ph/0008042}{{\tt
  astro-ph/0008042}}].

\bibitem{1996AA...315..243R}
M.~{Rosado}, P.~{Ambrocio-Cruz}, E.~{Le Coarer}, and M.~{Marcelin}, {\it
  {Kinematics of the galactic supernova remnants RCW 86, MSH 15-56 and MSH
  11-61A.}},  {\em \aap} {\bf 315} (Nov., 1996) 243--252.

\bibitem{2003AA...407..249S}
J.~{Sollerman}, P.~{Ghavamian}, P.~{Lundqvist}, and R.~C. {Smith}, {\it {High
  resolution spectroscopy of Balmer-dominated shocks in the RCW 86, Kepler and
  SN 1006 supernova remnants}},  {\em \aap} {\bf 407} (Aug., 2003) 249--257,
  [\href{http://xxx.lanl.gov/abs/astro-ph/0306196}{{\tt astro-ph/0306196}}].

\bibitem{1988ApJ...332..940R}
R.~S. {Roger}, D.~K. {Milne}, M.~J. {Kesteven}, K.~J. {Wellington}, and R.~F.
  {Haynes}, {\it {Symmetry of the radio emission from two high-latitude
  supernova remnants, G296.5 + 10.0 and G327.6 + 14.6 (SN 1006)}},  {\em \apj}
  {\bf 332} (Sept., 1988) 940--953.

\bibitem{2006ApJ...640L..55K}
E.~{Kalemci}, S.~E. {Boggs}, P.~A. {Milne}, and S.~P. {Reynolds}, {\it
  {Searching for Annihilation Radiation from SN 1006 with SPI on INTEGRAL}},
  {\em \apjl} {\bf 640} (Mar., 2006) L55--L57,
  [\href{http://xxx.lanl.gov/abs/astro-ph/0602233}{{\tt astro-ph/0602233}}].

\bibitem{2003ApJ...585..324W}
P.~F. {Winkler}, G.~{Gupta}, and K.~S. {Long}, {\it {The SN 1006 Remnant:
  Optical Proper Motions, Deep Imaging, Distance, and Brightness at Maximum}},
  {\em \apj} {\bf 585} (Mar., 2003) 324--335,
  [\href{http://xxx.lanl.gov/abs/astro-ph/0208415}{{\tt astro-ph/0208415}}].

\bibitem{2008PASJ...60S.153B}
A.~{Bamba}, Y.~{Fukazawa}, J.~S. {Hiraga}, J.~P. {Hughes}, H.~{Katagiri},
  M.~{Kokubun}, K.~{Koyama}, E.~{Miyata}, T.~{Mizuno}, K.~{Mori},
  H.~{Nakajima}, M.~{Ozaki}, R.~{Petre}, H.~{Takahashi}, T.~{Takahashi},
  T.~{Tanaka}, Y.~{Terada}, Y.~{Uchiyama}, S.~{Watanabe}, and H.~{Yamaguchi},
  {\it {Suzaku Wide-Band Observations of SN1006}},  {\em \pasj} {\bf 60} (Jan.,
  2008) 153, [\href{http://xxx.lanl.gov/abs/0708.0073}{{\tt arXiv:0708.0073}}].

\bibitem{1991ApJ...374..218L}
D.~A. {Leahy}, J.~{Nousek}, and A.~J.~S. {Hamilton}, {\it {HEAO 1 A-2
  low-energy detector X-ray spectra of the Lupus Loop and SN 1006}},  {\em
  \apj} {\bf 374} (June, 1991) 218--226.

\bibitem{2006ApJ...644L.189S}
J.-H. {Shinn}, K.~W. {Min}, C.-N. {Lee}, J.~{Edelstein}, E.~J. {Korpela}, B.~Y.
  {Welsh}, W.~{Han}, U.-W. {Nam}, H.~{Jin}, and D.-H. {Lee}, {\it {Diffuse
  Far-Ultraviolet Observations of the Lupus Loop Region}},  {\em \apjl} {\bf
  644} (June, 2006) L189--L192,
  [\href{http://xxx.lanl.gov/abs/astro-ph/0604454}{{\tt astro-ph/0604454}}].

\bibitem{2004ApJ...602..271L}
J.~S. {Lazendic}, P.~O. {Slane}, B.~M. {Gaensler}, S.~P. {Reynolds}, P.~P.
  {Plucinsky}, and J.~P. {Hughes}, {\it {A High-Resolution Study of Nonthermal
  Radio and X-Ray Emission from Supernova Remnant G347.3-0.5}},  {\em \apj}
  {\bf 602} (Feb., 2004) 271--285,
  [\href{http://xxx.lanl.gov/abs/astro-ph/0310696}{{\tt astro-ph/0310696}}].

\bibitem{1970ApJ...162L.181S}
C.~S. {Shen}, {\it {Pulsars and Very High-Energy Cosmic-Ray Electrons}},  {\em
  \apjl} {\bf 162} (Dec., 1970) L181.

\bibitem{1987ICRC....2...92H}
A.~K. {Harding} and R.~{Ramaty}, {\it {The Pulsar Contribution to Galactic
  Cosmic Ray Positrons}},  in {\em International Cosmic Ray Conference}, vol.~2
  of {\em International Cosmic Ray Conference}, p.~92, 1987.

\bibitem{1996SSRv...75..235A}
J.~{Arons}, {\it {Pulsars as Gamma-Rays Sources: Nebular Shocks and
  Magnetospheric Gaps}},  {\em \ssr} {\bf 75} (Jan., 1996) 235--255.

\bibitem{2001AA...368.1063Z}
L.~{Zhang} and K.~S. {Cheng}, {\it {Cosmic-ray positrons from mature gamma-ray
  pulsars}},  {\em \aap} {\bf 368} (Mar., 2001) 1063--1070.

\bibitem{Amato:2013fua}
E.~Amato, {\it {The theory of pulsar wind nebulae}},
  \href{http://xxx.lanl.gov/abs/1312.5945}{{\tt arXiv:1312.5945}}.

\bibitem{Ruderman:1975ju}
M.~Ruderman and P.~Sutherland, {\it {Theory of pulsars: Polar caps, sparks, and
  coherent microwave radiation}},  {\em {\aj}} {\bf 196} (1975) 51.

\bibitem{1976ApJ...203..209C}
A.~{Cheng}, M.~{Ruderman}, and P.~{Sutherland}, {\it {Current flow in pulsar
  magnetospheres}},  {\em \apj} {\bf 203} (Jan., 1976) 209--212.

\bibitem{Cheng:1986qt}
K.~Cheng, C.~Ho, and M.~A. Ruderman, {\it {Energetic Radiation from Rapidly
  Spinning Pulsars. 1. Outer Magnetosphere Gaps. 2. Vela and Crab}},  {\em
  {\aj}} {\bf 300} (1986) 500--539.

\bibitem{1974MNRAS.167....1R}
M.~J. {Rees} and J.~E. {Gunn}, {\it {The origin of the magnetic field and
  relativistic particles in the Crab Nebula}},  {\em \mnras} {\bf 167} (Apr.,
  1974) 1--12.

\bibitem{Grasso:2009ma}
{\bf Fermi-LAT} Collaboration, D.~Grasso et~al., {\it {On possible
  interpretations of the high energy electron-positron spectrum measured by the
  Fermi Large Area Telescope}},  {\em Astropart.Phys.} {\bf 32} (2009)
  140--151, [\href{http://xxx.lanl.gov/abs/0905.0636}{{\tt arXiv:0905.0636}}].

\bibitem{1996MNRAS.278..525A}
A.~M. {Atoyan} and F.~A. {Aharonian}, {\it {On the mechanisms of gamma
  radiation in the Crab Nebula}},  {\em \mnras} {\bf 278} (Jan., 1996)
  525--541.

\bibitem{2009PhRvD..80f3005M}
D.~{Malyshev}, I.~{Cholis}, and J.~{Gelfand}, {\it {Pulsars versus dark matter
  interpretation of ATIC/PAMELA}},  {\em \prd} {\bf 80} (Sept., 2009) 063005,
  [\href{http://xxx.lanl.gov/abs/0903.1310}{{\tt arXiv:0903.1310}}].

\bibitem{Aharonian:2006xx}
F.~{Aharonian}, A.~G. {Akhperjanian}, A.~R. {Bazer-Bachi}, M.~{Beilicke},
  W.~{Benbow}, D.~{Berge}, K.~{Bernl{\"o}hr}, C.~{Boisson}, O.~{Bolz},
  V.~{Borrel}, I.~{Braun}, F.~{Breitling}, A.~M. {Brown}, R.~{B{\"u}hler},
  I.~{B{\"u}sching}, S.~{Carrigan}, P.~M. {Chadwick}, L.-M. {Chounet},
  R.~{Cornils}, L.~{Costamante}, B.~{Degrange}, H.~J. {Dickinson},
  A.~{Djannati-Ata{\"i}}, L.~{O'C.~Drury}, G.~{Dubus}, K.~{Egberts},
  D.~{Emmanoulopoulos}, B.~{Epinat}, P.~{Espigat}, F.~{Feinstein},
  E.~{Ferrero}, G.~{Fontaine}, S.~{Funk}, S.~{Funk}, Y.~A. {Gallant},
  B.~{Giebels}, J.~F. {Glicenstein}, P.~{Goret}, C.~{Hadjichristidis},
  D.~{Hauser}, M.~{Hauser}, G.~{Heinzelmann}, G.~{Henri}, G.~{Hermann}, J.~A.
  {Hinton}, W.~{Hofmann}, M.~{Holleran}, D.~{Horns}, A.~{Jacholkowska}, O.~C.
  {de Jager}, B.~{Kh{\'e}lifi}, N.~{Komin}, A.~{Konopelko}, I.~J. {Latham},
  R.~{Le Gallou}, A.~{Lemi{\`e}re}, M.~{Lemoine-Goumard}, T.~{Lohse}, J.~M.
  {Martin}, O.~{Martineau-Huynh}, A.~{Marcowith}, C.~{Masterson}, T.~J.~L.
  {McComb}, M.~{de Naurois}, D.~{Nedbal}, S.~J. {Nolan}, A.~{Noutsos}, K.~J.
  {Orford}, J.~L. {Osborne}, M.~{Ouchrif}, M.~{Panter}, G.~{Pelletier},
  S.~{Pita}, G.~{P{\"u}hlhofer}, M.~{Punch}, B.~C. {Raubenheimer}, M.~{Raue},
  S.~M. {Rayner}, A.~{Reimer}, O.~{Reimer}, J.~{Ripken}, L.~{Rob},
  L.~{Rolland}, G.~{Rowell}, V.~{Sahakian}, L.~{Saug{\'e}}, S.~{Schlenker},
  R.~{Schlickeiser}, U.~{Schwanke}, H.~{Sol}, D.~{Spangler}, F.~{Spanier},
  R.~{Steenkamp}, C.~{Stegmann}, G.~{Superina}, J.-P. {Tavernet}, R.~{Terrier},
  C.~G. {Th{\'e}oret}, M.~{Tluczykont}, C.~{van Eldik}, G.~{Vasileiadis},
  C.~{Venter}, P.~{Vincent}, H.~J. {V{\"o}lk}, S.~J. {Wagner}, and M.~{Ward},
  {\it {First detection of a VHE gamma-ray spectral maximum from a cosmic
  source: HESS discovery of the Vela X nebula}},  {\em \aap} {\bf 448} (Mar.,
  2006) L43--L47, [\href{http://xxx.lanl.gov/abs/astro-ph/0601575}{{\tt
  astro-ph/0601575}}].

\bibitem{Atoyan11011996}
A.~M. Atoyan and F.~A. Aharonian, {\it On the mechanisms of gamma radiation in
  the crab nebula},  {\em \mnras} {\bf 278} (1996), no.~2 525--541,
  [\href{http://xxx.lanl.gov/abs/http://mnras.oxfordjournals.org/content/278/2/525.full.pdf+html}{{\tt
  http://mnras.oxfordjournals.org/content/278/2/525.full.pdf+html}}].

\bibitem{TheFermi-LAT:2013ssa}
{\bf Fermi-LAT} Collaboration, A.~Abdo et~al., {\it {The Second Fermi Large
  Area Telescope Catalog of Gamma-ray Pulsars}},  {\em {\apjs}} {\bf 208}
  (2013) 17, [\href{http://xxx.lanl.gov/abs/1305.4385}{{\tt arXiv:1305.4385}}].

\bibitem{Blasi:2010de}
P.~Blasi and E.~Amato, {\it {Positrons from pulsar winds}},
  \href{http://xxx.lanl.gov/abs/1007.4745}{{\tt arXiv:1007.4745}}.

\bibitem{proton_AMS02}
H.~{Haino} and {the Ams-02 Collaboration}, {\it {Precision measurement of the
  proton flux with AMS}},  {\em Talk at the 33rd ICRC Conference} (2013).

\bibitem{helium_AMS02}
{V. Choutko} and {the Ams-02 Collaboration}, {\it {Precision measurement of the
  helium flux with AMS}},  {\em Talk at the 33rd ICRC Conference} (2013).

\bibitem{2006ApJ...647..692K}
T.~{Kamae}, N.~{Karlsson}, T.~{Mizuno}, T.~{Abe}, and T.~{Koi}, {\it
  {Parameterization of {$\gamma$}, e$^{+/-}$, and Neutrino Spectra Produced by
  p-p Interaction in Astronomical Environments}},  {\em \apj} {\bf 647} (Aug.,
  2006) 692--708, [\href{http://xxx.lanl.gov/abs/astro-ph/0605581}{{\tt
  astro-ph/0605581}}].

\bibitem{2013arXiv1303.6482D}
C.~D. {Dermer}, J.~D. {Finke}, R.~J. {Murphy}, A.~W. {Strong}, F.~{Loparco},
  M.~N. {Mazziotta}, E.~{Orlando}, T.~{Kamae}, L.~{Tibaldo}, J.~{Cohen-Tanugi},
  M.~{Ackermann}, T.~{Mizuno}, and F.~W. {Stecker}, {\it {On the Physics
  Connecting Cosmic Rays and Gamma Rays: Towards Determining the Interstellar
  Cosmic Ray Spectrum}},  {\em ArXiv e-prints} (Mar., 2013)
  [\href{http://xxx.lanl.gov/abs/1303.6482}{{\tt arXiv:1303.6482}}].

\bibitem{2013arXiv1307.0497D}
C.~D. {Dermer}, A.~W. {Strong}, E.~{Orlando}, L.~{Tibaldo}, and {for the Fermi
  Collaboration}, {\it {Determining the Spectrum of Cosmic Rays in Interstellar
  Space from the Diffuse Galactic Gamma-Ray Emissivity}},  {\em ArXiv e-prints}
  (July, 2013) [\href{http://xxx.lanl.gov/abs/1307.0497}{{\tt
  arXiv:1307.0497}}].

\bibitem{2001ApJ...555..585M}
D.~{Maurin}, F.~{Donato}, R.~{Taillet}, and P.~{Salati}, {\it {Cosmic Rays
  below Z=30 in a Diffusion Model: New Constraints on Propagation Parameters}},
   {\em \apj} {\bf 555} (July, 2001) 585--596,
  [\href{http://xxx.lanl.gov/abs/astro-ph/0101231}{{\tt astro-ph/0101231}}].

\bibitem{2004PhRvD..69f3501D}
F.~{Donato}, N.~{Fornengo}, D.~{Maurin}, P.~{Salati}, and R.~{Taillet}, {\it
  {Antiprotons in cosmic rays from neutralino annihilation}},  {\em \prd} {\bf
  69} (Mar., 2004) 063501,
  [\href{http://xxx.lanl.gov/abs/astro-ph/0306207}{{\tt astro-ph/0306207}}].

\bibitem{1971JGR....76..221F}
L.~A. {Fisk}, {\it {Solar modulation of galactic cosmic rays, 2}},  {\em \jgr}
  {\bf 76} (1971) 221.

\bibitem{1987AA...184..119P}
J.~S. {Perko}, {\it {Solar modulation of galactic antiprotons}},  {\em \aap}
  {\bf 184} (Oct., 1987) 119--121.

\bibitem{Maccione:2012cu}
L.~Maccione, {\it {Low energy cosmic ray positron fraction explained by
  charge-sign dependent solar modulation}},  {\em Phys.Rev.Lett.} {\bf 110}
  (2013), no.~8 081101, [\href{http://xxx.lanl.gov/abs/1211.6905}{{\tt
  arXiv:1211.6905}}].

\bibitem{DellaTorre:2012zz}
S.~Della~Torre, P.~Bobik, M.~J. Boschini, C.~Consolandi, M.~Gervasi, et~al.,
  {\it {Effects of solar modulation on the cosmic ray positron fraction}},
  {\em Adv.Space Res.} {\bf 49} (2012) 1587--1592.

\bibitem{2011A&A...534A..54S}
A.~W. {Strong}, E.~{Orlando}, and T.~R. {Jaffe}, {\it {The interstellar
  cosmic-ray electron spectrum from synchrotron radiation and direct
  measurements}},  {\em \aap} {\bf 534} (Oct., 2011) A54,
  [\href{http://xxx.lanl.gov/abs/1108.4822}{{\tt arXiv:1108.4822}}].

\bibitem{2013MNRAS.436.2127O}
E.~{Orlando} and A.~{Strong}, {\it {Galactic synchrotron emission with cosmic
  ray propagation models}},  {\em \mnras} {\bf 436} (Dec., 2013) 2127--2142,
  [\href{http://xxx.lanl.gov/abs/1309.2947}{{\tt arXiv:1309.2947}}].

\bibitem{2004PhRvL..93x1102B}
J.~J. {Beatty}, A.~{Bhattacharyya}, C.~{Bower}, S.~{Coutu}, et~al., {\it {New
  Measurement of the Cosmic-Ray Positron Fraction from 5 to 15GeV}},  {\em
  Phys.Rev.Lett.} {\bf 93} (Dec., 2004) 241102,
  [\href{http://xxx.lanl.gov/abs/astro-ph/0412230}{{\tt astro-ph/0412230}}].

\bibitem{1998ApJ...498..779B}
S.~W. {Barwick}, J.~J. {Beatty}, C.~R. {Bower}, et~al., {\it {The Energy
  Spectra and Relative Abundances of Electrons and Positrons in the Galactic
  Cosmic Radiation}},  {\em \apj} {\bf 498} (May, 1998) 779,
  [\href{http://xxx.lanl.gov/abs/astro-ph/9712324}{{\tt astro-ph/9712324}}].

\bibitem{1997ApJ...482L.191B}
S.~W. {Barwick}, J.~J. {Beatty}, A.~{Bhattacharyya}, and {HEAT Collaboration},
  {\it {Measurements of the Cosmic-Ray Positron Fraction from 1 to 50 GeV}},
  {\em \apjl} {\bf 482} (June, 1997) L191,
  [\href{http://xxx.lanl.gov/abs/astro-ph/9703192}{{\tt astro-ph/9703192}}].

\bibitem{2001ApJ...559..296D}
M.~A. {DuVernois}, S.~W. {Barwick}, et~al., {\it {Cosmic-Ray Electrons and
  Positrons from 1 to 100 GeV: Measurements with HEAT and Their
  Interpretation}},  {\em \apj} {\bf 559} (Sept., 2001) 296--303.

\bibitem{2000ApJ...532..653B}
M.~{Boezio}, P.~{Carlson}, T.~{Francke}, et~al., {\it {The Cosmic-Ray Electron
  and Positron Spectra Measured at 1 AU during Solar Minimum Activity}},  {\em
  \apj} {\bf 532} (Mar., 2000) 653--669.

\bibitem{2001AdSpR..27..669B}
M.~{Boezio}, G.~{Barbiellini}, et~al., {\it {Measurements of cosmic-ray
  electrons and positrons by the Wizard/CAPRICE collaboration}},  {\em Advances
  in Space Research} {\bf 27} (2001) 669--674.

\bibitem{2008AdSpR..42.1670Y}
K.~{Yoshida}, S.~{Torii}, T.~{Yamagami}, et~al., {\it {Cosmic-ray electron
  spectrum above 100 GeV from PPB-BETS experiment in Antarctica}},  {\em
  Advances in Space Research} {\bf 42} (Nov., 2008) 1670--1675.

\bibitem{2001ApJ...559..973T}
S.~{Torii}, T.~{Tamura}, N.~{Tateyama}, et~al., {\it {The Energy Spectrum of
  Cosmic-Ray Electrons from 10 to 100 GeV Observed with a Highly Granulated
  Imaging Calorimeter}},  {\em \apj} {\bf 559} (Oct., 2001) 973--984.

\bibitem{2009AA...508..561A}
F.~{Aharonian}, A.~G. {Akhperjanian}, et~al., {\it {Probing the ATIC peak in
  the cosmic-ray electron spectrum with H.E.S.S.}},  {\em \aap} {\bf 508}
  (Dec., 2009) 561--564, [\href{http://xxx.lanl.gov/abs/0905.0105}{{\tt
  arXiv:0905.0105}}].

\bibitem{2005JGRA..11012108U}
I.~G. {Usoskin}, K.~{Alanko-Huotari}, G.~A. {Kovaltsov}, and K.~{Mursula}, {\it
  {Heliospheric modulation of cosmic rays: Monthly reconstruction for
  1951-2004}},  {\em Journal of Geophysical Research (Space Physics)} {\bf 110}
  (Dec., 2005) 12108.

\bibitem{2011JGRA..116.2104U}
I.~G. {Usoskin}, G.~A. {Bazilevskaya}, and G.~A. {Kovaltsov}, {\it {Solar
  modulation parameter for cosmic rays since 1936 reconstructed from
  ground-based neutron monitors and ionization chambers}},  {\em Journal of
  Geophysical Research (Space Physics)} {\bf 116} (Feb., 2011) 2104.

\bibitem{Yuksel:2008rf}
H.~Yuksel, M.~D. Kistler, and T.~Stanev, {\it {TeV Gamma Rays from Geminga and
  the Origin of the GeV Positron Excess}},  {\em Phys.Rev.Lett.} {\bf 103}
  (2009) 051101, [\href{http://xxx.lanl.gov/abs/0810.2784}{{\tt
  arXiv:0810.2784}}].

\bibitem{Cholis:2013psa}
I.~Cholis and D.~Hooper, {\it {Dark Matter and Pulsar Origins of the Rising
  Cosmic Ray Positron Fraction in Light of New Data From AMS}},  {\em
  Phys.Rev.} {\bf D88} (2013) 023013,
  [\href{http://xxx.lanl.gov/abs/1304.1840}{{\tt arXiv:1304.1840}}].

\end{thebibliography}\endgroup

\end{document}